# Comparison of water models for structure prediction

Bálint Soczó[1,2] and Ildikó Pethes[1,*]

[1] HUN-REN Wigner Research Centre for Physics, Konkoly Thege út 29-33., H-1121 Budapest, Hungary

[2] Faculty of Natural Sciences, Budapest University of Technology and Economics, Műegyetem rkp. 3., H-1111 Budapest, Hungary

[*] Corresponding author, e-mail address: pethes.ildiko@wigner.hun-ren.hu

Abstract

Describing the interactions of water molecules is one of the most common, yet critical, tasks in molecular dynamics simulations. Because of its unique properties, hundreds of attempts have been made to construct an ideal interaction potential model for water. In various studies, the models have been evaluated based on their ability to reproduce different properties of water. This work focuses on the atomic-scale structure in the liquid phase of water. Forty-four classical water potential models are compared to identify those that can accurately describe the structure in alignment with experimental results. In addition to some older models that are still popular today, new or re-parametrized classical models using effective pair-additive potentials that have appeared in recent years are examined.

Molecular dynamics simulations were performed over a wide range of temperatures and the resulting trajectories were used to calculate the partial radial distribution functions. The total scattering structure factors were compared with data from neutron and X-ray diffraction experiments. Our analysis indicates that models with more than four interaction sites, as well as flexible or polarizable models with higher computational requirements, do not provide a significant advantage in accurately describing the structure. On the other hand, recent three-site models have made considerable progress in this area, although the best agreement with experimental data over the entire temperature range was achieved with four-site, TIP4P-type models.

# 1. Introduction

Classical molecular dynamics (MD) simulations serve as an essential tool in investigating the behavior of complex systems at the atomic and molecular levels. By numerically solving Newton's equations of motion for a collection of interacting particles, MD simulations offer valuable insights into a wide range of phenomena.

Water, as the most essential and ubiquitous substance on Earth, is also one of the most frequently studied molecules in MD simulations. Understanding the properties and behavior of water at the molecular level is not only crucial in clarifying scientific principles, but also has practical implications in various disciplines, including biology, chemistry, materials science, and environmental science.

At the heart of MD simulations lies the choice of the potential model, which dictates the intermolecular interactions between atoms or molecules in the system. Selecting the appropriate potential model is crucial, since it directly affects the accuracy and reliability of the simulation results.

Since water has been one of the earliest targets of MD simulations, it is likely that more potential models have been developed for water molecules than for any other substance. These models range from simple empirical effective pair potentials to sophisticated, complex, *ab initio*-based approaches. However, it has not been successful (perhaps not even possible) to create such a model that that simultaneously reflects all of the unique properties (and anomalies) of water. Therefore, researchers need to find the model(s) most suitable for the subject of their research, assess and validate the suitability of the selected potential model to ensure meaningful and predictive simulations.

To facilitate this, numerous studies have been carried out over recent decades comparing water potential models from different perspectives. The first reviews, published 20-25 years ago, outlined the theoretical foundations, the milestones of modeling, the various types of models, and the results they can produce [1-5]. During the decades, new models and new families of models have emerged and collected in review works [6-16].

Several papers focus on testing popular pair-additive, rigid, and/or non-polarizable models. These studies compare simulation results with experimental values, evaluate the performance of the models, and rank their suitability for specific applications. For example, Refs. [17-25] provide such analyses. In many cases, previous models have been re-optimized using more recent experimental values, or over a wider range of parameters. More recently, artificial intelligence/machine learning techniques have been employed for such re-optimizations, as seen in Refs. [26-30]. For instance, the TIP4P-BGWT model [27] exhibits excellent predictive performance for the physical properties of stable and supercooled water. The TIP5P-BG and TIP5P-BGT models [30] demonstrate accurate and comprehensive performance for the physical properties of ice and liquid water. Additionally, reviews of advanced models, such as polarizable, many-body, machine learning-based models, and ab initio (density functional theory, Car-Parrinello MD) water studies can be found e.g., in Refs. [8, 9, 12, 14, 31-33].

In these studies, various properties (mainly thermodynamic) are examined along with the model parameters. These properties include density, the temperature of the maximum density, melting temperature, vapor-liquid coexistence properties (such as vaporization enthalpy, surface tension), static dielectric constant, self-diffusion, viscosity, compressibility, specific heat, heat capacity, etc. The target values were also selected from these properties when the models were developed.

The structure of liquids is typically analyzed using partial radial distribution functions (PRDFs) denoted as $g_{ij}(r)$), which represent the ratio of the local density to the average density of atoms of type $j$ at a distance $r$ from an atom of type $i$. In most cases, the comparison of potential models in terms of structural properties has been limited to studying the oxygen-oxygen PRDFs, mainly the position (nearest neighbor distances) and height of the first peak, or the coordination numbers [8,17,22,24,29,32,34-50]. In some studies, all PRDFs (O-O, O-H, and H-H) have been compared with (so-called) experimental curves [15,28,33,51-56].

Structural properties are often excluded from the set of quantities that are fitted during the development of potential models. In the last decade, a few potential models have been developed in which the target quantities include the oxygen-oxygen first neighbor distance (e.g., Refs. [50,57]) and also the height of the 1$^{st}$ peak of $g_{OO}(r)$ [29]. The effect of a slight modification of the potential parameters of the SPC model [58] on the fit of the 'experimental' PRDFs was investigated in Ref. [59]. It was found, that only marginal improvements could be achieved. In the models of Perrone et al. [28], Wade et al. [60], and Jiang et al. [56], the 'experimental' $g_{OO}(r)$ itself (as a microscopic property) was also included in the quantities to be fitted.

However, the partial radial distribution functions cannot be measured directly. The $g_{ij}(r)$ curves, used as 'experimental' reference, were usually derived from the total scattering structure factor (TSSF, $S(Q)$) using various theoretical and simulation techniques [61-66]. The TSSF can be measured in neutron (ND) or X-ray diffraction (XRD) experiments. The relationship between the PRDFs and the TSSF is given by the following equations: (The TSSF is the weighted sum of the Fourier-transformed PRDFs:)

$$S(Q) = \sum_{i \leq j} w_{ij}^{X,N}(Q) \left(1 + \frac{4\pi\rho}{Q} \int_0^\infty r(g_{ij}(r) - 1)\sin(Qr)dr\right) \quad (1)$$

where $Q$ is the amplitude of the scattering vector, $\rho$ is the average number density and $w_{ij}^X(Q)$ and $w_{ij}^N$ are the weighting parameters for X-ray and neutron scattering (depending on the concentrations and scattering properties of the constituents).

It follows from the above that at least 3 independent TSSFs are needed to determine the 3 PRDFs of water. It has been shown previously that more than one PRDF set can be equally consistent with a given TSSF set containing different isotopic ratios of $^{1}$H and $^{2}$H (D) in water [67]. Therefore, PRDFs should be only regarded as an interpretation of the measured data. TSSFs can provide a suitable experimental data set for comparing microscopic structures.

The most commonly used 'experimental' $g_{ij}(r)$ set [63] has been derived from TSSFs obtained in ND measurements by using the Empirical Potential Structure Refinement (EPSR) method. Steinczinger and Pusztai [68] investigated this PRDF set and found that it is consistent with the most reliable neutron diffraction data set; however, the $g_{OO}(r)$ of Ref. [63] is not entirely in agreement with the results of X-ray diffraction measurements.

TSSFs from ND measurements were the only target properties of the potential model in Ref. [69]. The author demonstrated the possibility of creating a pairwise additive intermolecular potential consistent with experimental data. Unfortunately, this rather complex potential model, which involves numerous parameters, has not been tested for thermodynamic properties. Furthermore, it was observed that the published experimental TSSFs and the generated PRDFs can only be fitted together if the intramolecular structure of water molecules is allowed to be unrealistically 'elastic' [70].

The experimental and simulated TSSFs were compared in a couple of publications. Some of them deal with only a few potential models, e.g., [25,71-78]. Several different potential models are analyzed in Refs. [79-82].

The intensity curves measured in the X-ray scattering experiments were compared with the curves calculated for 14 water models in Ref. [79]. The authors examined the non-polarizable models of the TIPnP family (TIP3P [83], TIP4P [83], TIP5P [84], TIP5P-Ew [44]), and found that they have shown steady improvement over the years. The polarizable TIP4P-Pol2 [85] model has been found to be the best of all the models tested over the 275 – 350 K temperature range investigated.

The compatibility of potential models and ND data was investigated for 8 water models in Ref. [80]. The authors attempted to simultaneously fit PRDFs available in the literature and experimental TSSFs using the reverse Monte Carlo [86] technique. They have found that although the models tested were quite different, none of them was in fatal disagreement with the measured TSSF.

This latter work was followed by Ref [77] by studying two additional models with the same technique. In this study, besides the TSSF of heavy water from ND, the data from the XRD experiment was also tested. While the neutron diffraction data is more sensitive to the O-H and H-H (or more precisely, the O-D and D-D) correlations, the TSSF from XRD measurement contains mainly the contribution of the O-O partial. Thus, the combination of the two types of experimental data provides a more rigorous testing method.

The comparison of their full set of potential models (10 models) with both types of data was made in two additional publications [81,82]. It was concluded that the best models among the tested are the BK3 [87] and TIP4P/2005 [45].

The water models examined in the above mentioned studies were mainly developed between 20 and 50 years ago, excluding the BK3 model. Since then, many new models have emerged, including simple classical models that use effective pair potentials and require minimal computational resources. However, their consistency with experimental TSSFs has not been investigated.

The aim of this study is to complement the previous studies by examining newer potential models published in the last fifteen years and extending the range of temperatures investigated. Classical molecular simulations were performed with 44 different water models with effective pair potentials. In 42 of these, the van der Waals forces are described by the 12-6 Lennard-Jones (LJ) potential. Two additional models using the Buckingham potential were also examined. Most of the models tested are rigid and non-polarizable, but some flexible and polarizable models were also studied. Some of the most widely used older models were also tested for comparison. The resulting trajectories were used to calculate the PRDFs and the neutron and X-ray weighted TSSFs.

Additionally, density and self-diffusion coefficients were also calculated and the obtained values were compared with those found in the literature to verify our simulations. The properties were tested at various temperatures within the liquid range as well as at lower temperatures, as many models have considerably lower freezing points. The models were validated by comparing the calculated TSSFs with experimental data. The PRDFs of different models were also compared. Models that perform well over the full temperature range have been identified.

## 2. Methods

*2.1 Molecular dynamics simulations*

Classical molecular dynamics simulations were conducted using the GROMACS software (version 2020.5) [88]. Unless otherwise stated, the same simulation protocol was applied to all tested potential models. The applied parameters were either typical for MD simulations or default values of the software. Although these parameters may differ from the ones used by the model developers, these differences lead to only small discrepancies. The partial radial distribution functions and structure factors under investigation are essentially insensitive to the parameters of the simulation [89].

Cubic simulation boxes with periodic boundary conditions were used. Initially, 4000 water molecules were randomly placed into the simulation box. To facilitate energy minimization, the initial box length was set to 5 nm (7 nm for the TIP7P and SWM6 models). Coulomb interactions were treated using the smoothed particle-mesh Ewald method (SPME) [90,91], using a 1 nm cutoff in direct space. Van der Waals interactions were truncated at 1 nm, with additional long-range corrections to energy and pressure [92]. For the SWM3, SWM4-HLJ, SWM4-dip, and SWM4-HD models, Lennard-Jones interactions were also treated using the SPME method.

The energy minimization process utilized the steepest descent method. It was carried out in two steps: initially, all molecules, including rigid ones, were allowed to be flexible to expedite energy minimization. From the second energy minimization step onward, rigid molecules remained constrained using the SETTLE algorithm [93].

Following energy minimization, the leapfrog algorithm was used to integrate the equations of motion during equilibration and production runs. The time step was typically set to 2 fs, except for flexible and polarizable models, where it was 0.5 fs and 1 fs, respectively.

The equilibration and production runs contained several steps, collected in Table 1.

Equilibration began at $T$ = 366 K and was done in 2 steps the system was equilibrated in the NVT ensemble (constant volume and temperature) for 0.1 ns; then, the NpT ensemble (constant pressure and temperature) was applied for 2.0 ns. The Nose-Hoover thermostat [94,95] was used with a $\tau_T$ = 1 ps time constant to maintain the set temperature, while in NpT simulations the Parrinello-Rahman barostat [96,97] was applied with $\tau_p$ = 2.0 ps coupling constant. After the high-temperature equilibration, the box was cooled down in several steps using the simulated annealing procedure of GROMACS, applying the NpT ensemble. In this procedure 1 ns long cooling phases (from $T_i$ to $T_{i+1}$) and 1 ns long equilibration steps (at $T_{i+1}$) were alternated. The final configurations of the equilibration steps were used later as starting configuration in the production run at $T_{i+1}$. The investigated temperatures ($T_i$) were: 366, 343, 324, 295, 284, 268, and 254 K. The density at each temperature was determined from the NpT equilibration steps. At each investigated temperature, a 1 ns long NVT production run was carried out, in which the trajectories were collected at every 10 ps.

*2.2 Calculation of the PRDFs and TSSFs*

Configurations spaced 10 ps apart (101 configurations in total) were used for calculating the PRDFs, using the 'gmx rdf' program of the GROMACS package. The TSSFs were calculated using Eq. (1), where the weighting parameters were:

$$w_{ij}^N = (2 - \delta_{ij}) \frac{c_i c_j b_i b_j}{\sum_{ij} c_i c_j b_i b_j} \qquad (2)$$

for neutron and

$$w_{ij}^X(Q) = (2 - \delta_{ij}) \frac{c_i c_j f_i(Q) f_j(Q)}{\sum_{ij} c_i c_j f_i(Q) f_j(Q)} \qquad (3)$$

for X-ray diffraction data. Here $\delta_{ij}$ is the Kronecker delta, $c_i$ represents the atomic concentration, $f_i(Q)$ denotes the atomic form factor (taken from Refs. [98,99]), $b_i$ is the coherent neutron scattering length (from Ref. [100]). The neutron and X-ray weighting parameters of water are shown in Fig. 1.

Since heavy water data provides the lowest uncertainty in neutron diffraction measurements, the corresponding structure functions were used with the appropriate weights.

The X-ray scattering intensities ($I^X(Q)$) were taken from Ref. [101], for the 254 – 366 K temperature range. The X-ray TSSF was calculated from the intensities according to Eq. (4):

$$S^X(Q) - 1 = \frac{I^X(Q) - \sum_i c_i f_i^2(Q)}{\sum_{ij} c_i c_j f_i(Q) f_j(Q)} \quad (4)$$

Experimental neutron TSSFs (of heavy water) were taken from three publications, Refs [65, 69, 102].

The agreement between measured and simulated TSSFs was quantified by calculating the *R*-factors (goodness of fit values) by:

$$R = \frac{\sqrt{\sum_i \left(S_{\text{mod}}(Q_i) - S_{\text{exp}}(Q_i)\right)^2}}{\sqrt{\sum_i S_{\text{exp}}^2(Q_i)}} \quad (5)$$

Here, 'mod' denotes the model and 'exp' the experimental curves, while $Q_i$ are the experimental points. The *R*-factors were calculated up to 18 Å$^{-1}$.

*2.3 Calculation of the density, self-diffusion coefficient*

Densities were calculated at each investigated temperature from the 1ns NpT equilibration run, based on the average of 101 values.

The self-diffusion coefficients were calculated from the mean-square displacement (MSD) using the Einstein relationship:

$$6Dt = \lim_{t \to \infty} \langle \|r_i(t) - r_i(0)\|^2 \rangle \quad (6)$$

The full trajectories from the (NVT) production run were used by restarting the MSD calculation every 10 ps. The linear fits of the obtained MSD-t curves were calculated in the 100 – 900 ps region. (The 'gmx msd' program of the GROMACS software was used.)

Accurately determining the self-diffusion coefficient requires additional corrections, such as accounting for finite-size effects through simulations with varying box sizes or viscosity calculations [89,103]. In this work, no finite-size corrections were applied; only approximate values were calculated. Moreover, more configurations (longer trajectories) should be used to get smaller uncertainty. The correct self-diffusion values will be investigated in a following publication. However, these approximate values serve as a useful reference for monitoring molecular mobility and confirming the independence of the tested configurations.

3. Pair potential models

*3.1 Interaction potentials, virtual sites*

The investigated potential models are listed in Table 2. Their potential parameters are provided in Tables S1 – S3 in the Supplementary Material (SM).

For all the models studied, electrostatic interactions were described by the Coulomb potential using point-like partial charges (Eq. 7):

$$V_{ij}^{C}(r_{ij}) = \frac{1}{4\pi\varepsilon_0} \frac{q_i q_j}{r_{ij}}, \tag{7}$$

where $r_{ij}$ is the distance between particles $i$ and $j$, $q_i$, and $q_j$ are their partial charges, and $\varepsilon_0$ is the vacuum permittivity.

Van der Waals interactions (dispersion and repulsion effects) were described either by the Buckingham potential (Eq. 8) for the TIP3P-Buck and TIP4P-Buck models or by the 12-6 Lennard-Jones (LJ) potential (Eq. 9) for all other models:

$$V_{ij}^{B}(r_{ij}) = A_{ij}\exp(-B_{ij}r_{ij}) - \frac{C_{ij}}{r_{ij}^6}, \tag{8}$$

$$V_{ij}^{LJ}(r_{ij}) = 4\varepsilon_{ij}\left[\left(\frac{\sigma_{ij}}{r_{ij}}\right)^{12} - \left(\frac{\sigma_{ij}}{r_{ij}}\right)^{6}\right], \tag{9}$$

where $\varepsilon_{ij}$ and $\sigma_{ij}$ represent the energy and distance parameters of the LJ potential, $A_{ij}$, $B_{ij}$, and $C_{ij}$ are the corresponding parameters of the Buckingham potential.

In most models considered here, van der Waals interactions are only taken into account between oxygen atoms ($i = j =$ O). For those models in which the LJ parameters of the hydrogen atom ($\varepsilon_{HH}$ and $\sigma_{HH}$) are non-zero, the $\varepsilon_{OH}$ and $\sigma_{OH}$ values are computed from the parameters of the like-pairs using geometric combination rules for TIP7P, and Lorentz-Berthelot for the other models.

The flexible molecules are kept together with so-called bonded potentials. The investigated models use harmonic potential for angle bending:

$$V_a(\theta_{ijk}) = \frac{1}{2}k_\theta(\theta_{ijk} - \theta_{HOH})^2. \tag{10}$$

Here $k_\theta$ is the force constant, and $\theta_{HOH}$ is the equilibrium H-O-H angle.

For bond stretching, the SPC/Fw [107], FBA/ε [41], and TIP4P/ε$_{Flex}$ [116] models apply harmonic potential (eq. 11) while the TIP4P/2005f [46] model uses a Morse potential (eq. 12)

$$V_b(r_{ij}) = \frac{1}{2}k_b(r_{ij} - d_{OH})^2 \tag{11}$$

and

$$V_b(r_{ij}) = D_r\{1 - \exp[-\beta(r_{ij} - d_{OH})]\}^2 \tag{12}$$

Here $d_{OH}$ is the equilibrium distance of the bonded O and H atoms, $k_b$ is the force constant, $D_r$ and $\beta$ are the parameters of the Morse potential that determine the bond strength and width.

The water potential models tested here consist of 3 – 7 interaction sites. 3 of them are the H and O atoms, and the others are virtual sites on which the charges of the atoms are partially or fully relocated.

Some models utilize simple virtual sites, whose positions are rigidly defined by geometric constraints, while others incorporate a special type of virtual site known as a shell particle. The position of the shell particles (Drude particles) is flexible as they are attached to the oxygen atom by a harmonic spring. These particles are responsible for the electronic polarizability of the model.

The LJ-parameters of the virtual sites are always zero. These particles are massless, except the shell particle of the OPC3-pol model [50]. In this model, the oxygen mass is split equally between the oxygen atom and the Drude particle. Moreover, the Drude particle is treated as an ordinary atom, with a flexible bond between it and the oxygen atom.

The typical geometries of these models are shown in Fig. 2.

*3.2 Three-site models*

In the case of the 3-site models, the H and O atoms constitute the 3 interaction sites, and the charges are assigned directly to them. In addition to the rather old, but still widely used, rigid, non-polarizable SPC [58], SPC/E [104], and TIP3P [83] models, a number of re-parametrized versions from the last decade were also tested: TIP3P-FB [49], SPC/ε [42], OPC3 [112], TIP3P-LJOPT [60], TIP3P-ST [115], ML-TIP3P [114], OPTI-1T and OPTI-3T [28]. A 3-site model (TIP3P-Buck [60]) using the Buckingham potential was also investigated.

Two flexible models, SPC/Fw [107] and FBA/ε [41], and two polarizable models, OPC3-pol [50] and SWM3 [118] (which include an additional shell particle) were also tested.

*3.3 Four-site models*

In the 4-site models, the charge of the oxygen atom is placed either fully or partially into a virtual site (denoted as M), which lies in the plane of the water molecule on the bisector of the H-O-H angle. The archetypal member of the TIP4 family is the TIP4P [83] model itself, which was re-parametrized several times. The twenty years old TIP4P-Ew [76] and TIP4P/2005 [45] models are still among the most popular 4-site models.

Several newer models, based on TIP4P, were also tested: TIP4Q [48], HUANG [109,110], OPC [57], TIP4P/ε [40], TIP4P-FB [49], TIP4P-D [111], TIP4P-LJOPT [60], TIP4P-ST [115], TIP4P-BG [35], and ECCw2024 [29]. An additional rigid, non-polarizable model, the TIP4P-Buck [60] model was also examined.

Besides the above listed rigid, non-polarizable models, two flexible models – TIP4P/2005f [46] and TIP4P/ε$_{Flex}$ [116] – were tested. Additionally, 5 polarizable models with similar geometries (1 simple virtual site and 1 shell particle) were also tested: SWM4-DP [105], SWM4-NDP [108], SWM4-HLJ [117], SWM4-dip [117], and SWM4-HD [117].

*3.4 Five and more site models*

The 5-site models include 2 virtual sites (denoted as L) located symmetrically along the lone-pair directions. From this family, 3 models were investigated: TIP5P [84], TIP5P-Ew [44], and TIP5P/2018 [54].

The 6-site models, TIP6P [106], and TIP6P-EW [113], consist of 3 simple virtual sites: 2 as in the 5-site model (L) and 1 as in the 4-site model (M). These models are rigid and non-polarizable.

The polarizable 6+1-site SWM6 model [43] uses the same 3 virtual sites as the TIP6P model with an additional Drude particle to account for polarizability.

A 7-site model, TIP7P [53], was also tested. This model includes 4 virtual sites: two along the lone-pair directions (L), similar to the 5-site models; and two additional virtual sites (M2) that are halfway between the oxygen and hydrogen atoms (see Fig. 2).

## 4. Results and discussions

*4.1 Density*

As a first step the density obtained by the different models was tested. Several water models had not been investigated using GROMACS software previously. The calculated density values, and their temperature dependence also served as validating quantities.

The calculated density values at the lowest and highest tested temperature and at room temperature are shown in Fig. 3 and for all temperatures listed in Table S4. For most models, these values agree well with previous data from the literature. For some models, small differences can be observed that are mainly caused by the different handling of the long range electrostatic and LJ forces.

The only notable exception is the ML-TIP3P model: the values presented in Ref. [114] are significantly lower. This model was parametrized to reproduce the experimental vapor-liquid properties of water, and the authors used Gibbs ensemble simulations where liquid water and vapor coexist in the same system. The discrepancy may arise from the fundamentally different simulation methodologies.

The simulated density values mostly agree with the experimental data within 1-2%. The best models are TIP4P-FB, TIP4P-ST, TIP4P/$\varepsilon_{Flex}$, TIP4P/2005f, ECCw2024, TIP3P-ST, TIP7P, TIP4P/2005, OPC, FBA/$\varepsilon$, TIP4P/$\varepsilon$, TIP4Q, SPC/$\varepsilon$ models. These models predict density with an accuracy of within 0.5% over the entire temperature range investigated. The largest deviations are observed for ML-TIP3P (10%), HUANG (6%), TIP3P (5%), TIP5P (5%), and SPC (5%) models. Three of these – TIP3P, TIP5P, and SPC – are relatively older models.

*4.2 Self-diffusion coefficient*

The calculated self-diffusion coefficient values at the lowest and highest tested temperature and at room temperature are shown in Fig. 4 and for all temperatures summarized in Table S5. It should be noted here that these values are only approximate values with high uncertainty and not corrected for the finite size effect. More precise calculations will be done in a following publication (see section 2.3). For most models, the values are in good agreement with previous simulation data provided differences in box size are considered. However, significantly lower values were obtained for the SWM6, TIP3P-Buck, TIP3P-LJOPT, TIP4P-LJOPT models compared to previously published results.

As the temperature decreases the mobility of the molecules decreases. With the simulation technique used, the system cannot freeze, at most it reaches a supercooled liquid state. The self-diffusion values at $T = 254$ K are around 20 % or more of the room temperature (simulated) value, except the SWM6 model, which yields a value of 0.0191 x$10^{-9}$ m$^2$/s, approximately 1% of the experimental value at $T = 295$ K (2.1 x$10^{-9}$ m$^2$/s).

For some models (TIP5P, TIP5P-Ew, TIP3P-ST, SWM4-HD), the self-diffusion coefficient at $T = 254$ K is 5-10% of its room temperature value. For the most models, this value is around 0.4 – 0.7 x$10^{-9}$ m$^2$/s. (The extrapolated value from the experimental data is 4.8 x$10^{-10}$ m$^2$/s [121].) The ML-TIP3P, TIP3P, HUANG, and SPC models gave the highest values at $T = 254$ K (2.7 – 1.6 x$10^{-9}$ m$^2$/s). These models overestimate self-diffusion across the entire investigated temperature range.

*4.3 Total scattering structure factors*

*4.3.1 XRD structure factor*

The TSSFs were calculated and compared with ones derived from experimental data at all investigated temperatures. Some typical example of the temperature dependence and the agreement/disagreement with experimental data are presented in Figs. 5, 6 (the other models are shown in Figs S1 – S7). The values of the *R*-factor (the quantity quantifying the fit terms, see Eq. 5) are given in Table 3 and shown in Fig. S8).

In the XRD experimental data, the most striking change with temperature is observed at the first two peaks (around 2 and 3 Å$^{-1}$). As the temperature increases, the position of the 1$^{st}$ peak shifts toward higher $Q$ values, while the 2$^{nd}$ peak shifts to lower $Q$ values. The amplitude of the 2$^{nd}$ peak decreases with increasing temperature, with the ratio of the first two peaks also changing in favor of peak 1. At the highest temperature, the two peaks begin to merge, with peak 2 becoming more of a shoulder after the 1$^{st}$ peak. A slight shift towards lower $Q$ values is also observed in the position of the 3$^{rd}$ – 5$^{th}$ peaks (around 4.8, 7.2, and 9.4 Å$^{-1}$)

Most of the simulated curves reproduce the experimentally observed trends. Apart from two models, the positions of the first 5 peaks are similar and behave identically as the temperature changes. The two exceptions are the HUANG and the ML-TIP3P models: for these, instead of two main peaks of 2 and 3 Å$^{-1}$, there is only one peak at $Q = 2.25 – 2.35$ Å$^{-1}$, which remains unchanged in the studied temperature range.

For TIP3P, the first two peaks are significantly closer together (2.15 and 2.75 Å$^{-1}$) than in the experiment and merge into a single peak already at room temperature. For several models, especially older ones, the two peaks are inseparable at high temperatures – besides HUANG and ML-TIP3P, this behavior is observed in the SPC, TIP3P, TIP4P, TIP3P-LJOPT, TIP5P, TIP6P-Ew, and SWM4-HLJ models. In most models, the amplitude of the 2$^{nd}$ peak decreases with increasing temperature. Exceptions include HUANG, ML-TIP3P, and TIP3P, where the amplitude does not decrease, and OPC3-POL and SPC, where the decrease is only slight.

The largest difference between the models is in the amplitude of the first two peaks and their relative magnitudes. In the experimental data, the second peak is higher at 254 K, while the first peak is higher at 366 K. This behavior is accurately reproduced by the ECCw2024, FBA/ε, OPC, TIP4P/2005, TIP4P/2005f, TIP4P/ε, TIP4P/ε$_{Flex}$, TIP4P-Ew, TIP4P-FB, TIP4P-ST, TIP4Q, TIP6P, and SWM4-dip models (Group A). In another group of models (Group B), the low temperature behavior is similar, with the amplitude of the 2$^{nd}$ peak decreasing as the temperature increases, but at the highest temperature they deviate from the experimental observations: the two peaks are no longer separable, their magnitudes become equal or the 2$^{nd}$ peak remains higher. These models are OPC3, OPTI-1T, OPTI-3T, SPC/E, SPC/ε, SPC/Fw, SWM6, TIP3P-FB, TIP3P-LJOPT, TIP3P-ST, TIP4P, TIP4P-BG, TIP4P-LJOPT TIP5P, TIP5P-Ew, TIP5P/2018, TIP6P-Ew, TIP7P, SWM4-HD, and SWM4-HLJ. For OPC3-POL, SWM4-DP, SWM4-NDP, TIP4P-D, TIP3P-Buck, TIP4P-Buck, and SWM3 models (Group C), peak 1 is higher even at 254 K.

The agreement with the experimental data can be quantified by $R$-factors, whose values reflect the degree of agreement and divergence between models and observations. In addition to general trends in structural behavior, the amplitude of the peaks also strongly influences the quality of the fit.

The OPC model consistently produced the lowest $R$-factors for the XRD data across all temperatures, with values ranging from 8% to 17%. Notably, at room temperature and above, this model exhibited excellent agreement with experimental data, yielding $R$-factor values below 10% for T ≥ 284 K across all investigated temperatures.

Since the $R$-factor values depend not only on the model but also on the quality of the experimental data, it is useful to compare $R$-factors across models. To facilitate this comparison, we introduce the relative $R$-factor ($R_{rel}$), defined as $R_{rel} = R / R_{best}$, which expresses the goodness of fit relative to the $R$-factor of the best-performing model. For the XRD data, the best model at all temperatures is OPC, meaning $R_{best} = R_{OPC}$.

In terms of agreement with the XRD data, the ECCw2024 model ranked second, with $R$-factor values on average 23% higher than those of the OPC model. Following this, 16 models exhibited $R$-factor values 28%–50% higher than OPC. These include primarily models classified in Group A, which accurately reproduce the relationship between peaks 1 and 2 at both low and high temperatures.

Two models from Group A exhibited notably higher $R$-factors: TIP4P/$\varepsilon_{Flex}$ ($R_{rel}$ = 1.66) and FBA/$\varepsilon$ ($R_{rel}$ = 2.09). Among the models with better $R$-factors, we also find models from Groups B and C, including TIP4P-Buck ($R_{rel}$ = 1.28), TIP5P/2018 (1.36), SWM4-DP (1.37), TIP4P-D (1.40), TIP4P-Ew (1.47), SWM4-NDP (1.47), and TIP7P (1.48).

The weakest fit was observed in the HUANG (4.82) and ML-TIP3P (4.76) models, which could not even reproduce the density well.

It is also worth noting the striking similarity of the $S^X(Q)$ curves across the entire temperature range for some models, namely the TIP4P/2005, TIP4P/$\varepsilon$, TIP4P-Ew, TIP4P-FB, TIP4P-ST, and TIP4Q. These models have the same water molecule geometry (TIPnP geometry), the same $\sigma_{OO}$ = 0.316 nm, similar $\varepsilon_{OO}$ values (0.68 – 0.77 kJ/mol) and H charge values ranging from 0.52 to 0.55 e. Among models which retain the TIPnP geometry but have lower $\varepsilon_{OO}$ values, the $R$-factor values tend to be higher, indicating a weaker fit. Conversely among models with $\varepsilon_{OO}$ values above this range, there are still models with good $R$-factors, despite their $S^X(Q)$ curves showing greater deviation from models falling within the reference $\varepsilon_{OO}$ range.

It is noteworthy that most models reproduced the 295 K and 324 K data the best, with the fit quality deteriorating with temperatures further from these values. This is presumably due to room temperature experimental data being chosen as the target properties during model development.

When examining the temperature dependence of the simulated TSSF curves, it is observed that TIP3P models produce curves similar to experimental curves at higher temperatures or to those of models that better align with experimental data. By shifting the temperature scale upward by approximately 40 K (i.e., comparing the simulated curve at 254 K with the experimental curve at 295 K), we obtain curves more in line with the experimental data. Similarly, for a well-performing model such as OPC3, its curves at 380 K and 400 K closely resemble the TSSF curves of TIP3P at 343 K and 366 K, respectively (Fig. 7). This temperature shift effect is also observed e.g., for the self-diffusion coefficients calculated from the TIP3P model.

*4.3.2 ND structure factor*

There is no neutron diffraction structure factor data available in the literature that covers a temperature range similar to the XRD data in Ref [101]. Neutron diffraction with isotopic substitution measurements (H/D substitution, with varying H/D ratio) were published at room temperature in Ref. [65], and at 283 K in Ref. [69]. Since the $D_2O$ curve has the smallest uncertainty, this structure factor was used for comparison with the simulated data. In an earlier publication [102], the structure factor was investigated at four temperatures (298 K, 323 K, 343 K, and 368 K), these curves were also compared with the calculated ones. The 298 K data of Ohtomo et al. is compared with the room temperature Soper's data on Fig. 8. As clearly shown in the figure, there are evident differences between the two curves. The amplitude of the peaks and the position of the 4$^{th}$ peak at about 14.5 Å$^{-1}$ exhibit significant differences, possibly due to experimental error and differences in handling the corrections. The temperature dependence of the experimental data is weak, especially compared to the differences of the XRD structure factor.

The position of the first peak (around 2 Å$^{-1}$) shifts to higher $Q$ values as the temperature increases, its amplitude slightly decreases (see also [122]). At higher temperatures a small shoulder develops at 2.5 Å$^{-1}$. The small peak around 4 Å$^{-1}$ becomes less significant at higher temperatures.

Examples of ND structure factors are presented in Figs. 9, 10, with additional models shown in Figs. S9 – S15. The $R$-factors of the fits are collected in Table 4 and Fig. S16.

For some models (e.g., TIP5P, SWM6), it is observed that at low temperatures, the main peak around 2 Å$^{-1}$ is followed by a 2$^{nd}$ peak around 2.7 Å$^{-1}$. As the temperature increases, the valley between the two peaks decreases, with peak 2 gradually becoming the shoulder of peak 1, this can also be observed in the experimental data around 2.5 Å$^{-1}$.

For most models, the position of the first peak shifts towards higher $Q$ values at higher temperatures, the amplitude at approximately 2.5 Å$^{-1}$ (initially the valley, later the shoulder) increases, in agreement with the experimental data. Again, the exceptions are the HUANG, ML-TIP3P, and the 3-site polarizable models (OPC3-POL and SWM3). For the simulated curves, a decrease in the amplitude of the small peak around 4 Å$^{-1}$ and a shift toward higher $Q$ values are observed (except in the HUANG, ML-TIP3P, OPC3-POL, SWM3 models).

Comparing the fits of the three data sets around room temperature (Ref. [69] 283 K, Ref. [65] 295 K and Ref. [102] 298 K), Soper's 295 K data shows the best agreement with the simulated curves. The $R$-factors for this data are typically around 14 – 25%, while the 283 K data produces $R$-factors of 24 – 35% and the Ohtomo et al. data results in $R$-factors of 30 – 45% for most of the models. The agreement with higher temperature data also yields $R$-factor values of 30 – 45%.

The TIP4P-BG model has the lowest $R$-factor value with the 283 K data ($R$ = 24.3%), while Soper's 295 K data is in the best agreement with the TIP6P-Ew model ($R$ = 14.2%). The TIP4P/2005f model has the lowest $R$-factor for the 298 and 323 K data of Ohtomo et al. ($R$ = 30.3 and 27.5%), while the SWM6 model presented the best fit of the 343 and 366 K data ($R$ = 28.2 and 31.5%). The relative $R$-factors, normalized to the $R$-factor values of these models, are collected in Table 4.

Differences in fit quality among the three data sets result from variations in the accuracy and reliability of the experimental data. For this reason, it is beneficial to compare the models using the three data sources separately.

Based on the 295 K Soper data, the TIP4P/$\varepsilon_{Flex}$ and TIP6P models perform comparably to the TIP6P-Ew model ($R_{rel}$ = 1.04, 1.05), but the modern 3-site OPC3 and OPTI-3T models are only slightly less accurate ($R_{rel}$ = 1.12, 1.15). 27 other models have $R_{rel}$ values between 1.2 and 1.55, including most of the TIPnP and SPC-family models. The OPTI-1T, TIP3P-ST, TIP4P-Buck, TIP3P, ECCw2024, FBA/$\varepsilon$, TIP4P-LJOPT models are less successful, with $R_{rel}$ values of 1.7 – 2.1. The OPC, SWM3, ML-TIP3P, HUANG and OPC3-POL models performed the worst (3.13 – 4.16).

In terms of 283 K data from [69], 19 other models performed similarly to the TIP4P-BG model ($R_{rel}$ ≤ 1.2), including TIP4P-Ew, TIP4P, TIP4P/2005f, and TIP4P-FB, which have the lowest values. $R_{rel}$ values between 1.21 and 1.62 were obtained by 18 other models, while values around 1.9 were observed for FBA/$\varepsilon$ and OPC. The highest values were obtained by the same models as previously.

When compared to Ohtomo's data, the models from the TIP4P-family came out on top at room temperature, with $R_{rel}$ ≤ 1.2 for a total of 28 models. 11 models had relative $R$-factors between 1.25 and 1.7, while $R_{rel}$ ≥ 2 was obtained for FBA/$\varepsilon$, ML-TIP3P, HUANG, SWM3, and OPC3-POL models. When looking at the full temperature range (average of the 4 relative $R$-factors), the models with the TIP4P geometry are the most consistent with the data of Ohtomo et al, among them TIP4P/2005f, SWM6, TIP4P-BG, SWM4-dip, TIP4P-Ew, TIP4P/$\varepsilon_{Flex}$, TIP4P/2005, TIP4P-ST, TIP4P-FB, TIP4P, TIP4P/$\varepsilon$, TIP4Q are the top performers. In contrast, OPC3-POL, SWM3, HUANG, ML-TIP3P, and FBA/$\varepsilon$ show the worst agreement.

As with the X-ray diffraction TSSF, several models generate nearly identical curves: TIP4P/2005, TIP4P/$\varepsilon$, TIP4P-Ew, TIP4P-FB, TIP4P-ST, TIP4Q.

The effect of temperature shift can also be observed for neutron TSSF. When the OPC3 model is tested at lower temperatures (see Fig S17), the 2$^{nd}$ peak around 2.7 Å$^{-1}$ becomes more pronounced in the resulting TSSFs. The curves of the previously mentioned TIP5P, SWM6 etc. models at 254 K and that of the OPC3 model at 200 K show a greater agreement.

### 4.3.3 *Combined ranking by XRD and ND structure factors*

The different weights of OO, OH, and HH partials in the neutron and XRD data mean that the combination of the two types of experimental data provides a more rigorous method of analysis. Some models fit the XRD data well but fail to accurately reproduce the neutron diffraction data, such as the OPC model. The most accurate description of the structure is given by models that fit both types of measurement data well. Therefore, it's worth ranking the models based on the sum of the relative $R$-factors from XRD and ND.

At room temperature, the fit to both Soper's 295 K data and to the 295 K XRD data were examined together ($R_{rel}^{RT} = R_{rel}$ (Soper, 295K) + $R_{rel}$ (XRD, 295K), see, Fig. 11a, and Table S6). This suggests that the 6-site TIP6P and TIP6P-Ew models are the most suitable for describing the structure of water at room temperature. Similarly good fits were obtained with the 3-site OPTI-3T and OPC3 models and the TIP4P family of models, with TIP4P/2005 being the most accurate. Among the models tested, 30 had fits that were, on average, no more than 25% worse at this temperature. It is worth noting that among the models that are in principle optimized directly for the structure, the OPTI-3T performed particularly well, whereas OPTI-1T, TIP3P-LJOPT, and TIP4P-Buck models were in the middle of the range. Meanwhile, the TIP3P-Buck and TIP4P-LJOPT ranked among the weaker models at room temperature.

For the performance over the full temperature range, the average of the relative $R$-factors of the XRD data (Fig. S18a) plus the average of the relative $R$-factors of the ND data (Fig. S18b) was calculated (denoted hereafter as $R_{tot}$, see, Fig. 11b, Table S6). Five out of the six ND sets were included in this average: From the two room-temperature ND data sets, we selected only Soper's dataset, which had lower $R$-factors (and probably smaller uncertainties). Using different ND dataset combinations from the three publications does not significantly affect model rankings, as the relative R-factors vary by only a few hundredths, and the model ranking remains largely unchanged. In this comparison the TIP4P/2005 model ranked highest, followed by other 4-site models, mostly from the TIP4P family: TIP4P/ε, TIP4Q, TIP4P/2005f, TIP4P-FB, SWM4-dip, TIP4P-Ew, SWM4-DP, TIP4P-ST, and ECCw2024.

It is important to point out that the relative $R$-factors of the top 17 models differ by less than 10%, indicating that the differences in fit quality are minor. Given the experimental uncertainties, this variation is not considered significant. Thus, any of these models can be reliably used for structural analysis. The best-performing 3-site models are OPTI-1T, OPTI-3T, and OPC3 (ranking 17 – 19), with relative $R$-factors approximately about 10 – 13% higher than the best 4-site model. The other 3-site models performed significantly worse.

In these comparisons, the HUANG, ML-TIP3P, OPC3-POL, and SWM3 models ranked the lowest, exhibiting notably poor fits. Apart from these, the worst-performing models were the flexible three-site models (FBA/ε, SPC/Fw) and the TIP3P-like models (TIP3P-ST, TIP3P-Buck, TIP3P-FB, TIP3P). Notably, re-parameterized TIP3P models did not show significant improvements over their predecessors in terms of structural accuracy. Among them, the best-performing model was the structurally optimized TIP3P-LJOPT, while TIP3P-Buck, despite using a more complex potential, did not provide a better structural match.

Among the models optimized for the structure, the TIP4P-Buck model performed best across the full range of temperatures tested. However, despite its higher computational cost, it does not outperform the best models using the Lennard-Jones potential. Moreover, the 3-site models OPTI-1T and OPTI-3T, which are also optimized for structure, achieved nearly comparable performance.

It is also worth mentioning the OPC model, which resulted the best fit to the XRD data but reproduced the ND data poorly. This is probably due to the intramolecular geometry used in this model, which is significantly different from the gas-phase water molecule geometry. To investigate this, a hypothetical model was tested: mixed PRDSs were constructed from the intermolecular PRDFs of the OPC model and the intramolecular PRDSs of the TIP4P/2005 model. The neutron TSSF calculated from these 'mix' PRDFs is compared with the $S^N(Q)$-s of the OPC and TIP4P/2005 models, as well as the experimental data in Fig. 12. As can be seen from the figure, the 'mix' model shows significantly better agreement with the experimental data than the OPC model. The poor performance of the OPC model is therefore largely due to its intramolecular geometry.

This highlights that when designing models, it is not enough to consider the O-O distance (or the O-O PRDF) to reproduce structural properties accurately.

*4.4. Partial radial distribution functions*

The intramolecular PRDFs are determined by the geometry of the molecule as defined in the model. Most of the models investigated here are rigid, meaning the O-H and H-H intramolecular distances are fixed. Therefore, the intramolecular peaks of these models are sharp typically spanning one or two bins. By contrast, flexible models produce broader and lower amplitude intramolecular peaks. Most models follow the geometry of the water molecule in the vapor phase, using an O-H distance around 0.96 Å and a H-O-H angle of about 104.5° (e.g., models from the TIPn family). Another widely used geometry is the tetrahedral arrangement (SPC geometry), which features an H-O-H angle of 109.5° and an O-H distance of around 1 Å. Some models use significantly different parameters: the OPC model has a shorter O-H distance (0.8724 Å), while the OPC3-POL (1.211 Å), SWM3 (1.1686 Å), and HUANG (1.1549 Å) models exhibit longer O-H distances. Similarly, flexible models often have larger H-O-H angles, including FBA/ε (114.7°), SPC/Fw (113.24°), and TIP4P/$\varepsilon_{Flex}$ (111.5°).

The intramolecular O-H and H-H PRDFs in rigid models change only slightly with temperature: their peak amplitudes increase slightly at higher temperatures. In contrast the intramolecular O-H and H-H peaks of flexible models are wider and decrease in amplitude as temperature rises (see Figs. S19, S20).

A general characteristic of intermolecular PRDFs is that at low temperatures the peaks appear sharper and the minima more pronounced. At high temperatures, essentially only the first one or two peaks can be observed. A typical example of the PRDFs is shown in Fig. 13. For a complete set of the PRDFs, see the SM (Figs. S21 – S35). Several models result similar curves, with a slight variation of the position and amplitude of the peaks. For some models (e.g. SWM6, TIP6P) the peaks are sharper with higher amplitude first peak, particularly at low temperatures. Conversely, models such as TIP3P, exhibit broader, less distinct peaks with lower amplitudes. The sharpness of the peaks and minima, and the magnitude of the peak amplitudes, are a function of temperature, and such differences between the models are due to the shift of their temperature scales. In Fig. S36, the PRDFs of the SMW6 and TIP3P models (in the 254 – 366 K range) are compared with the PRDFs of the OPC3 model for lower and higher temperatures. It can be clearly observed that the 200 K OPC3 curves show a greater similarity with the 254 K SWM6 curves, while the 400 K OPC3 curves show a greater similarity with the 366 K TIP3P curves.

The first O-O peak (or the first two peaks for O-H and H-H) remains visible even at the highest temperatures and characterizes the first shell, i.e., the region around each water molecule where hydrogen bonding occurs. The position of the peaks varies only slightly with temperature: $r$ increases by 0.02 – 0.04 Å for O-O and O-H, and by 0.08 – 0.15 Å for H-H at the highest temperature compared to the lowest. Room temperature values are given in Table S7.

At room temperature the O-O first neighbor distance ranges from 2.72 – 2.82 Å, except for the ML-TIP3P (2.94 Å) and HUANG (2.96 Å) models, which deviate significantly. For most models it is around 2.76 – 2.78 Å. These two outliers also exhibit poor fit quality. Among the models that best match the XRD data, $r_{OO}$ ranges from 2.76 to 2.8 Å. However, it should be noted that an accurate O-O distance is a necessary but not sufficient condition for a good fit; for example, the OPC3-POL model correctly predicts $r_{OO}$ but exhibits high $R$-factor values. The first intermolecular O-H distance corresponds to the hydrogen bonding length. At room temperature, most models yield values between 1.7 – 1.88 Å, with a shorter value for OPC3-POL and SWM3 (1.58 Å) and a longer value for ML-TIP3P (2.02 Å) and OPC (1.96 Å). The lowest ND $R$-factors correspond to models with $r_{OH}$ = 1.78 – 1.82 Å.

The first intermolecular H-H peak represents the distance between two hydrogen atoms, one of which participates in a hydrogen bond. This peak is broader than those of O-O and O-H pairs, and its position increases by up to 0.14 Å at higher temperatures in some models. Certain models (e.g., OPC3-POL, SWM3) exhibit partial overlap between intramolecular and intermolecular H-H distances, meaning that in some cases, the H-H distance between molecules is shorter than the H-H distance within a single molecule. For most models, the first intermolecular H-H peak falls between 2.32 and 2.44 Å, except for ML-TIP3P, where it is significantly higher (2.66 Å). Apart from this extreme case, no clear correlation is observed between fit quality and the position of the first H-H peak. However, the broad peak shape and its contribution to the total scattering structure factors (TSSFs) vary across models.

The contributions of the hydrogen-bonding molecules appear in the first intermolecular O-H and O-O peaks. The corresponding coordination number (Eq. 13) can give an approximation of the number of hydrogen bonds:

$$N_{ij} = 4\pi\rho c_j \int_0^{r_{\min}} g_{ij} r^2 \mathrm{d}r, \tag{13}$$

where $r_{\min}$ is the upper limit, which is usually the minimum of the corresponding PRDF.

The $N_{\text{OH}}$ coordination numbers calculated up to the first minima are given in Table S8. This gives the H atoms around the O atoms thus the pairs in which the molecule participates as proton acceptors. The total number of H-bonds is roughly (see below) twice this number. The number of H-bonds decreases with increasing temperature, being close to 4 (3.78 – 3.98) at the lowest temperature tested and around 3.6 (3.4 – 3.8) at the highest temperature (the OPC3-POL and SWM3 model values are also out of line here, 3.5 – 3.4 and 3.3 – 3.15, at 254 K and 366 K, respectively). However, this method is not sufficiently precise. While the first O-H peak is typically well separated from the subsequent peaks, the first two O-O peaks often overlap. Furthermore, as temperature increases, the minima following these peaks become shallower, even for the O-H PRDF. This results in contributions from both hydrogen-bonding molecules and more distant molecules merging together. Consequently, the total number of hydrogen bonds – and its dependence on model and temperature – should be analyzed using more advanced methods, which will be explored in a forthcoming paper.

4.5 *Quality of models by year of publication and complexity*

As we have seen in the previous chapters, one of the best models in terms of structure is TIP4P/2005, which was published 20 years ago. In Fig. 14 the goodness of fit is compared with the year of publication of the models. The metric used here is $R_{\text{tot}}$, the sum of the average relative *R*-factors from XRD and ND data (see Section 4.3). While the figure does not reveal a clear trend, it does show that most of the models created in the last decade fit the experimental TSSFs as well as the best models did in the past. The most striking improvement in the last decade can be found in the rigid 3-site models. OPC3 and OPTI-3T models exhibit fit quality nearly on par with the best 4-site models. In addition, the OPC3 model reproduces many other properties (e.g., density and self-diffusion coefficient) significantly better over a wide temperature range than the early SPC/E or TIP3P models.

In many cases, models with additional interaction sites or more complex potentials are introduced to improve agreement with experimental data. However, these models also impose a greater computational cost. The simulation speed of a simple three-site model can be up to an order of magnitude faster than that of a flexible or polarizable model. Additionally, flexible and shell particle models require a smaller time step, while models using the Buckingham potential are constrained by limited software support in GROMACS, particularly for GPU acceleration. The simulation speeds of the investigated models are summarized in Fig. S37. All of the simulations were performed on the same hardware (CPU: Intel® Core (TM) i7-8700K CPU @ 3.70GHz. GPU: NVIDIA GeForce RTX 3080 10GB).

One would expect a more complex, multi-parameter model to match the experimental data better. However, the main aim of creating these models was not necessarily to reproduce the structure more accurately.

In Fig. 15 we plot the average relative goodness-of-fit values of the models ($R_{tot}$) as a function of their simulation speed. As shown, there is no observable correlation between these two quantities. The best fits are achieved by four-site models, which have a simulation speed of approximately 700 ns/day. Of course, the selection of models is not exhaustive. However, if a direct correlation between computational cost and fit quality existed, it should still be detectable within this dataset, even if not fully representative. Thus, it can be concluded that increasing model complexity does not necessarily improve structural accuracy. For studying pure, liquid water at ambient pressure, no clear advantage is observed in using models more complex than four-site models.

## 5. Conclusions

A total of 44 classical, pairwise-additive interaction models of water were analyzed using molecular dynamics simulations. Their predictions regarding the structure of pure liquid water at ambient pressure were examined. Additionally, the density and self-diffusion coefficients obtained from these models were compared with both experimental and previous simulation data. The partial radial distribution functions (PRDFs) and the total scattering structure factors (TSSFs), weighted for neutron or X-ray diffraction, were calculated from the obtained trajectories. These TSSFs were then compared with experimental data from the literature.

Among the models tested, the OPC model provided the best fit to the XRD data across all investigated temperatures. The different experimental ND datasets were best reproduced by the TIP4P/2005f, SWM6, TIP4P-BG, and TIP6P-Ew models. When considering both data types together, the TIP4P/2005 model yielded the best overall agreement, closely followed by TIP4P/ε, TIP4Q, TIP4P/2005f, TIP4P-FB, SWM4-dip, TIP4P-Ew, SWM4-DP, TIP4P-ST, and ECCw2024. We have found that models in which the O-O first neighbor distance is significantly greater than the experimental 2.78 Å, or where the geometry of the water molecule is significantly different from that of the gas-phase water molecule, fail to reproduce XRD or ND data accurately. It was demonstrated that the agreement in the nearest neighbor oxygen-oxygen distance is necessary but not sufficient for accurately reproducing the total scattering structure factors from experimental neutron and X-ray diffraction simultaneously.

7 out of the top 10 models are reparametrized versions of the TIP4P model, indicating that increasing model complexity does not necessarily improve structural fits. In addition, significant improvements have been observed in newer three-site models, such as OPTI-1T, OPTI-3T, and OPC3, which were developed in the last decade.

While newer models incorporating polarizability, flexibility, or additional interaction sites may improve the description of certain properties, these enhancements do not necessarily translate into a better structural agreement. In particular, recent three-site polarizable models were found to be inadequate for accurately describing the structure of water meanwhile simple 3-site models (such as OPC3) perform significantly better. Since structure is a fundamental property, model developers should ensure that it is well-reproduced.

This comparative analysis provides valuable insights for future MD simulation studies, aiding in the selection of appropriate potential models. Although the present findings focus on pure water, further studies are required to assess model performance in solutions and mixtures. Nonetheless, the results presented here serve as a strong foundation for such investigations.

Acknowledgments

I.P. is grateful to the National Research, Development and Innovation Office (NKFIH, Hungary) for financial support through Grant No K142429. B.S. was supported by the National Research, Development and Innovation Office (NKFIH, Hungary) under the University Research Scholarship Programme (EKÖP-2024) Grant No EKÖP-24-1-BME-77 and the HUN-REN Wigner RCP under the Wigner Internship Programme. The authors would like to thank László Temleitner for helpful discussions.

# References

[1] A. Brodsky, Is there predictive value in water computer simulations?, Chem. Phys. Lett. 261 (1996) 563–568. doi:10.1016/0009-2614(96)00997-9.

[2] I. Nezbeda, Simple short-ranged models of water and their application. A review, J. Mol. Liq. 73–74 (1997) 317–336. doi:10.1016/S0167-7322(97)00076-7.

[3] A. Wallqvist, R.D. Mountain, Molecular models of water: Derivation and description, in: K.B. Lipkowitz, D.B. Boyd (Eds.), Rev. Comput. Chem. Vol. 13, Wiley-VCH, John Wiley and Sons, Inc., New York, US, 1999: pp. 183–247. doi:10.1002/9780470125908.ch4.

[4] J.L. Finney, The water molecule and its interactions: the interaction between theory, modelling, and experiment, J. Mol. Liq. 90 (2001) 303–312. doi:10.1016/S0167-7322(01)00134-9.

[5] B. Guillot, A reappraisal of what we have learnt during three decades of computer simulations on water, J. Mol. Liq. 101 (2002) 219–260. doi:10.1016/S0167-7322(02)00094-6.

[6] P.K. Yuet, D. Blankschtein, Molecular dynamics simulation study of water surfaces: Comparison of flexible water models, J. Phys. Chem. B 114 (2010) 13786–13795. doi:10.1021/jp1067022.

[7] J.F. Ouyang, R.P.A. Bettens, Modelling water: A lifetime enigma, Chimia (Aarau). 69 (2015) 104–111. doi:10.2533/chimia.2015.104.

[8] G.A. Cisneros, K.T. Wikfeldt, L. Ojamäe, J. Lu, Y. Xu, H. Torabifard, et al., Modeling molecular interactions in water: From pairwise to many-body potential energy functions, Chem. Rev. 116 (2016) 7501–7528. doi:10.1021/acs.chemrev.5b00644.

[9] I. Shvab, R.J. Sadus, Atomistic water models: Aqueous thermodynamic properties from ambient to supercritical conditions, Fluid Phase Equilib. 407 (2016) 7–30. doi:10.1016/j.fluid.2015.07.040.

[10] M. Fugel, V.C. Weiss, A corresponding-states analysis of the liquid-vapor equilibrium properties of common water models, J. Chem. Phys. 146 (2017) 64505. doi:10.1063/1.4975778.

[11] A. V Onufriev, S. Izadi, Water models for biomolecular simulations, WIREs Comput. Mol. Sci. 8 (2018) e1347. doi:10.1002/wcms.1347.

[12] O. Demerdash, L. Wang, T. Head-Gordon, Advanced models for water simulations, WIREs Comput. Mol. Sci. 8 (2018) e1355. doi:10.1002/wcms.1355.

[13] I.N. Tsimpanogiannis, O.A. Moultos, L.F.M. Franco, M.B. de M. Spera, M. Erdős, I.G. Economou, Self-diffusion coefficient of bulk and confined water: a critical review of classical molecular simulation studies, Mol. Simul. 45 (2019) 425–453. doi:10.1080/08927022.2018.1511903.

[14] E. Lambros, F. Paesani, How good are polarizable and flexible models for water: Insights from a many-body perspective, J. Chem. Phys. 153 (2020) 60901. doi:10.1063/5.0017590.

[15] S.P. Kadaoluwa Pathirannahalage, N. Meftahi, A. Elbourne, A.C.G. Weiss, C.F. McConville, A. Padua, et al., Systematic comparison of the structural and dynamic properties of commonly used water models for molecular dynamics simulations, J. Chem. Inf. Model. 61 (2021) 4521–4536. doi:10.1021/acs.jcim.1c00794.

[16] P.K. Quoika, M. Zacharias, Liquid–vapor coexistence and spontaneous evaporation at atmospheric pressure of common rigid three-point water models in molecular simulations, J. Phys. Chem. B 128 (2024) 2457–2468. doi:10.1021/acs.jpcb.3c08183.

[17] J.V.L. Valle, B.H.S. Mendonça, M.C. Barbosa, H. Chacham, E.E. de Moraes, Accuracy of TIP4P/2005 and SPC/Fw water models, J. Phys. Chem. B 128 (2024) 1091–1097. doi:10.1021/acs.jpcb.3c07044.

[18] M.-L. Tan, J.R. Cendagorta, T. Ichiye, The molecular charge distribution, the hydration shell, and the unique properties of liquid water, J. Chem. Phys. 141 (2014) 244504. doi:10.1063/1.4904263.

[19] Y. Mao, Y. Zhang, Thermal conductivity, shear viscosity and specific heat of rigid water models, Chem. Phys. Lett. 542 (2012) 37–41. doi:10.1016/j.cplett.2012.05.044.

[20] S.H. Lee, J. Kim, Transport properties of bulk water at 243–550 K: A comparative molecular dynamics simulation study using SPC/E, TIP4P, and TIP4P/2005 water models, Mol. Phys. 117 (2019) 1926–1933. doi:10.1080/00268976.2018.1562123.

[21] C. Vega, J.L.F. Abascal, Simulating water with rigid non-polarizable models: a general perspective, Phys. Chem. Chem. Phys. 13 (2011) 19663. doi:10.1039/c1cp22168j.

[22] R. Capelli, F. Muniz-Miranda, G.M. Pavan, Ephemeral ice-like local environments in classical rigid models of liquid water, J. Chem. Phys. 156 (2022) 214503. doi:10.1063/5.0088599.

[23] M. Agarwal, M.P. Alam, C. Chakravarty, Thermodynamic, diffusional, and structural anomalies in rigid-body water models, J. Phys. Chem. B 115 (2011) 6935–6945. doi:10.1021/jp110695t.

[24] C. Vega, J.L.F. Abascal, M.M. Conde, J.L. Aragones, What ice can teach us about water interactions: a critical comparison of the performance of different water models, Faraday Discuss. 141 (2009) 251–276. doi:10.1039/B805531A.

[25] T.I. Morozova, N.A. García, J.-L. Barrat, Temperature dependence of thermodynamic, dynamical, and dielectric properties of water models, J. Chem. Phys. 156 (2022) 126101. doi:10.1063/5.0079003.

[26] A. Kulkarni, M. Bortz, K.-H. Küfer, M. Kohns, H. Hasse, Multicriteria optimization of molecular models of water using a reduced units approach, J. Chem. Theory Comput. 16 (2020) 5127–5138. doi:10.1021/acs.jctc.0c00301.

[27] J. Wang, Y. Zheng, H. Zhang, H. Ye, Machine learning-generated TIP4P-BGWT model for liquid and supercooled water, J. Mol. Liq. 367 (2022) 120459. doi:10.1016/j.molliq.2022.120459.

[28] M. Perrone, R. Capelli, C. Empereur-mot, A. Hassanali, G.M. Pavan, Lessons learned from multiobjective automatic optimizations of classical three-site rigid water models using microscopic and macroscopic target experimental observables, J. Chem. Eng. Data. 68 (2023) 3228–3241. doi:10.1021/acs.jced.3c00538.

[29] V. Cruces Chamorro, P. Jungwirth, H. Martinez-Seara, Building water models compatible with charge scaling molecular dynamics, J. Phys. Chem. Lett. 15 (2024) 2922–2928. doi:10.1021/acs.jpclett.4c00344.

[30] J. Wang, H. Hei, Y. Zheng, H. Zhang, H. Ye, Five-site water models for ice and liquid water generated by a series–parallel machine learning strategy, J. Chem. Theory Comput. 20 (2024) 7533–7545. doi:10.1021/acs.jctc.4c00440.

[31] R.A. DiStasio, B. Santra, Z. Li, X. Wu, R. Car, The individual and collective effects of exact exchange and dispersion interactions on the ab initio structure of liquid water, J. Chem. Phys. 141 (2014) 84502. doi:10.1063/1.4893377.

[32] K. Szalewicz, C. Leforestier, A. van der Avoird, Towards the complete understanding of water by a first-principles computational approach, Chem. Phys. Lett. 482 (2009) 1–14. doi:10.1016/j.cplett.2009.09.029.

[33] M.J. Gillan, D. Alfè, A. Michaelides, Perspective: How good is DFT for water?, J. Chem. Phys. 144 (2016) 130901. doi:10.1063/1.4944633.

[34] X. Wang, Y.-L.S. Tse, Flexible Polarizable water model parameterized via Gaussian process regression, J. Chem. Theory Comput. 18 (2022) 7155–7165. doi:10.1021/acs.jctc.2c00529.

[35] H. Ye, J. Wang, Y. Zheng, H. Zhang, Z. Chen, Machine learning for reparameterization of four-site water models: TIP4P-BG and TIP4P-BGT, Phys. Chem. Chem. Phys. 23 (2021) 10164–10173. doi:10.1039/D0CP05831A.

[36] X. Zhu, M. Riera, E.F. Bull-Vulpe, F. Paesani, MB-pol(2023): Sub-chemical accuracy for water simulations from the gas to the liquid phase, J. Chem. Theory Comput. 19 (2023) 3551–3566. doi:10.1021/acs.jctc.3c00326.

[37] G. Camisasca, H. Pathak, K.T. Wikfeldt, L.G.M. Pettersson, Radial distribution functions of water: Models vs experiments, J. Chem. Phys. 151 (2019) 44502. doi:10.1063/1.5100871.

[38] W.L. Jorgensen, J. Tirado-Rives, Potential energy functions for atomic-level simulations of water and organic and biomolecular systems, Proc. Natl. Acad. Sci. 102 (2005) 6665–6670. doi:10.1073/pnas.0408037102.

[39] S.K. Reddy, S.C. Straight, P. Bajaj, C. Huy Pham, M. Riera, D.R. Moberg, et al., On the accuracy of the MB-pol many-body potential for water: Interaction energies, vibrational frequencies, and classical thermodynamic and dynamical properties from clusters to liquid water and ice, J. Chem. Phys. 145 (2016) 194504. doi:10.1063/1.4967719.

[40] R. Fuentes-Azcatl, J. Alejandre, Non-polarizable force field of water based on the dielectric constant: TIP4P/ε, J. Phys. Chem. B 118 (2014) 1263–1272. doi:10.1021/jp410865y.

[41] R. Fuentes-Azcatl, M.C. Barbosa, Flexible bond and angle, FBA/ε model of water, J. Mol. Liq. 303 (2020) 112598. doi:10.1016/j.molliq.2020.112598.

[42] R. Fuentes-Azcatl, N. Mendoza, J. Alejandre, Improved SPC force field of water based on the dielectric constant: SPC/ε, Physica A. 420 (2015) 116–123. doi:10.1016/j.physa.2014.10.072.

[43] W. Yu, P.E.M. Lopes, B. Roux, A.D. MacKerell, Six-site polarizable model of water based on the classical Drude oscillator, J. Chem. Phys. 138 (2013) 34508. doi:10.1063/1.4774577.

[44] S.W. Rick, A reoptimization of the five-site water potential (TIP5P) for use with Ewald sums, J. Chem. Phys. 120 (2004) 6085–6093. doi:10.1063/1.1652434.

[45] J.L.F. Abascal, C. Vega, A general purpose model for the condensed phases of water: TIP4P/2005, J. Chem. Phys. 123 (2005) 234505. doi:10.1063/1.2121687.

[46] M.A. González, J.L.F. Abascal, A flexible model for water based on TIP4P/2005, J. Chem. Phys. 135 (2011). doi:10.1063/1.3663219.

[47] D. van der Spoel, P.J. van Maaren, H.J.C. Berendsen, A systematic study of water models for molecular simulation: Derivation of water models optimized for use with a reaction field, J. Chem. Phys. 108 (1998) 10220–10230. doi:10.1063/1.476482.

[48] J. Alejandre, G.A. Chapela, H. Saint-Martin, N. Mendoza, A non-polarizable model of water that yields the dielectric constant and the density anomalies of the liquid: TIP4Q, Phys. Chem. Chem. Phys. 13 (2011) 19728–19740. doi:10.1039/c1cp20858f.

[49] L.-P. Wang, T.J. Martinez, V.S. Pande, Building force fields: An automatic, systematic, and reproducible approach, J. Phys. Chem. Lett. 5 (2014) 1885–1891. doi:10.1021/jz500737m.

[50] Y. Xiong, S. Izadi, A. V. Onufriev, Fast polarizable water model for atomistic simulations, J. Chem. Theory Comput. 18 (2022) 6324–6333. doi:10.1021/acs.jctc.2c00378.

[51] H.-S. Lee, M.E. Tuckerman, Structure of liquid water at ambient temperature from ab initio molecular dynamics performed in the complete basis set limit, J. Chem. Phys. 125 (2006) 154507. doi:10.1063/1.2354158.

[52] I. V. Leontyev, A.A. Stuchebrukhov, Electronic polarizability and the effective pair potentials of water, J. Chem. Theory Comput. 6 (2010) 3153–3161. doi:10.1021/ct1002048.

[53] C.-L. Zhao, D.-X. Zhao, C.-C. Bei, X.-N. Meng, S. Li, Z.-Z. Yang, Seven-site effective pair potential for simulating liquid water, J. Phys. Chem. B 123 (2019) 4594–4603. doi:10.1021/acs.jpcb.9b03149.

[54] Y. Khalak, B. Baumeier, M. Karttunen, Improved general-purpose five-point model for water: TIP5P/2018, J. Chem. Phys. 149 (2018) 224507. doi:10.1063/1.5070137.

[55] D.J. Huggins, Correlations in liquid water for the TIP3P-Ewald, TIP4P-2005, TIP5P-Ewald, and SWM4-NDP models, J. Chem. Phys. 136 (2012) 64518. doi:10.1063/1.3683447.

[56] H. Jiang, O.A. Moultos, I.G. Economou, A.Z. Panagiotopoulos, Hydrogen-bonding polarizable intermolecular potential model for water, J. Phys. Chem. B 120 (2016) 12358–12370. doi:10.1021/acs.jpcb.6b08205.

[57] S. Izadi, R. Anandakrishnan, A. V Onufriev, Building water models: A different approach, J. Phys. Chem. Lett. 5 (2014) 3863–3871. doi:10.1021/jz501780a.

[58] H.J.C. Berendsen, J.P.M. Postma, W.F. van Gunsteren, J. Hermans, Interaction models for water in relation to protein hydration, in: B. Pullman (Ed.), Intermol. Forces. Jerusalem Symp. Quantum Chem. Biochem. Vol. 14, Springer Netherlands, Dordrecht, Holland, 1981: pp. 331–342. doi:10.1007/978-94-015-7658-1_21.

[59] D. O'Connor, N.J. English, Systematic design-of-experiments, factorial-design approaches for tuning simple empirical water models, Mol. Simul. 47 (2021) 119–130. doi:10.1080/08927022.2019.1626987.

[60] A.D. Wade, L.-P. Wang, D.J. Huggins, Assimilating radial distribution functions to build water models with improved structural properties, J. Chem. Inf. Model. 58 (2018) 1766–1778. doi:10.1021/acs.jcim.8b00166.

[61] G. Hura, J.M. Sorenson, R.M. Glaeser, T. Head-Gordon, A high-quality x-ray scattering experiment on liquid water at ambient conditions, J. Chem. Phys. 113 (2000) 9140–9148. doi:10.1063/1.1319614.

[62] J.M. Sorenson, G. Hura, R.M. Glaeser, T. Head-Gordon, What can x-ray scattering tell us about the radial distribution functions of water?, J. Chem. Phys. 113 (2000) 9149–9161. doi:10.1063/1.1319615.

[63] A.K. Soper, The radial distribution functions of water and ice from 220 to 673 K and at pressures up to 400 MPa, Chem. Phys. 258 (2000) 121–137. doi:10.1016/S0301-0104(00)00179-8.

[64] A.K. Soper, Joint structure refinement of x-ray and neutron diffraction data on disordered materials: application to liquid water, J. Phys. Condens. Matter. 19 (2007) 335206. doi:10.1088/0953-8984/19/33/335206.

[65] A.K. Soper, The radial distribution functions of water as derived from radiation total scattering experiments: Is there anything we can say for sure?, ISRN Phys. Chem. 2013 (2013) 279463. doi:10.1155/2013/279463.

[66] L.B. Skinner, C. Huang, D. Schlesinger, L.G.M. Pettersson, A. Nilsson, C.J. Benmore, Benchmark oxygen-oxygen pair-distribution function of ambient water from x-ray diffraction measurements with a wide Q-range, J. Chem. Phys. 138 (2013) 74506. doi:10.1063/1.4790861.

[67] L. Pusztai, Partial pair correlation functions of liquid water, Phys. Rev. B 60 (1999) 11851–11854. doi:10.1103/PhysRevB.60.11851.

[68] Z. Steinczinger, L. Pusztai, An independent, general method for checking consistency between diffraction data and partial radial distribution functions derived from them: the example of liquid water, Condens. Matter Phys. 15 (2012) 23606. doi:10.5488/CMP.15.23606.

[69] A.K. Soper, Disordered atom molecular potential for water parameterized against neutron diffraction data. Application to the structure of ice Ih, J. Phys. Chem. B 119 (2015) 9244–9253. doi:10.1021/jp509909w.

[70] I. Pethes, L. Pusztai, Reverse Monte Carlo investigations concerning recent isotopic substitution neutron diffraction data on liquid water, J. Mol. Liq. 212 (2015) 111–116. doi:10.1016/j.molliq.2015.08.050.

[71] V.C. Chamorro, C. Tempra, P. Jungwirth, Heavy water models for classical molecular dynamics: Effective inclusion of nuclear quantum effects, J. Phys. Chem. B 125 (2021) 4514–4519. doi:10.1021/acs.jpcb.1c02235.

[72] D. Dhabal, K.T. Wikfeldt, L.B. Skinner, C. Chakravarty, H.K. Kashyap, Probing the triplet correlation function in liquid water by experiments and molecular simulations, Phys. Chem. Chem. Phys. 19 (2017) 3265–3278. doi:10.1039/C6CP07599A.

[73] R. Fuentes-Azcatl, M.C. Barbosa, Thermodynamic and dynamic anomalous behavior in the TIP4P/ε water model, Physica A 444 (2016) 86–94. doi:10.1016/j.physa.2015.10.027.

[74] Y. Katayama, T. Hattori, H. Saitoh, T. Ikeda, K. Aoki, H. Fukui, et al., Structure of liquid water under high pressure up to 17 GPa, Phys. Rev. B 81 (2010) 14109. doi:10.1103/PhysRevB.81.014109.

[75] F. Paesani, G.A. Voth, The properties of water: Insights from quantum simulations, J. Phys. Chem. B 113 (2009) 5702–5719. doi:10.1021/jp810590c.

[76] H.W. Horn, W.C. Swope, J.W. Pitera, J.D. Madura, T.J. Dick, G.L. Hura, et al., Development of an improved four-site water model for biomolecular simulations: TIP4P-Ew, J. Chem. Phys. 120 (2004) 9665–9678. doi:10.1063/1.1683075.

[77] Z. Steinczinger, L. Pusztai, Comparison of the TIP4P-2005, SWM4-DP and BK3 interaction potentials of liquid water with respect to their consistency with neutron and X-ray diffraction data of pure water, Condens. Matter Phys. 16 (2013) 43604. doi:10.5488/CMP.16.43604.

[78] I. Pethes, L. Pusztai, Reverse Monte Carlo modeling of liquid water with the explicit use of the SPC/E interatomic potential, J. Chem. Phys. 146 (2017) 64506. doi:10.1063/1.4975987.

[79] G. Hura, D. Russo, R.M. Glaeser, T. Head-Gordon, M. Krack, M. Parrinello, Water structure as a function of temperature from X-ray scattering experiments and ab initio molecular dynamics, Phys. Chem. Chem. Phys. 5 (2003) 1981–1991. doi:10.1039/b301481a.

[80] L. Pusztai, O. Pizio, S. Sokolowski, Comparison of interaction potentials of liquid water with respect to their consistency with neutron diffraction data of pure heavy water, J. Chem. Phys. 129 (2008) 184103. doi:10.1063/1.2976578.

[81] Z. Steinczinger, P. Jóvári, L. Pusztai, Comparison of interatomic potentials of water via structure factors reconstructed from simulated partial radial distribution functions: a reverse Monte Carlo based approach, Phys. Scr. 92 (2017) 14001. doi:10.1088/0031-8949/92/1/014001.

[82] Z. Steinczinger, P. Jóvári, L. Pusztai, Comparison of 9 classical interaction potentials of liquid water: Simultaneous reverse Monte Carlo modeling of X-ray and neutron diffraction results and partial radial distribution functions from computer simulations, J. Mol. Liq. 228 (2017) 19–24. doi:10.1016/j.molliq.2016.09.068.

[83] W.L. Jorgensen, J. Chandrasekhar, J.D. Madura, R.W. Impey, M.L. Klein, Comparison of simple potential functions for simulating liquid water, J. Chem. Phys. 79 (1983) 926–935. doi:10.1063/1.445869.

[84] M.W. Mahoney, W.L. Jorgensen, A five-site model for liquid water and the reproduction of the density anomaly by rigid, nonpolarizable potential functions, J. Chem. Phys. 112 (2000) 8910. doi:10.1063/1.481505.

[85] B. Chen, J. Xing, J.I. Siepmann, Development of polarizable water force fields for phase equilibrium calculations, J. Phys. Chem. B 104 (2000) 2391–2401. doi:10.1021/jp993687m.

[86] R.L. McGreevy, L. Pusztai, Reverse Monte Carlo simulation: A new technique for the determination of disordered structures, Mol. Simul. 1 (1988) 359–367. doi:10.1080/08927028808080958.

[87] P.T. Kiss, A. Baranyai, A systematic development of a polarizable potential of water, J. Chem. Phys. 138 (2013) 204507. doi:10.1063/1.4807600.

[88] M.J. Abraham, T. Murtola, R. Schulz, S. Páll, J.C. Smith, B. Hess, et al., GROMACS: High performance molecular simulations through multi-level parallelism from laptops to supercomputers, SoftwareX. 1–2 (2015) 19–25. doi:10.1016/j.softx.2015.06.001.

[89] O. Gereben, L. Pusztai, System size and trajectory length dependence of the static structure factor and the diffusion coefficient as calculated from molecular dynamics simulations: The case of SPC/E water, J. Mol. Liq. 161 (2011) 36–40. doi:10.1016/j.molliq.2011.04.004.

[90] T. Darden, D. York, L. Pedersen, Particle mesh Ewald: An N·log(N) method for Ewald sums in large systems, J. Chem. Phys. 98 (1993) 10089–10092. doi:10.1063/1.464397.

[91] U. Essmann, L. Perera, M.L. Berkowitz, T. Darden, H. Lee, L.G. Pedersen, A smooth particle mesh Ewald method, J. Chem. Phys. 103 (1995) 8577–8593. doi:10.1063/1.470117.

[92] M.P. Allen, D.J. Tildesley, Computer Simulation of Liquids, Oxford University Press, Oxford, 1987. doi:10.1093/oso/9780198803195.001.0001.

[93] S. Miyamoto, P.A. Kollman, SETTLE: An analytical version of the SHAKE and RATTLE algorithm for rigid water models, J. Comput. Chem. 13 (1992) 952–962. doi:10.1002/jcc.540130805.

[94] S. Nosé, A molecular dynamics method for simulations in the canonical ensemble, Mol. Phys. 52 (1984) 255–268. doi:10.1080/00268978400101201.

[95] W.G. Hoover, Canonical dynamics: Equilibrium phase-space distributions, Phys. Rev. A 31 (1985) 1695–1697. doi:10.1103/PhysRevA.31.1695.

[96] M. Parrinello, A. Rahman, Polymorphic transitions in single crystals: A new molecular dynamics method, J. Appl. Phys. 52 (1981) 7182–7190. doi:10.1063/1.328693.

[97] S. Nosé, M.L. Klein, Constant pressure molecular dynamics for molecular systems, Mol. Phys. 50 (1983) 1055–1076. doi:10.1080/00268978300102851.

[98] D. Waasmaier, A. Kirfel, New analytical scattering-factor functions for free atoms and ions, Acta Crystallogr. Sect. A Found. Crystallogr. 51 (1995) 416–431. doi:10.1107/S0108767394013292.

[99] P.J. Brown, A.G. Fox, E.N. Maslen, M.A. O'Keefe, B.T.M. Willis, Intensity of diffracted intensities, in: Int. Tables Crystallogr. Vol. C, International Union of Crystallography, Chester, England, 2006: pp. 554–595. doi:10.1107/97809553602060000600.

[100] V.F. Sears, Neutron scattering lengths and cross sections, Neutron News. 3 (1992) 26–37. doi:10.1080/10448639208218770.

[101] L.B. Skinner, C.J. Benmore, J.C. Neuefeind, J.B. Parise, The structure of water around the compressibility minimum, J. Chem. Phys. 141 (2014) 214507. doi:10.1063/1.4902412.

[102] N. Ohtomo, K. Tokiwano, K. Arakawa, The structure of liquid water by neutron scattering. II. Temperature dependence of the liquid structure, Bull. Chem. Soc. Jpn. 55 (1982) 2788–2795. doi:10.1246/bcsj.55.2788.

[103] I.-C. Yeh, G. Hummer, System-size dependence of diffusion coefficients and viscosities from molecular dynamics simulations with periodic boundary conditions, J. Phys. Chem. B 108 (2004) 15873–15879. doi:10.1021/jp0477147.

[104] H.J.C. Berendsen, J.R. Grigera, T.P. Straatsma, The missing term in effective pair potentials, J. Phys. Chem. 91 (1987) 6269–6271. doi:10.1021/j100308a038.

[105] G. Lamoureux, A.D. MacKerell, B. Roux, A simple polarizable model of water based on classical Drude oscillators, J. Chem. Phys. 119 (2003) 5185–5197. doi:10.1063/1.1598191.

[106] H. Nada, J.P.J.M. van der Eerden, An intermolecular potential model for the simulation of ice and water near the melting point: A six-site model of H2O, J. Chem. Phys. 118 (2003) 7401–7413. doi:10.1063/1.1562610.

[107] Y. Wu, H.L. Tepper, G.A. Voth, Flexible simple point-charge water model with improved liquid-state properties, J. Chem. Phys. 124 (2006) 24503. doi:10.1063/1.2136877.

[108] G. Lamoureux, E. Harder, I. V Vorobyov, B. Roux, A.D. MacKerell, A polarizable model of water for molecular dynamics simulations of biomolecules, Chem. Phys. Lett. 418 (2006) 245–249. doi:10.1016/j.cplett.2005.10.135.

[109] Y.-L. Huang, T. Merker, M. Heilig, H. Hasse, J. Vrabec, Molecular modeling and simulation of vapor–liquid equilibria of ethylene oxide, ethylene glycol, and water as well as their binary mixtures, Ind. Eng. Chem. Res. 51 (2012) 7428–7440. doi:10.1021/ie300248z.

[110] A. Kulkarni, R. Fingerhut, M. Kohns, H. Hasse, J. Vrabec, Correction to "Molecular modeling and simulation of vapor–liquid equilibria of ethylene oxide, ethylene glycol, and water as well as their binary mixtures," Ind. Eng. Chem. Res. 59 (2020) 20232–20234. doi:10.1021/acs.iecr.0c04937.

[111] S. Piana, A.G. Donchev, P. Robustelli, D.E. Shaw, Water dispersion interactions strongly influence simulated structural properties of disordered protein states, J. Phys. Chem. B 119 (2015) 5113–5123. doi:10.1021/jp508971m.

[112] S. Izadi, A. V. Onufriev, Accuracy limit of rigid 3-point water models, J. Chem. Phys. 145 (2016) 74501. doi:10.1063/1.4960175.

[113] H. Nada, Anisotropy in geometrically rough structure of ice prismatic plane interface during growth: Development of a modified six-site model of H$_2$O and a molecular dynamics simulation, J. Chem. Phys. 145 (2016) 244706. doi:10.1063/1.4973000.

[114] T.D. Loeffler, H. Chan, K. Sasikumar, B. Narayanan, M.J. Cherukara, S. Gray, et al., Teaching an old dog new tricks: Machine learning an improved TIP3P potential model for liquid–vapor phase phenomena, J. Phys. Chem. C 123 (2019) 22643–22655. doi:10.1021/acs.jpcc.9b06348.

[115] Y. Qiu, P.S. Nerenberg, T. Head-Gordon, L.-P. Wang, Systematic optimization of water models using liquid/vapor surface tension data, J. Phys. Chem. B 123 (2019) 7061–7073. doi:10.1021/acs.jpcb.9b05455.

[116] R. Fuentes-Azcatl, Flexible model of water based on the dielectric and electromagnetic spectrum properties: TIP4P/ε$_{Flex}$, J. Mol. Liq. 338 (2021) 116770. doi:10.1016/j.molliq.2021.116770.

[117] X. Teng, W. Yu, A.D. MacKerell Jr, Revised 4-Point water model for the classical Drude oscillator polarizable force field: SWM4-HLJ, J. Chem. Theory Comput. 20 (2024) 10034–10044. doi:10.1021/acs.jctc.4c00966.

[118] X. Teng, W. Yu, A.D. MacKerell, Computationally efficient polarizable MD simulations: A simple water model for the classical Drude oscillator polarizable force field, J. Phys. Chem. Lett. 16 (2025) 1016–1023. doi:10.1021/acs.jpclett.4c03451.

[119] G.S. Kell, Density, thermal expansivity, and compressibility of liquid water from 0° to 150°C. Correlations and tables for atmospheric pressure and saturation reviewed and expressed on 1968 temperature scale, J. Chem. Eng. Data. 20 (1975) 97–105. doi:10.1021/je60064a005.

[120] D.E. Hare, C.M. Sorensen, The density of supercooled water. II. Bulk samples cooled to the homogeneous nucleation limit, J. Chem. Phys. 87 (1987) 4840–4845. doi:10.1063/1.453710.

[121] M. Holz, S.R. Heil, A. Sacco, Temperature-dependent self-diffusion coefficients of water and six selected molecular liquids for calibration in accurate 1H NMR PFG measurements, Phys. Chem. Chem. Phys. 2 (2000) 4740–4742. doi:10.1039/b005319h.

[122] L. Bosio, J. Teixeira, J.C. Dore, D.C. Steytler, P. Chieux, Neutron diffraction studies of water, Mol. Phys. 50 (1983) 733–740. doi:10.1080/00268978300102651.

Tables

**Table 1** Simulation stages.

| Step | Procedure | Algorithm | Temperature [K] | Ensemble | Run time [ns] / steps |
|---|---|---|---|---|---|
| 1 | Energy minimization 1 | Steepest descent | | | 1000 |
| 2 | Energy minimization 2 | Steepest descent | | | 50000 |
| 3 | Equilibration | Leapfrog | 366 K | NVT | 0.1 |
| 4 | Equilibration | Leapfrog | 366 K | NpT | 2.0 |
| 5 | Cooling down | Leapfrog | $T_i \to T_{i+1}$ | NpT | 1.0 |
| 6 | Equilibration (and density determination) | Leapfrog | $T_{i+1}$ | NpT | 1.0 |
| Repeating Steps 5 and 6 with $T_i$ = 366 K, 343 K, 324 K, 295 K, 284 K, 268 K, 254 K | | | | | |
| 7 | Production | Leap-frog | $T_i$ | NVT | 1.0 |

**Table 2** Investigated water potential models

| Name | Year published | Sites | Specialty | Time step [fs] | Reference |
|---|---|---|---|---|---|
| SPC | 1981 | 3 | no | 2 | [58] |
| TIP3P | 1983 | 3 | no | 2 | [83] |
| TIP4P | 1983 | 4 | no | 2 | [83] |
| SPC/E | 1987 | 3 | no | 2 | [104] |
| TIP5P | 2000 | 5 | no | 2 | [84] |
| SWM4-DP | 2003 | 4+1 | polarizable | 1 | [105] |
| TIP6P | 2003 | 6 | LJ on H | 2 | [106] |
| TIP4P-Ew | 2004 | 4 | no | 2 | [76] |
| TIP5P-Ew | 2004 | 5 | no | 2 | [44] |
| TIP4P/2005 | 2005 | 4 | no | 2 | [45] |
| SPC/Fw | 2006 | 3 | flexible | 0.5 | [107] |
| SWM4-NDP | 2006 | 4+1 | polarizable | 1 | [108] |
| TIP4P/2005f | 2011 | 4 | flexible | 0.5 | [46] |
| TIP4Q | 2011 | 4 | no | 2 | [48] |
| HUANG | 2012 | 4 | no | 2 | [109,110] |
| SWM6 | 2013 | 6+1 | polarizable | 1 | [43] |
| TIP3P-FB | 2014 | 3 | no | 2 | [49] |
| OPC | 2014 | 4 | no | 2 | [57] |
| TIP4P/ε | 2014 | 4 | no | 2 | [40] |
| TIP4P-FB | 2014 | 4 | no | 2 | [49] |
| SPC/ε | 2015 | 3 | no | 2 | [42] |
| TIP4P-D | 2015 | 4 | no | 2 | [111] |
| OPC3 | 2016 | 3 | no | 2 | [112] |
| TIP6P-Ew | 2016 | 6 | LJ on H | 2 | [113] |
| TIP3P-LJOPT | 2018 | 3 | no | 2 | [60] |
| TIP3P-Buck | 2018 | 3 | Buckingham | 2 | [60] |
| TIP4P-LJOPT | 2018 | 4 | no | 2 | [60] |
| TIP4P-Buck | 2018 | 4 | Buckingham | 2 | [60] |
| TIP5P/2018 | 2018 | 5 | no | 2 | [54] |
| ML-TIP3P | 2019 | 3 | LJ on H | 2 | [114] |
| TIP3P-ST | 2019 | 3 | no | 2 | [115] |
| TIP4P-ST | 2019 | 4 | no | 2 | [115] |
| TIP7P | 2019 | 7 | LJ on H | 2 | [53] |
| FBA/ε | 2020 | 3 | flexible | 0.5 | [41] |
| TIP4P/$ε_{Flex}$ | 2021 | 4 | flexible | 0.5 | [116] |
| TIP4P-BG | 2021 | 4 | no | 2 | [35] |
| OPC3-pol | 2022 | 3+1 | polarizable | 1 | [50] |
| OPTI-1T | 2023 | 3 | no | 2 | [28] |
| OPTI-3T | 2023 | 3 | no | 2 | [28] |
| ECCw2024 | 2024 | 4 | no | 2 | [29] |
| SWM4-HLJ | 2024 | 4+1 | polarizable, LJ on H | 1 | [117] |
| SWM4-dip | 2024 | 4+1 | polarizable | 1 | [117] |
| SWM4-HD | 2024 | 4+1 | polarizable, LJ on H | 1 | [117] |
| SWM3 | 2025 | 3+1 | polarizable | 1 | [118] |

**Table 3** Relative $R$-factors ($R_{rel} = R/R_{best}$) of the XRD structure factors ($S^X(Q)$) obtained with models at different temperatures. The best $R$-factor values ($R_{best}$) are also presented (in bold).

| model | 254 K | 268 K | 284 K | 295 K | 324 K | 343 K | 366 K |
|---|---|---|---|---|---|---|---|
| **$R_{best}$** | **16.7** | **15.1** | **9.6** | **9.3** | **8.1** | **8.2** | **9.3** |
| OPC | 1.00 | 1.00 | 1.00 | 1.00 | 1.00 | 1.00 | 1.00 |
| ECCw2024 | 1.04 | 1.19 | 1.32 | 1.12 | 1.20 | 1.40 | 1.33 |
| TIP4P-Buck | 1.13 | 1.24 | 1.34 | 1.18 | 1.25 | 1.50 | 1.35 |
| TIP4Q | 1.18 | 1.30 | 1.38 | 1.27 | 1.32 | 1.56 | 1.45 |
| TIP5P/2018 | 1.20 | 1.32 | 1.44 | 1.26 | 1.31 | 1.53 | 1.42 |
| TIP4P/ε | 1.18 | 1.31 | 1.44 | 1.30 | 1.37 | 1.58 | 1.39 |
| SWM4-DP | 1.30 | 1.32 | 1.42 | 1.29 | 1.35 | 1.49 | 1.40 |
| TIP4P/2005 | 1.16 | 1.30 | 1.41 | 1.31 | 1.36 | 1.58 | 1.47 |
| TIP4P-D | 1.27 | 1.32 | 1.46 | 1.37 | 1.46 | 1.55 | 1.42 |
| TIP4P-FB | 1.19 | 1.33 | 1.51 | 1.35 | 1.37 | 1.70 | 1.51 |
| SWM4-dip | 1.23 | 1.33 | 1.54 | 1.43 | 1.45 | 1.67 | 1.51 |
| TIP4P-ST | 1.20 | 1.36 | 1.51 | 1.43 | 1.47 | 1.67 | 1.52 |
| TIP4P-Ew | 1.22 | 1.37 | 1.56 | 1.36 | 1.49 | 1.68 | 1.60 |
| SWM4-NDP | 1.46 | 1.45 | 1.55 | 1.43 | 1.44 | 1.54 | 1.45 |
| TIP7P | 1.27 | 1.39 | 1.51 | 1.41 | 1.49 | 1.70 | 1.56 |
| TIP4P | 1.29 | 1.38 | 1.47 | 1.33 | 1.42 | 1.79 | 1.70 |
| OPTI-1T | 1.29 | 1.37 | 1.49 | 1.38 | 1.48 | 1.82 | 1.58 |
| TIP4P/2005f | 1.25 | 1.40 | 1.58 | 1.48 | 1.57 | 1.68 | 1.50 |
| TIP5P-Ew | 1.75 | 1.62 | 1.61 | 1.38 | 1.36 | 1.76 | 1.61 |
| OPTI-3T | 1.40 | 1.49 | 1.62 | 1.47 | 1.62 | 1.85 | 1.76 |
| TIP6P | 1.91 | 1.60 | 1.62 | 1.46 | 1.48 | 1.72 | 1.56 |
| OPC3 | 1.36 | 1.51 | 1.67 | 1.54 | 1.68 | 2.02 | 1.70 |
| TIP4P/εFlex | 1.39 | 1.53 | 1.84 | 1.65 | 1.67 | 1.89 | 1.66 |
| TIP4P-LJOPT | 1.26 | 1.41 | 1.71 | 1.58 | 1.74 | 2.04 | 2.03 |
| TIP6P-Ew | 1.54 | 1.63 | 1.67 | 1.58 | 1.70 | 1.90 | 1.80 |
| SWM6 | 1.50 | 1.62 | 1.93 | 1.74 | 1.66 | 1.91 | 1.67 |
| TIP4P-BG | 1.36 | 1.60 | 1.93 | 1.72 | 1.82 | 2.13 | 1.86 |
| SWM4-HLJ | 1.51 | 1.58 | 1.83 | 1.76 | 1.86 | 2.08 | 1.82 |
| SWM4-HD | 1.44 | 1.59 | 1.85 | 1.84 | 1.84 | 2.08 | 1.81 |
| TIP3P-LJOPT | 1.48 | 1.59 | 1.81 | 1.68 | 1.83 | 2.15 | 1.93 |
| SPC/E | 1.47 | 1.60 | 1.84 | 1.75 | 1.85 | 2.16 | 1.94 |
| SPC/ε | 1.39 | 1.64 | 1.93 | 1.78 | 1.96 | 2.16 | 1.90 |
| TIP5P | 1.93 | 1.87 | 1.94 | 1.65 | 1.65 | 2.03 | 1.87 |
| SPC | 1.69 | 1.73 | 1.93 | 1.76 | 1.90 | 2.24 | 1.92 |
| TIP3P-Buck | 1.56 | 1.68 | 1.93 | 1.82 | 1.98 | 2.28 | 2.03 |
| TIP3P-FB | 1.47 | 1.64 | 1.95 | 1.79 | 2.19 | 2.28 | 2.15 |
| TIP3P | 1.93 | 1.86 | 2.16 | 1.92 | 1.95 | 2.07 | 1.98 |
| SPC/Fw | 1.58 | 1.77 | 2.13 | 2.01 | 2.22 | 2.56 | 2.20 |
| FBA/ε | 1.58 | 1.81 | 2.17 | 2.10 | 2.31 | 2.50 | 2.16 |
| TIP3P-ST | 1.65 | 1.98 | 2.52 | 2.42 | 2.63 | 2.87 | 2.49 |
| OPC3-POL | 2.13 | 2.24 | 2.79 | 2.70 | 3.01 | 3.19 | 2.79 |
| SWM3 | 2.01 | 2.20 | 2.76 | 2.66 | 3.09 | 3.30 | 3.02 |
| ML-TIP3P | 4.08 | 4.04 | 5.68 | 5.34 | 5.44 | 4.86 | 3.89 |
| HUANG | 4.03 | 4.00 | 5.62 | 5.43 | 5.46 | 5.09 | 4.13 |

**Table 4** Relative $R$-factors ($R_{rel} = R/R_{best}$) of the ND structure factors ($S^N(Q)$) obtained with models at different temperatures. The labels (O) and (S) denote experimental data from Ref. [102] and Refs. [65, 69], respectively. The best $R$-factor values ($R_{best}$) are also presented (in bold).

| model | 284 K (S) | 295 K (S) | 295 K (O) | 324 K (O) | 343 K (O) | 366 K (O) |
|---|---|---|---|---|---|---|
| $R_{best}$ | **24.3** | **14.2** | **30.3** | **27.5** | **28.2** | **31.5** |
| TIP4P/2005f | 1.06 | 1.21 | 1.00 | 1.00 | 1.01 | 1.01 |
| TIP4P/εFlex | 1.12 | 1.04 | 1.10 | 1.09 | 1.04 | 1.01 |
| TIP4P-BG | 1.00 | 1.27 | 1.01 | 1.06 | 1.02 | 1.05 |
| TIP4P-Ew | 1.03 | 1.31 | 1.02 | 1.07 | 1.06 | 1.04 |
| SWM4-dip | 1.10 | 1.34 | 1.03 | 1.08 | 1.02 | 1.03 |
| SWM6 | 1.16 | 1.38 | 1.06 | 1.08 | 1.00 | 1.00 |
| TIP4P | 1.06 | 1.33 | 1.04 | 1.13 | 1.06 | 1.06 |
| TIP4P/2005 | 1.10 | 1.33 | 1.03 | 1.10 | 1.06 | 1.05 |
| TIP4P-FB | 1.09 | 1.34 | 1.04 | 1.14 | 1.04 | 1.06 |
| TIP4P-ST | 1.11 | 1.34 | 1.02 | 1.10 | 1.07 | 1.07 |
| TIP4P/ε | 1.10 | 1.36 | 1.04 | 1.11 | 1.07 | 1.13 |
| TIP4Q | 1.16 | 1.39 | 1.06 | 1.16 | 1.08 | 1.08 |
| TIP6P-Ew | 1.14 | 1.00 | 1.16 | 1.22 | 1.28 | 1.26 |
| OPC3 | 1.16 | 1.12 | 1.17 | 1.25 | 1.22 | 1.24 |
| SWM4-DP | 1.21 | 1.45 | 1.11 | 1.16 | 1.12 | 1.11 |
| OPTI-3T | 1.21 | 1.15 | 1.19 | 1.25 | 1.24 | 1.21 |
| TIP6P | 1.15 | 1.05 | 1.17 | 1.27 | 1.29 | 1.33 |
| SWM4-HD | 1.13 | 1.47 | 1.08 | 1.21 | 1.16 | 1.18 |
| SWM4-NDP | 1.25 | 1.53 | 1.16 | 1.21 | 1.11 | 1.10 |
| TIP7P | 1.17 | 1.46 | 1.13 | 1.25 | 1.19 | 1.21 |
| TIP4P-D | 1.25 | 1.50 | 1.10 | 1.22 | 1.14 | 1.18 |
| SWM4-HLJ | 1.20 | 1.46 | 1.16 | 1.25 | 1.19 | 1.20 |
| TIP5P | 1.15 | 1.48 | 1.12 | 1.27 | 1.22 | 1.20 |
| OPTI-1T | 1.22 | 1.70 | 1.11 | 1.21 | 1.12 | 1.12 |
| TIP5P/2018 | 1.19 | 1.51 | 1.14 | 1.30 | 1.22 | 1.19 |
| TIP4P-Buck | 1.28 | 1.87 | 1.16 | 1.25 | 1.18 | 1.15 |
| ECCw2024 | 1.28 | 1.94 | 1.16 | 1.25 | 1.19 | 1.14 |
| TIP5P-Ew | 1.25 | 1.55 | 1.20 | 1.41 | 1.32 | 1.30 |
| SPC/ε | 1.34 | 1.28 | 1.38 | 1.45 | 1.47 | 1.43 |
| SPC | 1.36 | 1.25 | 1.40 | 1.48 | 1.47 | 1.43 |
| SPC/E | 1.32 | 1.21 | 1.38 | 1.49 | 1.51 | 1.46 |
| TIP3P-LJOPT | 1.36 | 1.30 | 1.40 | 1.52 | 1.52 | 1.47 |
| TIP3P | 1.50 | 1.89 | 1.35 | 1.41 | 1.29 | 1.20 |
| TIP4P-LJOPT | 1.39 | 2.09 | 1.25 | 1.39 | 1.33 | 1.25 |
| TIP3P-FB | 1.41 | 1.31 | 1.50 | 1.59 | 1.63 | 1.55 |
| SPC/Fw | 1.58 | 1.48 | 1.67 | 1.79 | 1.82 | 1.73 |
| TIP3P-Buck | 1.59 | 1.50 | 1.67 | 1.80 | 1.84 | 1.75 |
| TIP3P-ST | 1.62 | 1.74 | 1.70 | 1.78 | 1.82 | 1.71 |
| OPC | 1.93 | 3.13 | 1.67 | 1.87 | 1.77 | 1.60 |
| FBA/ε | 1.91 | 2.04 | 1.98 | 2.14 | 2.18 | 2.04 |
| ML-TIP3P | 2.89 | 3.95 | 2.44 | 2.55 | 2.30 | 2.00 |
| HUANG | 3.24 | 4.02 | 3.21 | 3.51 | 3.47 | 3.20 |
| SWM3 | 3.17 | 3.85 | 3.33 | 3.93 | 4.01 | 3.78 |
| OPC3-POL | 3.40 | 4.16 | 3.57 | 4.25 | 4.34 | 4.08 |

Figures

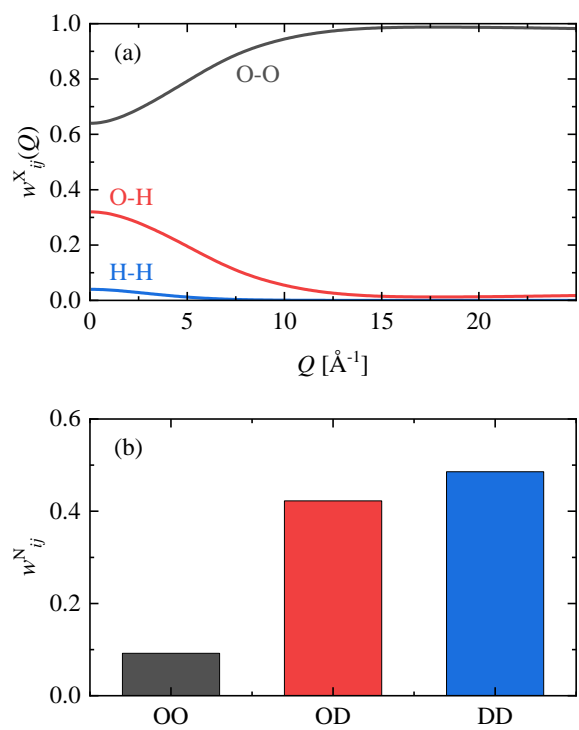

**Figure 1** (a) X-ray and (b) neutron weighting parameters used for the calculations of the X-ray and neutron total structure factors.

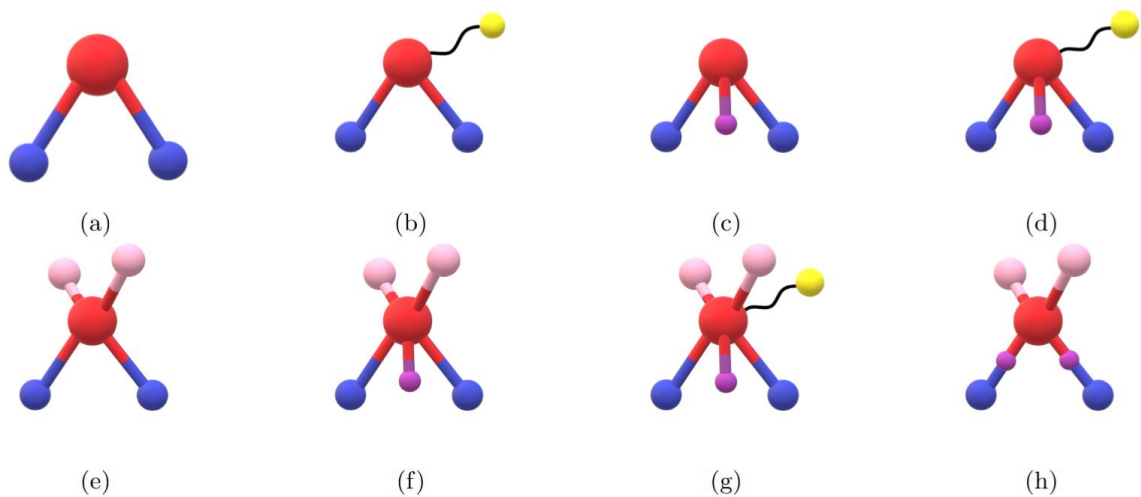

**Figure 2** Schematic drawing of the different type of water models investigated. (a) 3-site, (b) 3-site with shell, (c) 4-site, (d) 4-site with shell, (e) 5-site, (f) 6-site, (g) 6-site with shell, and (h) 7-site models. The large red sphere represents the oxygen atom, the blue spheres are the hydrogen atoms, small purple spheres are the M or M2 virtual sites, light pink spheres denote the L virtual sites and yellow spheres with a spring are the shell particles.

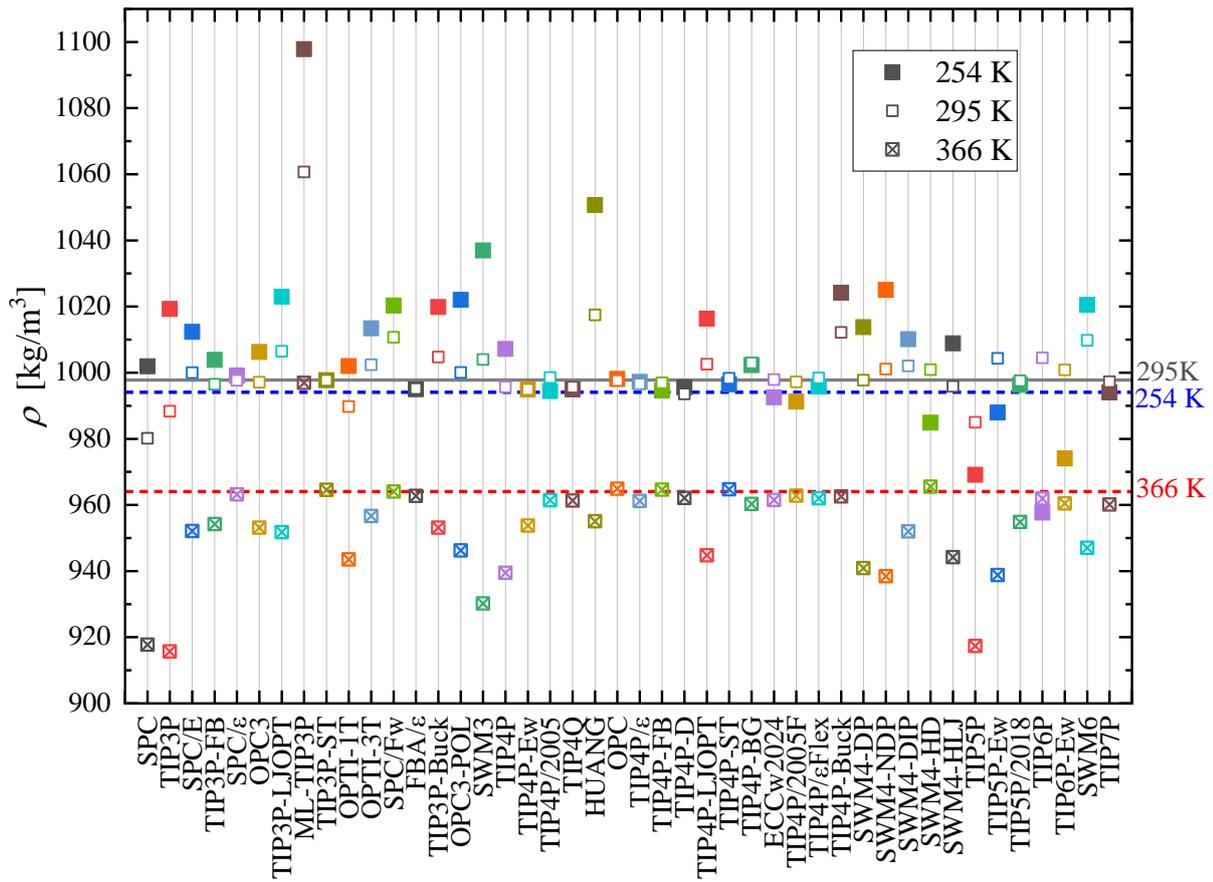

**Figure 3** Simulated density values obtained by different potential models at 254, 295, and 366 K. The experimental values [119, 120] are also shown for comparison as horizontal lines.

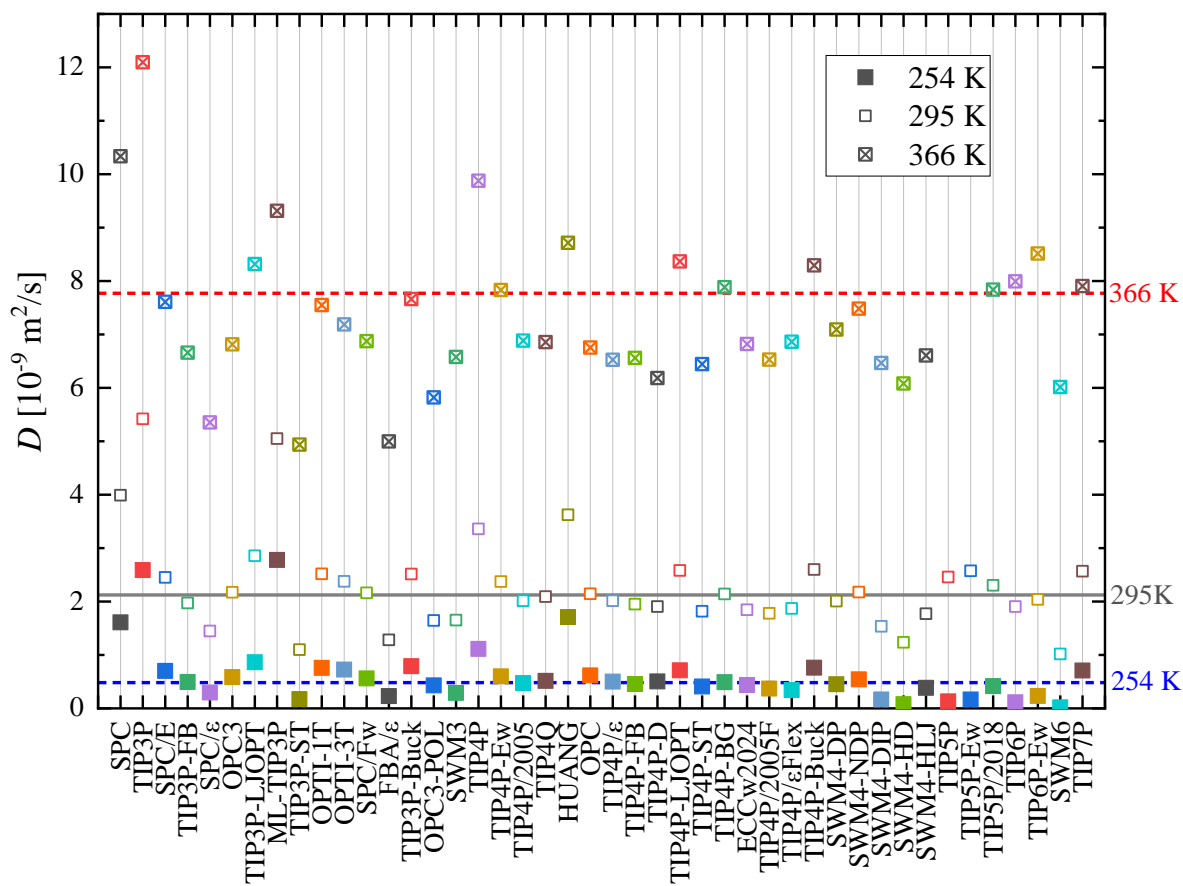

**Figure 4** Self-diffusion values obtained by using different potential models at 254, 295, and 366 K. The experimental values for 295, and 366 K, and an extrapolated value for 254 K [121] are also shown for comparison as horizontal lines.

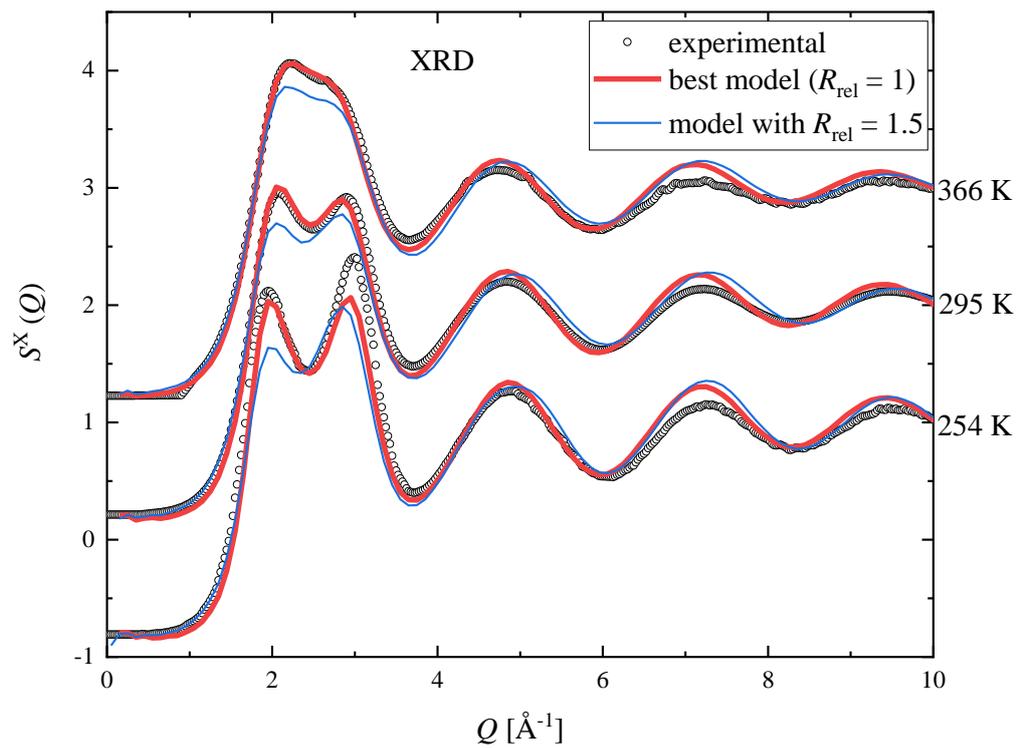

**Figure 5** Comparison of the simulated and experimental X-ray diffraction total structure factor at 3 different temperatures for models with $R_{rel} = 1$ (best fit) and $R_{rel} = 1.5$.

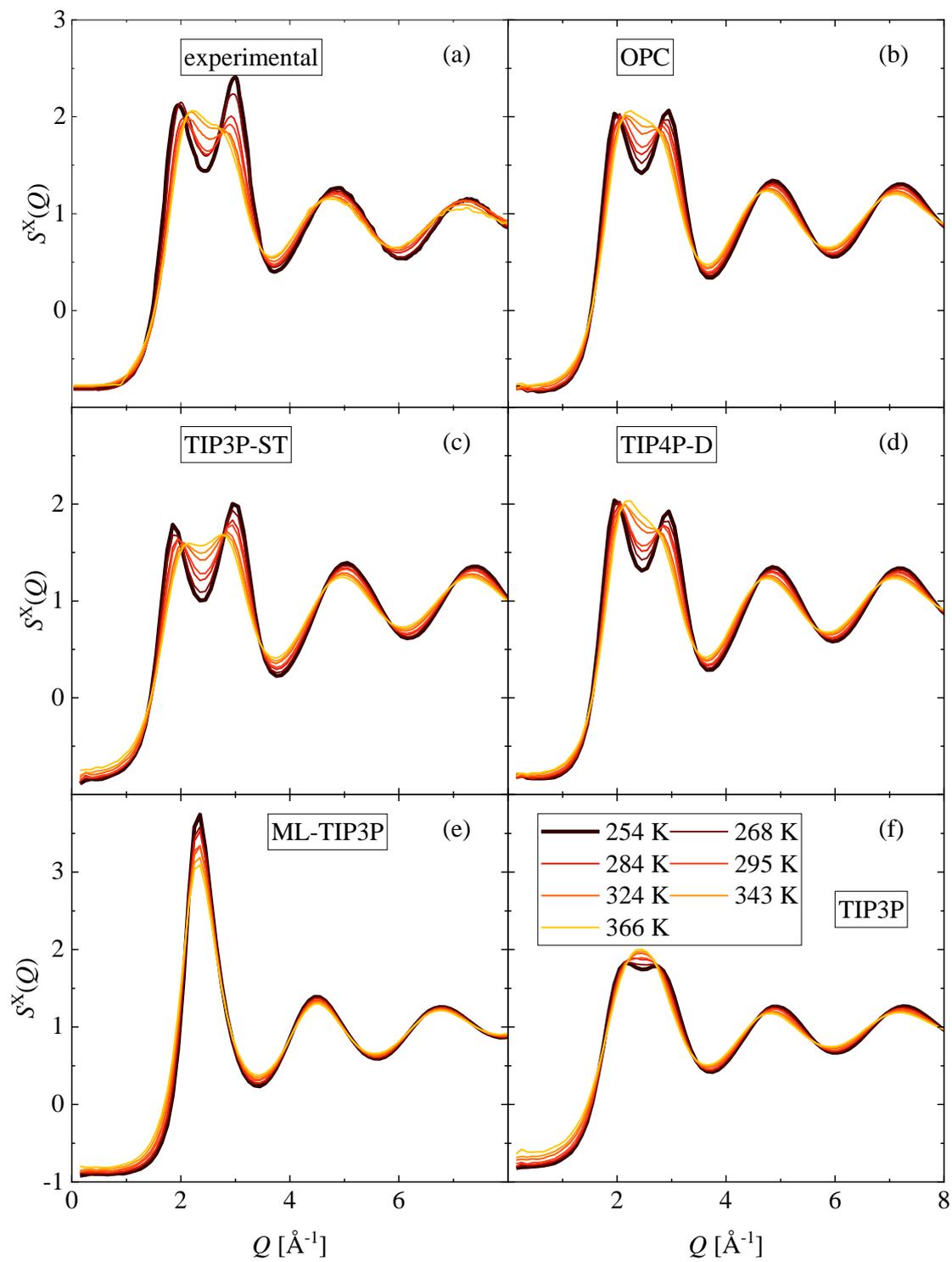

**Figure 6** Temperature dependence of the (a) experimental and (b-f) some simulated XRD total structure factor. Typical trends for (b) Group A, (c) Group B, and (d) Group C models.

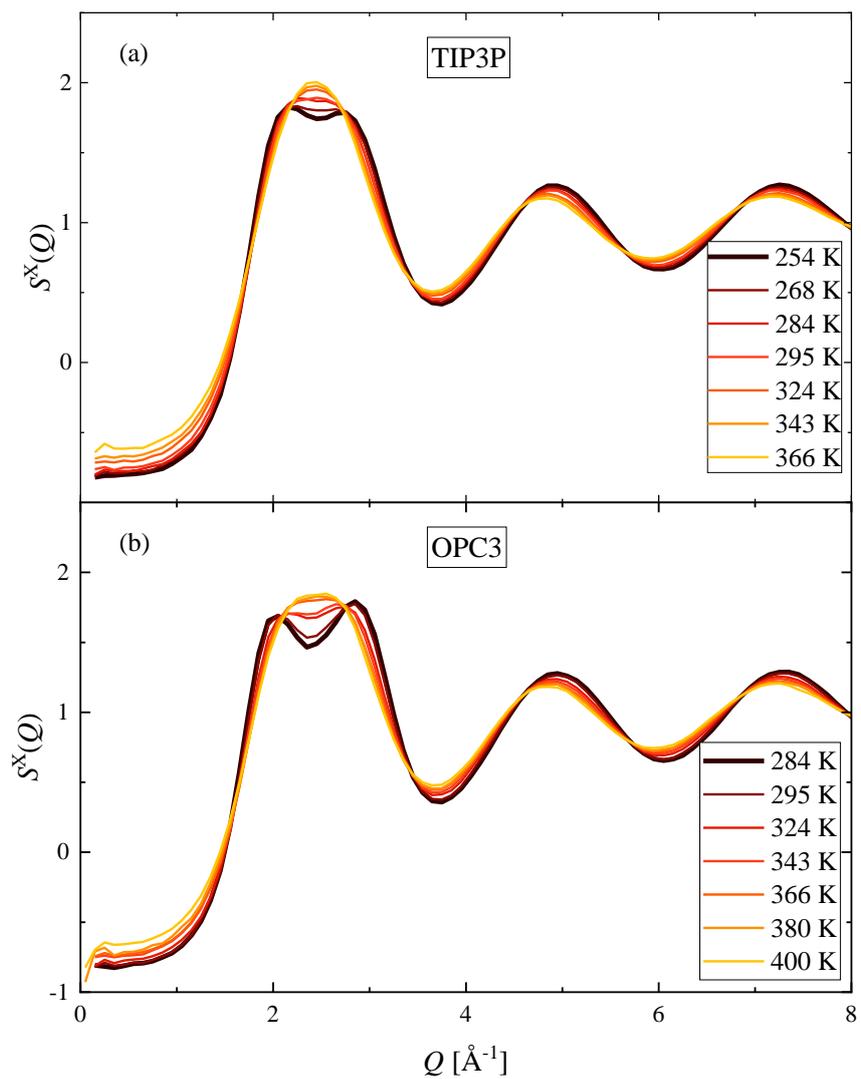

**Figure 7** Comparison of the X-ray structure factors of (a) TIP3P and (b) OPC3 models. The curves of the OPC3 model are shown at higher temperatures.

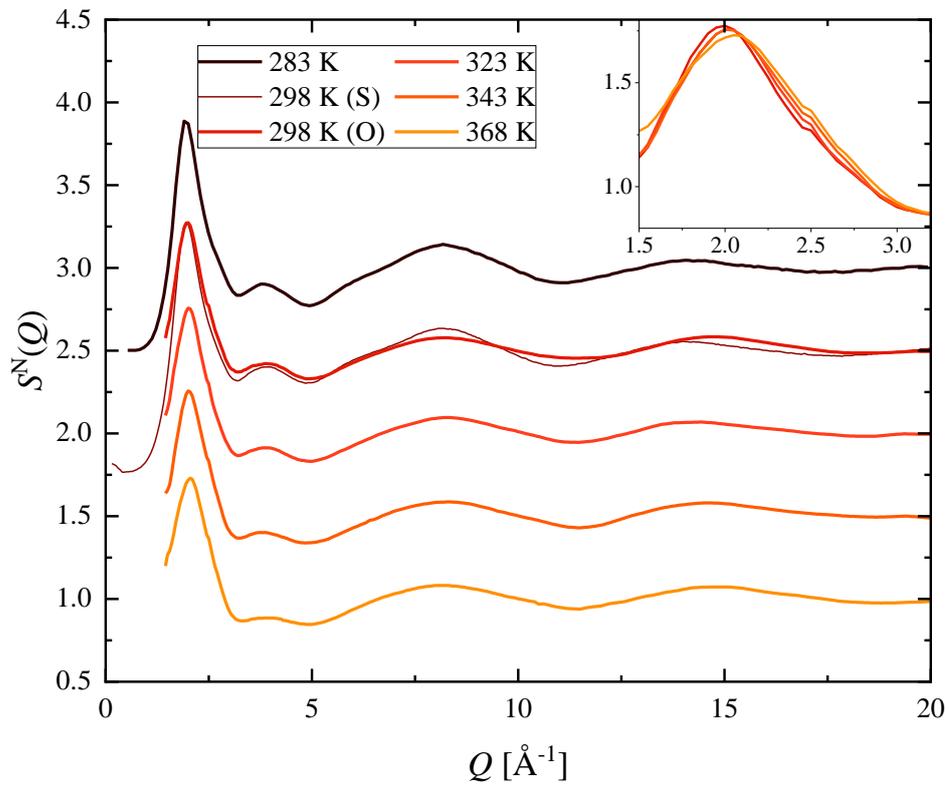

**Figure 8** Experimental neutron diffraction total structure factors. The labels (O) and (S) denote experimental data from Ref. [102] and Ref. [65], respectively. The inset is an enlargement of the main peak region (only for the data from Ref. [102]).

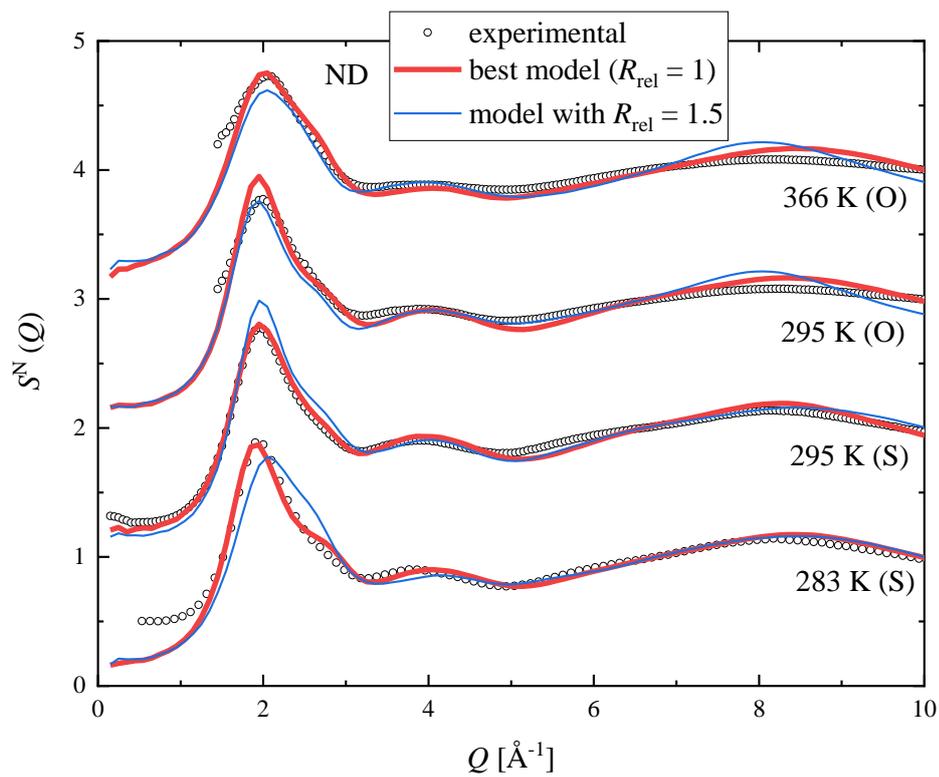

**Figure 9** Comparison of the simulated and experimental neutron diffraction total structure factor for models with $R_{rel} = 1$ (best fit) and $R_{rel} = 1.5$. The labels (O) and (S) denote experimental data from Ref. [102] and Refs. [65, 69], respectively.

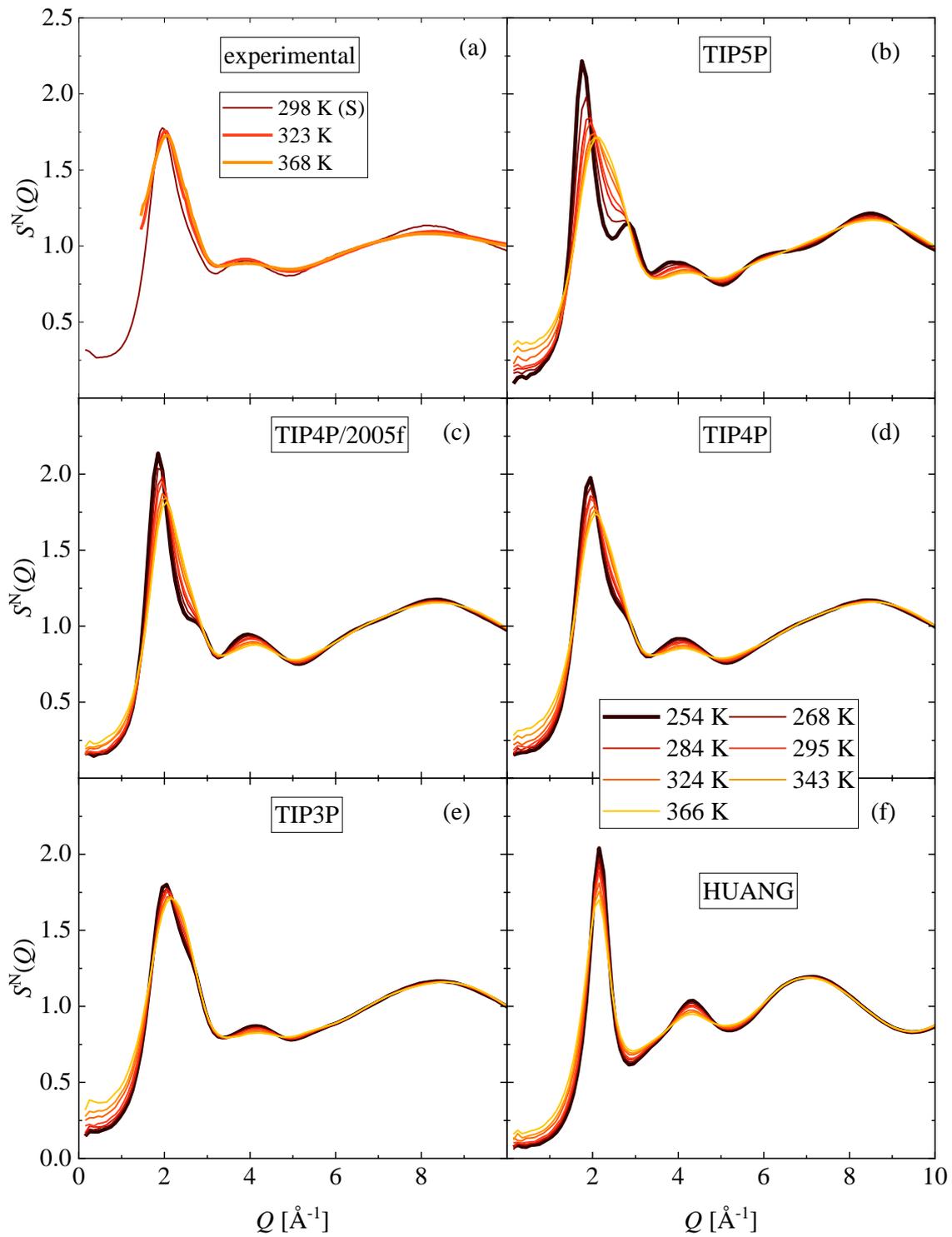

**Figure 10** Temperature dependence of the (a) experimental and (b-f) some simulated ND total structure factor.

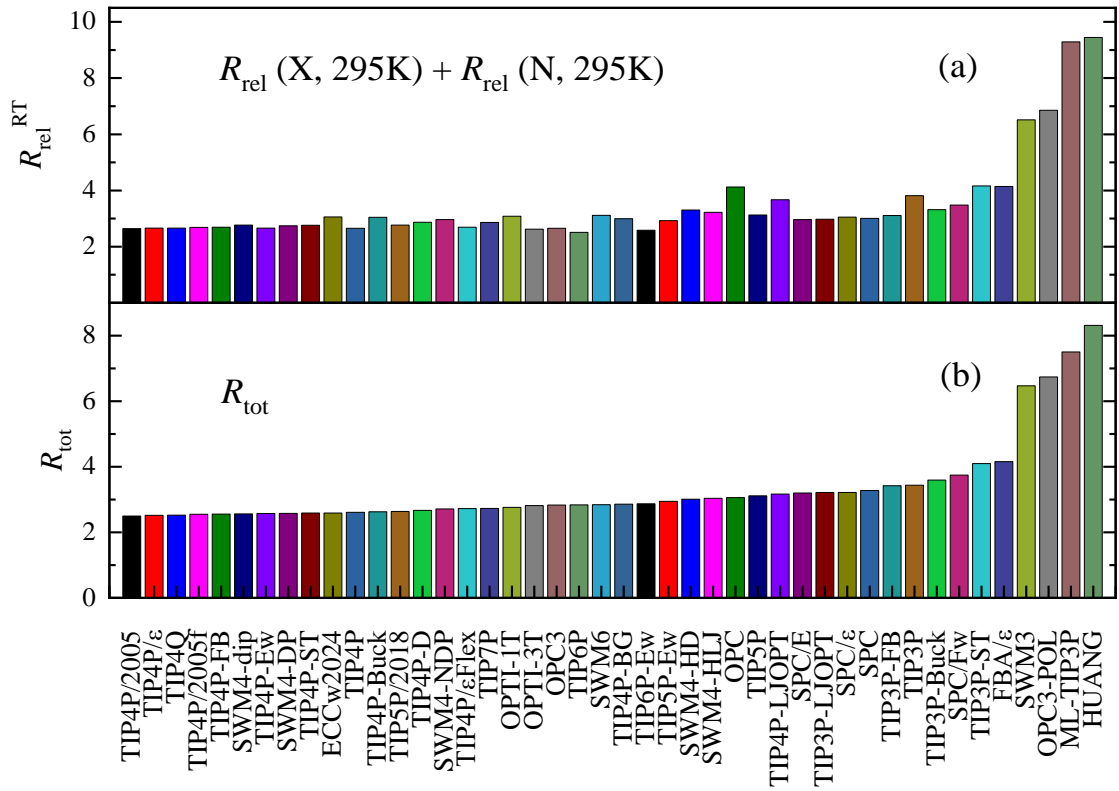

**Figure 11** Combined relative *R*-factors of the tested models (a) at room temperature and (b) for the full temperature range. For the definition of the *R*-factors, see the text.

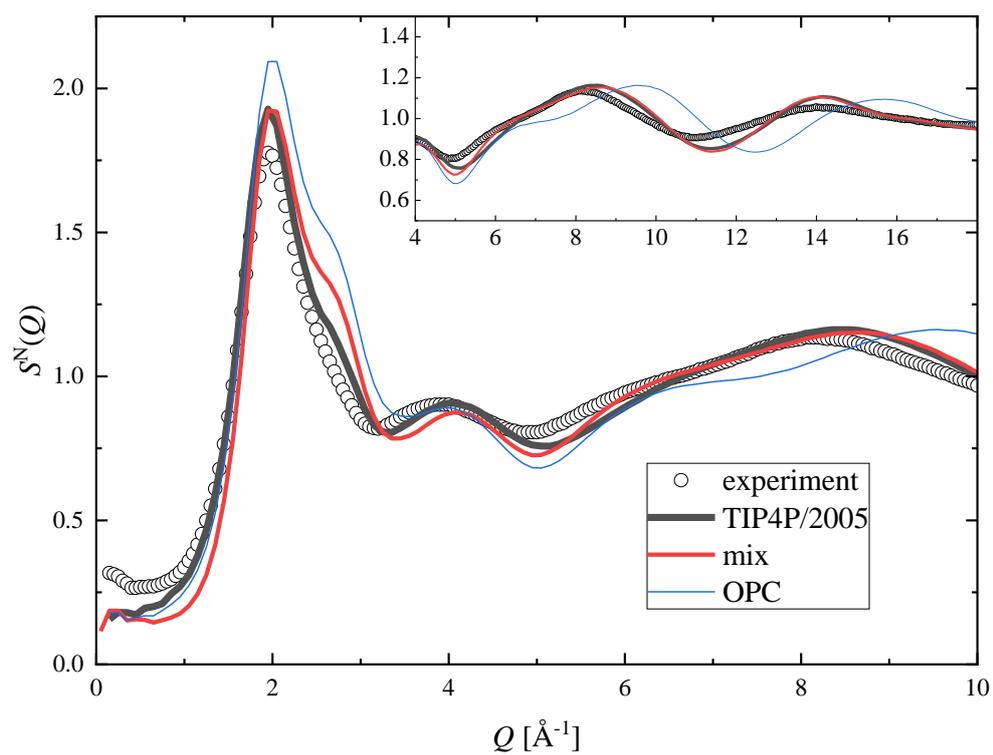

**Figure 12** Comparison of the experimental and simulated ND structure factors of TIP4P/2005, OPC and "mix" models at 295 K. The structure factor called "mix" is calculated from a hypothetical PRDF that combines the intramolecular part of the TIP4P/2005 model and the intramolecular part of the OPC model. The inset shows the curves at higher $Q$ range.

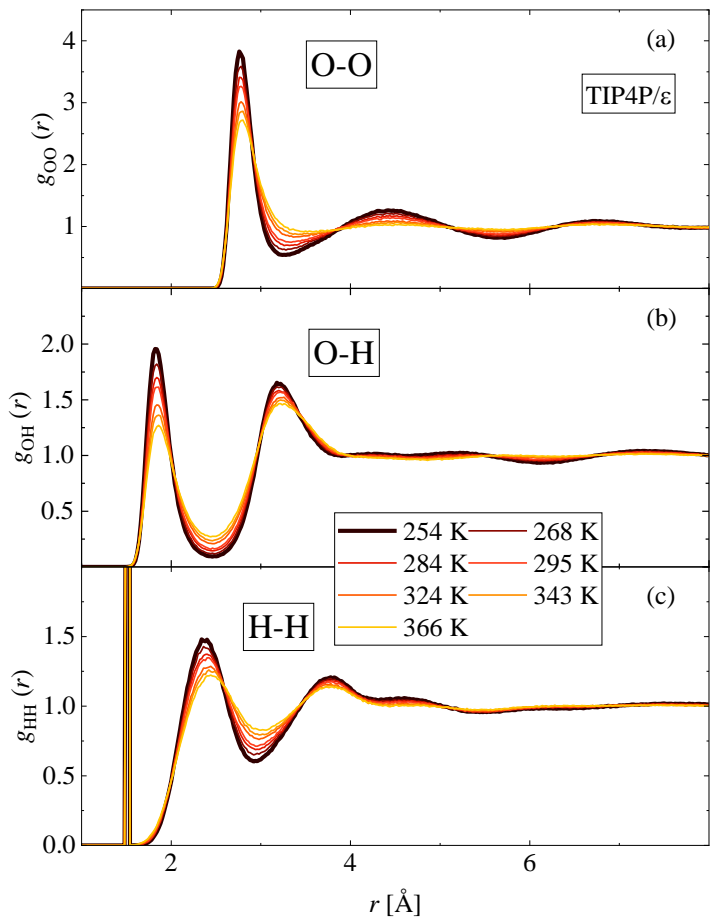

**Figure 13** Partial pair distribution functions at different temperatures obtained by using the TIP4P/ε model.

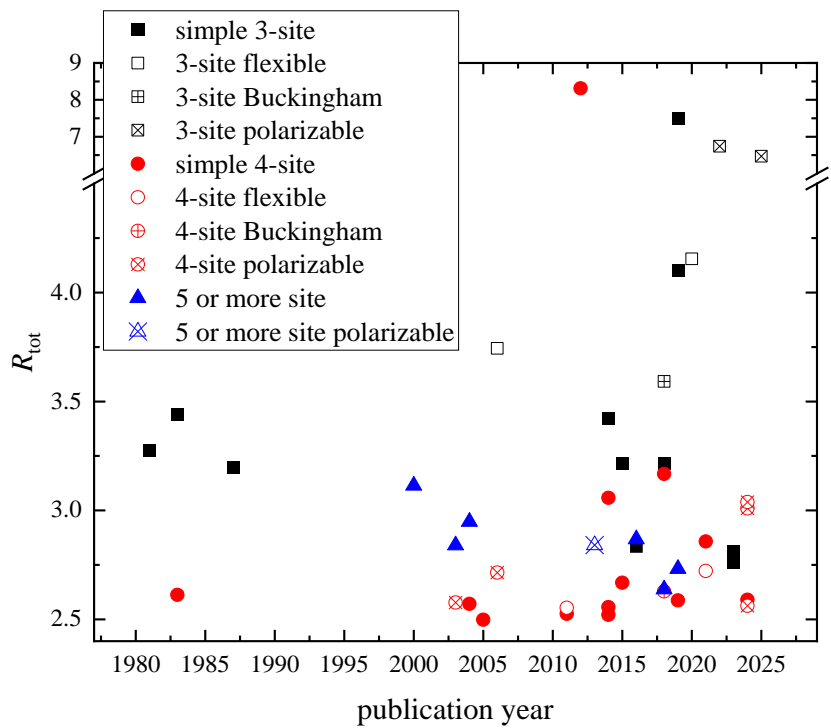

**Figure 14** Combined relative *R*-factors of the tested models as a function of the publication year of the models.

**Figure 15** Combined relative *R*-factors of the tested models as a function of the simulation speed.

**Supplementary Material**

for

# Comparison of water models for structure prediction

Bálint Soczó[1,2] and Ildikó Pethes[1,*]

[1]HUN-REN Wigner Research Centre for Physics, Konkoly Thege út 29-33., H-1121 Budapest, Hungary

[2] Faculty of Natural Sciences, Budapest University of Technology and Economics, Műegyetem rkp. 3., H-1111 Budapest, Hungary

[*] Corresponding author, e-mail address: pethes.ildiko@wigner.hun-ren.hu

# Tables

**Table S1** Parameters of the tested models using the Lennard-Jones potential. For the TIP7P models $\varepsilon_{OH}$ and $\sigma_{OH}$ are calculated according to the geometric combination rule: they are the geometric average of the parameters of the like pairs ($\varepsilon_{HH}$ and $\varepsilon_{OO}$, and $\sigma_{HH}$ and $\sigma_{OO}$, respectively). For the ML-TIP3P, SWM4-HLJ, SWM4-HD, TIP6P, and TIP6P-Ew models $\varepsilon_{OH}$ is calculated as the geometric average of $\varepsilon_{HH}$ and $\varepsilon_{OO}$, and $\sigma_{OH}$ is the arithmetic average of $\sigma_{HH}$ and $\sigma_{OO}$ (Lorentz-Berthelot combination rule). In the case of the other models $\sigma_{OH}$ and $\varepsilon_{OH}$ are zero.

| model | $d_{OH}$ [nm] | $\theta_{HOH}$ [°] | $\sigma_{OO}$ [nm] | $\varepsilon_{OO}$ [kJ/mol] | $\varepsilon_{HH}$ [kJ/mol] | $\sigma_{HH}$ [nm] | $q_O$ [e] | $q_H$ [e] | $d_{OM}$ [nm] | $q_M$ [e] |
|---|---|---|---|---|---|---|---|---|---|---|
| SPC | 0.10000 | 109.47 | 0.31656 | 0.65017 | 0 | 0 | -0.82000 | 0.41000 | | |
| TIP3P | 0.09572 | 104.52 | 0.31507 | 0.63627 | 0 | 0 | -0.83400 | 0.41700 | | |
| SPC/E | 0.10000 | 109.47 | 0.31656 | 0.65017 | 0 | 0 | -0.84760 | 0.42380 | | |
| SPC/Fw | 0.10120 | 113.24 | 0.31655 | 0.65030 | 0 | 0 | -0.82000 | 0.41000 | | |
| TIP3P-FB | 0.10118 | 108.15 | 0.31780 | 0.65214 | 0 | 0 | -0.84844 | 0.42422 | | |
| SPC/ε | 0.10000 | 109.47 | 0.31785 | 0.70591 | 0 | 0 | -0.89000 | 0.44500 | | |
| OPC3 | 0.09789 | 109.47 | 0.31743 | 0.68369 | 0 | 0 | -0.89520 | 0.44760 | | |
| TIP3P-LJOPT | 0.09899 | 114.13 | 0.31480 | 0.65103 | 0 | 0 | -0.85060 | 0.42530 | | |
| ML-TIP3P | 0.09572 | 104.52 | 0.30910 | 1.48552 | 0.06641 | 0.05634 | -0.86076 | 0.43038 | | |
| TIP3P-ST | 0.10230 | 108.11 | 0.31926 | 0.60190 | 0 | 0 | -0.85112 | 0.42556 | | |
| FBA/ε | 0.10270 | 114.70 | 0.31776 | 0.79232 | 0 | 0 | -0.84500 | 0.42250 | | |
| OPTI-1T | 0.09408 | 107.56 | 0.31922 | 0.61796 | 0 | 0 | -0.93501 | 0.46751 | | |
| OPTI-3T | 0.09771 | 108.94 | 0.31657 | 0.69750 | 0 | 0 | -0.88924 | 0.44462 | | |
| OPC3-POL | 0.12110 | 109.47 | 0.31760 | 0.87236 | 0 | 0 | -1.93100 | 0.30500 | | |
| SWM3 | 0.11686 | 109.47 | 0.32078 | 0.63095 | 0 | 0 | 0.77040 | 0.30620 | | |
| TIP4P | 0.09572 | 104.52 | 0.31507 | 0.63627 | 0 | 0 | 0 | 0.52000 | 0.01500 | -1.04000 |
| TIP4P-Ew | 0.09572 | 104.52 | 0.31644 | 0.68095 | 0 | 0 | 0 | 0.52422 | 0.01250 | -1.04844 |
| TIP4P/2005 | 0.09572 | 104.52 | 0.31589 | 0.77492 | 0 | 0 | 0 | 0.55640 | 0.01546 | -1.11280 |
| TIP4P/2005F | 0.09419 | 107.40 | 0.31644 | 0.77492 | 0 | 0 | 0 | 0.55640 | 0.01471 | -1.11280 |
| TIP4Q | 0.09572 | 104.52 | 0.31666 | 0.77492 | 0 | 0 | 0.5 | 0.52500 | 0.00690 | -1.55000 |
| HUANG | 0.11549 | 104.52 | 0.31183 | 1.73009 | 0 | 0 | 0 | 0.41955 | 0.02048 | -0.83910 |
| OPC | 0.08724 | 103.60 | 0.31666 | 0.89036 | 0 | 0 | 0 | 0.67910 | 0.01594 | -1.35820 |
| TIP4P/ε | 0.09572 | 104.52 | 0.31650 | 0.77325 | 0 | 0 | 0 | 0.52700 | 0.01050 | -1.05400 |
| TIP4P-FB | 0.09572 | 104.52 | 0.31655 | 0.74928 | 0 | 0 | 0 | 0.52587 | 0.01053 | -1.05174 |
| TIP4P-D | 0.09572 | 104.52 | 0.31650 | 0.93655 | 0 | 0 | 0 | 0.58000 | 0.01546 | -1.16000 |
| TIP4P-LJOPT | 0.09116 | 118.00 | 0.31702 | 0.54936 | 0 | 0 | 0 | 0.49120 | 0.00133 | -0.98240 |
| TIP4P-ST | 0.09572 | 104.52 | 0.31661 | 0.74030 | 0 | 0 | 0 | 0.52172 | 0.00989 | -1.04344 |
| TIP4P-BG | 0.09572 | 104.52 | 0.31640 | 0.60514 | 0 | 0 | 0 | 0.51244 | 0.01210 | -1.02488 |
| TIP4P/$\varepsilon_{Flex}$ | 0.09300 | 111.50 | 0.31734 | 0.79404 | 0 | 0 | 0 | 0.51000 | 0.00830 | -1.02000 |
| ECCw2024 | 0.09208 | 108.74 | 0.31548 | 0.76115 | 0 | 0 | 0 | 0.60569 | 0.01639 | -1.21138 |
| SWM4-DP | 0.09572 | 104.52 | 0.31803 | 0.86057 | 0 | 0 | -1.77185 | 0.55370 | 0.02381 | -1.10740 |
| SWM4-NDP | 0.09572 | 104.52 | 0.31840 | 0.88257 | 0 | 0 | 1.71636 | 0.55733 | 0.02403 | -1.11466 |
| SWM4-HLJ | 0.09572 | 108.12 | 0.32217 | 0.49246 | 0.12473 | 0.02092 | 1.66260 | 0.52855 | 0.02170 | -0.95710 |
| SWM4-dip | 0.09572 | 104.52 | 0.31891 | 0.72174 | 0 | 0 | 1.44460 | 0.54635 | 0.02075 | -1.09270 |
| SWM4-HD | 0.09572 | 104.52 | 0.32215 | 0.43388 | 0.12473 | 0.02092 | 1.53750 | 0.55665 | 0.02150 | -1.11330 |
| TIP5P | 0.09572 | 104.52 | 0.31200 | 0.66944 | 0 | 0 | 0.00000 | 0.24100 | | |
| TIP5P-Ew | 0.09572 | 104.52 | 0.30970 | 0.74475 | 0 | 0 | 0.00000 | 0.24100 | | |
| TIP5P/2018 | 0.09572 | 104.52 | 0.31450 | 0.79000 | 0 | 0 | -0.64111 | 0.39414 | | |
| TIP6P | 0.09800 | 108.00 | 0.31150 | 0.71485 | 0.06730 | 0.11542 | 0.00000 | 0.47700 | 0.02300 | -0.86600 |
| TIP6P-Ew | 0.98000 | 108.00 | 0.31330 | 0.59864 | 0.07850 | 0.07483 | 0.00000 | 0.49575 | 0.02300 | -0.90850 |
| SWM6 | 0.09572 | 104.52 | 0.31983 | 0.67781 | 0 | 0 | 1.91589 | 0.53070 | 0.02470 | -1.13340 |
| TIP7P | 0.09572 | 104.52 | 0.31560 | 0.62844 | 0.14476 | 0.01046 | 0.11094 | 0.58014 | 0.04786 | -0.45837 |

**Table S2** Parameters of the tested models with Buckingham potential.

| model | $d_{OH}$ [nm] | $\theta_{HOH}$ [°] | $A_{OO}$ [kJ/mol] | $B_{OO}$ [1/nm] | $C_{OO}$ [nm$^6$ kJ/mol] | $q_O$ [e] | $q_H$ [e] | $d_{OM}$ [nm] | $q_M$ [e] |
|---|---|---|---|---|---|---|---|---|---|
| TIP3P-Buck | 0.10284 | 109.28 | 1762638.3 | 42.54660 | 0.00252 | -0.81890 | 0.40945 | | |
| TIP4P-Buck | 0.09251 | 109.93 | 829184.7 | 39.48104 | 0.00301 | 0 | 0.54140 | 0.00982 | -1.08280 |

**Table S3** Additional parameters of polarizable and flexible models, and 5 and more site models.

| model | $d_{OL}$ [nm] | $\theta_{LOL}$ [°] | $q_L$ [e] | $q_S$ [e] | $k_D$ [kJ/mol nm$^2$] | $k_b$ [kJ/ mol nm$^2$] | $D_r$ [kJ/mol] | $\beta$ [1/nm] | $k_\theta$ [kJ/mol rad$^2$] |
|---|---|---|---|---|---|---|---|---|---|
| SPC/Fw | | | | | | 443153 | | | 317.5656 |
| FBA/ε | | | | | | 300000 | | | 383 |
| OPC3-pol | | | | 1.321 | 418400 | | | | |
| SWM3 | | | | -1.38 | 418400 | | | | |
| TIP4P/2005f | | | | | | | 432.581 | 22.87 | 367.81 |
| TIP4P/ε$_{Flex}$ | | | | | | 157000 | | | 212 |
| SWM4-DP | | | | 1.77000 | 418400 | | | | |
| SWM4-NDP | | | | -1.71636 | 418400 | | | | |
| SWM4-HLJ | | | | -1.76 | 418400 | | | | |
| SWM4-dip | | | | -1.44 | 418400 | | | | |
| SWM4-HD | | | | -1.54 | 418400 | | | | |
| TIP5P | 0.07000 | 109.47 | -0.24000 | | | | | | |
| TIP5P-Ew | 0.07000 | 109.47 | -0.24100 | | | | | | |
| TIP5P/2018 | 0.07000 | 109.47 | -0.07358 | | | | | | |
| TIP6P | 0.08892 | 111.0 | -0.044 | | | | | | |
| TIP6P-Ew | 0.08 | 111.0 | -0.0415 | | | | | | |
| SWM6 | 0.03150 | 101.098 | -0.10800 | -1.62789 | 418400 | | | | |
| TIP7P | 0.04100 | 109.47 | -0.17724 | | | | | | |

**Table S4** Densities at different temperatures obtained in MD simulations using different water models (in kg/m$^3$). Experimental values are also presented (in bold) for comparison.

| model | 254 K | 268 K | 284 K | 295 K | 324 K | 343 K | 366 K |
|---|---|---|---|---|---|---|---|
| **Experimental [1,2]** | **994.1** | **999.2** | **999.6** | **997.8** | **988.1** | **978.4** | **964.1** |
| SPC | 1001.9 | 996.0 | 986.5 | 980.1 | 956.7 | 940.0 | 917.7 |
| TIP3P | 1019.3 | 1009.6 | 997.6 | 988.4 | 962.0 | 943.1 | 915.7 |
| SPC/E | 1012.3 | 1010.3 | 1005.2 | 1000.0 | 984.1 | 970.9 | 952.1 |
| TIP3P-FB | 1003.9 | 1003.3 | 999.4 | 996.5 | 983.1 | 971.3 | 954.2 |
| SPC/ε | 999.1 | 1000.3 | 998.9 | 997.7 | 987.3 | 977.5 | 963.2 |
| OPC3 | 1006.2 | 1005.2 | 1001.0 | 997.1 | 982.5 | 970.5 | 953.1 |
| TIP3P-LJOPT | 1023.0 | 1019.7 | 1012.8 | 1006.5 | 987.6 | 972.8 | 951.8 |
| ML-TIP3P | 1097.8 | 1084.6 | 1070.2 | 1060.7 | 1035.7 | 1018.8 | 996.9 |
| TIP3P-ST | 997.7 | 1000.4 | 999.9 | 997.7 | 989.1 | 978.9 | 964.6 |
| OPTI-1T | 1002.0 | 1000.1 | 994.1 | 989.8 | 973.9 | 961.5 | 943.5 |
| OPTI-3T | 1013.4 | 1011.2 | 1006.0 | 1002.4 | 987.2 | 973.5 | 956.7 |
| SPC/Fw | 1020.3 | 1019.0 | 1014.5 | 1010.7 | 995.3 | 981.4 | 964.1 |
| FBA/ε | 995.0 | 997.6 | 996.8 | 995.1 | 985.0 | 975.9 | 962.7 |
| TIP3P-Buck | 1019.8 | 1016.8 | 1009.6 | 1004.7 | 986.4 | 972.6 | 953.1 |
| OPC3-POL | 1022.0 | 1015.7 | 1005.7 | 1000.0 | 980.5 | 965.1 | 946.3 |
| SWM3 | 1036.9 | 1026.6 | 1013.9 | 1004.0 | 977.6 | 957.5 | 930.2 |
| TIP4P | 1007.2 | 1006.4 | 1000.1 | 995.7 | 977.1 | 961.2 | 939.5 |
| TIP4P-Ew | 994.9 | 997.8 | 997.4 | 995.0 | 982.7 | 971.4 | 953.7 |
| TIP4P/2005 | 994.4 | 1000.0 | 998.8 | 998.5 | 987.0 | 976.7 | 961.4 |
| TIP4Q | 994.9 | 998.2 | 997.5 | 995.6 | 984.9 | 975.0 | 961.3 |
| HUANG | 1050.7 | 1039.7 | 1025.9 | 1017.4 | 992.1 | 975.4 | 955.1 |
| OPC | 998.1 | 1001.9 | 999.9 | 997.6 | 988.0 | 978.3 | 964.9 |
| TIP4P/ε | 997.2 | 998.9 | 998.2 | 996.6 | 986.5 | 976.0 | 961.2 |
| TIP4P-FB | 994.6 | 998.2 | 999.4 | 996.9 | 987.7 | 978.7 | 964.6 |
| TIP4P-D | 995.5 | 997.6 | 995.1 | 993.6 | 982.9 | 973.4 | 962.1 |
| TIP4P-LJOPT | 1016.3 | 1013.6 | 1008.3 | 1002.5 | 983.2 | 967.1 | 944.8 |
| TIP4P-ST | 996.5 | 999.4 | 999.5 | 998.3 | 989.2 | 980.0 | 964.7 |
| TIP4P-BG | 1002.3 | 1005.0 | 1005.0 | 1003.0 | 989.1 | 979.5 | 960.3 |
| ECCw2024 | 992.5 | 999.3 | 1000.7 | 997.9 | 988.4 | 976.8 | 961.5 |
| TIP4P/2005f | 991.3 | 997.8 | 999.3 | 997.2 | 988.1 | 978.5 | 962.8 |
| TIP4P/εFlex | 995.8 | 999.6 | 1000.5 | 998.4 | 988.4 | 977.7 | 962.0 |
| TIP4P-Buck | 1024.2 | 1021.8 | 1016.4 | 1012.2 | 995.1 | 981.5 | 962.6 |
| SWM4-DP | 1013.7 | 1009.2 | 1002.9 | 997.7 | 978.2 | 963.0 | 940.9 |
| SWM4-NDP | 1025.0 | 1017.8 | 1007.6 | 1001.1 | 978.6 | 960.4 | 938.5 |
| SWM4-dip | 1010.1 | 1011.1 | 1006.6 | 1002.1 | 985.6 | 971.9 | 952.0 |
| SWM4-HD | 984.9 | 996.8 | 1000.8 | 1000.9 | 992.3 | 981.8 | 965.5 |
| SWM4-HLJ | 1008.9 | 1005.8 | 1000.7 | 995.8 | 978.7 | 964.5 | 944.2 |
| TIP5P | 969.1 | 983.2 | 987.6 | 985.0 | 965.9 | 946.7 | 917.3 |
| TIP5P-Ew | 988.0 | 1003.1 | 1007.4 | 1004.3 | 985.4 | 967.4 | 938.8 |
| TIP5P/2018 | 996.3 | 1001.0 | 1000.2 | 997.6 | 984.4 | 972.5 | 954.8 |
| TIP6P | 957.7 | 989.1 | 1002.8 | 1004.5 | 994.7 | 982.4 | 962.0 |
| TIP6P-Ew | 974.0 | 992.0 | 1000.8 | 1000.8 | 991.5 | 979.7 | 960.4 |
| SWM6 | 1020.5 | 1020.7 | 1015.6 | 1009.8 | 989.1 | 971.3 | 947.0 |
| TIP7P | 994.0 | 998.9 | 1000.3 | 997.2 | 987.1 | 977.2 | 960.1 |

**Table S5** Self-diffusion coefficients of the investigated models at different temperatures (in $10^{-9}$ m$^2$/s). (The simulated values shown are not corrected to finite size effect.) Experimental values are also presented (in bold) for comparison (the 254 and 268 K values are extrapolated values).

| model | 254 K | 268 K | 284 K | 295 K | 324 K | 343 K | 366 K |
|---|---|---|---|---|---|---|---|
| **Experimental [3]** | **0.48** | **0.91** | **1.56** | **2.12** | **3.94** | **5.51** | **7.77** |
| SPC | 1.61 | 2.30 | 3.29 | 3.99 | 6.29 | 7.80 | 10.34 |
| TIP3P | 2.59 | 3.36 | 4.46 | 5.42 | 7.87 | 9.48 | 12.09 |
| SPC/E | 0.70 | 1.22 | 1.94 | 2.45 | 4.12 | 5.32 | 7.61 |
| TIP3P-FB | 0.49 | 0.89 | 1.42 | 1.97 | 3.58 | 4.75 | 6.66 |
| SPC/ε | 0.30 | 0.56 | 1.01 | 1.45 | 2.84 | 3.72 | 5.35 |
| OPC3 | 0.59 | 1.01 | 1.61 | 2.17 | 3.81 | 4.97 | 6.81 |
| TIP3P-LJOPT | 0.86 | 1.48 | 2.23 | 2.86 | 4.43 | 6.39 | 8.32 |
| ML-TIP3P | 2.78 | 3.48 | 4.29 | 5.05 | 6.38 | 7.64 | 9.31 |
| TIP3P-ST | 0.17 | 0.35 | 0.76 | 1.10 | 2.37 | 3.47 | 4.93 |
| OPTI-1T | 0.76 | 1.31 | 1.93 | 2.52 | 4.11 | 5.73 | 7.55 |
| OPTI-3T | 0.73 | 1.18 | 1.85 | 2.38 | 4.02 | 5.20 | 7.19 |
| SPC/Fw | 0.56 | 0.98 | 1.57 | 2.16 | 3.70 | 5.01 | 6.87 |
| FBA/ε | 0.23 | 0.49 | 0.92 | 1.28 | 2.47 | 3.61 | 5.00 |
| TIP3P-Buck | 0.79 | 1.33 | 2.12 | 2.51 | 4.31 | 5.77 | 7.66 |
| OPC3-POL | 0.43 | 0.73 | 1.26 | 1.64 | 3.02 | 4.37 | 5.82 |
| SWM3 | 0.29 | 0.63 | 1.20 | 1.65 | 3.15 | 4.63 | 6.58 |
| TIP4P | 1.11 | 1.77 | 2.74 | 3.36 | 5.84 | 7.63 | 9.88 |
| TIP4P-Ew | 0.60 | 1.06 | 1.75 | 2.37 | 4.17 | 5.50 | 7.83 |
| TIP4P/2005 | 0.47 | 0.90 | 1.56 | 2.01 | 3.57 | 4.93 | 6.89 |
| TIP4Q | 0.51 | 0.96 | 1.58 | 2.09 | 3.74 | 5.00 | 6.86 |
| HUANG | 1.71 | 2.32 | 3.08 | 3.62 | 5.40 | 6.64 | 8.72 |
| OPC | 0.61 | 1.00 | 1.60 | 2.15 | 3.65 | 4.91 | 6.75 |
| TIP4P/ε | 0.50 | 0.93 | 1.50 | 2.01 | 3.60 | 5.00 | 6.52 |
| TIP4P-FB | 0.46 | 0.83 | 1.36 | 1.95 | 3.41 | 4.94 | 6.56 |
| TIP4P-D | 0.50 | 0.87 | 1.37 | 1.90 | 3.30 | 4.57 | 6.18 |
| TIP4P-LJOPT | 0.71 | 1.27 | 1.98 | 2.58 | 4.43 | 5.98 | 8.37 |
| TIP4P-ST | 0.41 | 0.81 | 1.40 | 1.82 | 3.29 | 4.68 | 6.44 |
| TIP4P-BG | 0.49 | 0.90 | 1.66 | 2.14 | 4.12 | 5.43 | 7.88 |
| ECCw2024 | 0.44 | 0.81 | 1.41 | 1.85 | 3.56 | 5.00 | 6.82 |
| TIP4P/2005f | 0.37 | 0.76 | 1.30 | 1.78 | 3.47 | 4.91 | 6.53 |
| TIP4P/εFlex | 0.34 | 0.73 | 1.31 | 1.87 | 3.38 | 4.72 | 6.86 |
| TIP4P-Buck | 0.76 | 1.25 | 1.96 | 2.60 | 4.54 | 5.96 | 8.29 |
| SWM4-DP | 0.45 | 0.86 | 1.51 | 2.01 | 3.79 | 5.24 | 7.09 |
| SWM4-NDP | 0.54 | 0.98 | 1.72 | 2.18 | 4.01 | 5.63 | 7.48 |
| SWM4-dip | 0.16 | 0.51 | 1.03 | 1.53 | 3.16 | 4.63 | 6.46 |
| SWM4-HD | 0.08 | 0.31 | 0.86 | 1.24 | 2.85 | 4.14 | 6.08 |
| SWM4-HLJ | 0.39 | 0.71 | 1.36 | 1.77 | 3.44 | 4.85 | 6.61 |
| TIP5P | 0.13 | 0.60 | 1.69 | 2.46 | 5.49 | 7.95 | 11.00 |
| TIP5P-Ew | 0.17 | 0.75 | 1.75 | 2.57 | 5.31 | 7.76 | 10.70 |
| TIP5P/2018 | 0.42 | 0.88 | 1.54 | 2.30 | 4.06 | 5.51 | 7.84 |
| TIP6P | 0.11 | 0.60 | 1.40 | 1.90 | 4.09 | 5.60 | 7.99 |
| TIP6P-Ew | 0.23 | 0.66 | 1.50 | 2.04 | 3.97 | 5.56 | 8.51 |
| SWM6 | 0.02 | 0.15 | 0.57 | 1.02 | 2.58 | 4.00 | 6.01 |
| TIP7P | 0.71 | 1.23 | 1.94 | 2.57 | 4.34 | 5.95 | 7.90 |

**Table S6** Average relative R-factors and combined $R$-factors: average of the 7 XRD relative $R$-factors (av$R_{rel}^{XRD}$), 5 ND relative $R$-factors (the Ohtomo 295 K values are left) (av$R_{rel}^{ND}$), sum of the room temperature XRD and ND $R$-factors ($R_{rel}^{RT} = R_{rel}$(XRD, 295K) + $R_{rel}$(ND-Soper, 295K)), and the sum of the average XRD and ND relative $R$-factors, $R_{tot}$ ($R_{tot}$ = av$R_{rel}^{XRD}$ + av$R_{rel}^{ND}$).

| | av$R_{rel}^{XRD}$ | av$R_{rel}^{ND}$ | $R_{rel}^{RT}$ | $R_{tot}$ |
|---|---|---|---|---|
| TIP4P/2005 | 1.37 | 1.13 | 2.64 | 2.50 |
| TIP4P/ε | 1.37 | 1.16 | 2.66 | 2.52 |
| TIP4Q | 1.35 | 1.17 | 2.66 | 2.53 |
| TIP4P/2005f | 1.50 | 1.06 | 2.69 | 2.55 |
| TIP4P-FB | 1.42 | 1.13 | 2.69 | 2.56 |
| SWM4-dip | 1.45 | 1.11 | 2.77 | 2.56 |
| TIP4P-Ew | 1.47 | 1.10 | 2.66 | 2.57 |
| SWM4-DP | 1.37 | 1.21 | 2.74 | 2.58 |
| TIP4P-ST | 1.45 | 1.14 | 2.76 | 2.59 |
| ECCw2024 | 1.23 | 1.36 | 3.06 | 2.59 |
| TIP4P | 1.48 | 1.13 | 2.66 | 2.61 |
| TIP4P-Buck | 1.28 | 1.34 | 3.05 | 2.63 |
| TIP5P/2018 | 1.36 | 1.28 | 2.77 | 2.64 |
| TIP4P-D | 1.41 | 1.26 | 2.87 | 2.67 |
| SWM4-NDP | 1.47 | 1.24 | 2.96 | 2.71 |
| TIP4P/εFlex | 1.66 | 1.06 | 2.69 | 2.72 |
| TIP7P | 1.48 | 1.26 | 2.87 | 2.73 |
| OPTI-1T | 1.49 | 1.27 | 3.08 | 2.76 |
| OPTI-3T | 1.60 | 1.21 | 2.62 | 2.81 |
| OPC3 | 1.64 | 1.20 | 2.65 | 2.83 |
| TIP6P | 1.62 | 1.22 | 2.51 | 2.84 |
| SWM6 | 1.72 | 1.12 | 3.12 | 2.84 |
| TIP4P-BG | 1.78 | 1.08 | 2.99 | 2.86 |
| TIP6P-Ew | 1.69 | 1.18 | 2.58 | 2.87 |
| TIP5P-Ew | 1.58 | 1.36 | 2.92 | 2.95 |
| SWM4-HD | 1.78 | 1.23 | 3.31 | 3.01 |
| SWM4-HLJ | 1.78 | 1.26 | 3.22 | 3.04 |
| OPC | 1.00 | 2.06 | 4.13 | 3.06 |
| TIP5P | 1.85 | 1.26 | 3.13 | 3.11 |
| TIP4P-LJOPT | 1.68 | 1.49 | 3.67 | 3.17 |
| SPC/E | 1.80 | 1.40 | 2.96 | 3.20 |
| TIP3P-LJOPT | 1.78 | 1.43 | 2.98 | 3.22 |
| SPC/ε | 1.82 | 1.39 | 3.05 | 3.22 |
| SPC | 1.88 | 1.40 | 3.01 | 3.28 |
| TIP3P-FB | 1.92 | 1.50 | 3.11 | 3.42 |
| TIP3P | 1.98 | 1.46 | 3.81 | 3.44 |
| TIP3P-Buck | 1.90 | 1.69 | 3.32 | 3.59 |
| SPC/Fw | 2.07 | 1.68 | 3.48 | 3.74 |
| TIP3P-ST | 2.37 | 1.73 | 4.16 | 4.10 |
| FBA/ε | 2.09 | 2.07 | 4.14 | 4.15 |
| SWM3 | 2.72 | 3.75 | 6.51 | 6.47 |
| OPC3-POL | 2.69 | 4.05 | 6.86 | 6.74 |
| ML-TIP3P | 4.76 | 2.74 | 9.29 | 7.50 |
| HUANG | 4.82 | 3.49 | 9.45 | 8.31 |

**Table S7** Position of the first (intermolecular) maximum in the O-O, O-H, and H-H partial radial distribution functions at room temperature (in Å).

|  | O-O | O-H | H-H |
|---|---|---|---|
| SPC | 2.76 | 1.8 | 2.44 |
| TIP3P | 2.78 | 1.84 | 2.44 |
| SPC/E | 2.76 | 1.76 | 2.38 |
| TIP3P-FB | 2.76 | 1.76 | 2.36 |
| SPC/ε | 2.76 | 1.76 | 2.36 |
| OPC3 | 2.76 | 1.8 | 2.4 |
| TIP3P-LJOPT | 2.74 | 1.76 | 2.36 |
| ML-TIP3P | 2.94 | 2.02 | 2.66 |
| TIP3P-ST | 2.72 | 1.72 | 2.34 |
| OPTI-1T | 2.76 | 1.84 | 2.4 |
| OPTI-3T | 2.76 | 1.8 | 2.4 |
| SPC/Fw | 2.74 | 1.7 | 2.34 |
| FBA/ε | 2.76 | 1.72 | 2.36 |
| TIP3P-Buck | 2.76 | 1.72 | 2.36 |
| OPC3-POL | 2.78 | 1.58 | 2.3 |
| SWM3 | 2.74 | 1.58 | 2.32 |
| TIP4P | 2.76 | 1.82 | 2.36 |
| TIP4P-Ew | 2.76 | 1.82 | 2.36 |
| TIP4P/2005 | 2.78 | 1.84 | 2.36 |
| TIP4Q | 2.78 | 1.84 | 2.4 |
| HUANG | 2.96 | 1.82 | 2.42 |
| OPC | 2.8 | 1.96 | 2.44 |
| TIP4P/ε | 2.78 | 1.86 | 2.42 |
| TIP4P-FB | 2.76 | 1.84 | 2.4 |
| TIP4P-D | 2.82 | 1.88 | 2.42 |
| TIP4P-LJOPT | 2.72 | 1.84 | 2.34 |
| TIP4P-ST | 2.76 | 1.82 | 2.36 |
| TIP4P-BG | 2.72 | 1.8 | 2.36 |
| ECCw2024 | 2.78 | 1.86 | 2.34 |
| TIP4P/2005f | 2.76 | 1.8 | 2.32 |
| TIP4P/εFlex | 2.76 | 1.78 | 2.34 |
| TIP4P-Buck | 2.76 | 1.84 | 2.34 |
| SWM4-DP | 2.78 | 1.84 | 2.42 |
| SWM4-NDP | 2.78 | 1.84 | 2.42 |
| SWM4-dip | 2.76 | 1.82 | 2.38 |
| SWM4-HD | 2.76 | 1.86 | 2.4 |
| SWM4-HLJ | 2.78 | 1.86 | 2.4 |
| TIP5P | 2.74 | 1.82 | 2.36 |
| TIP5P-Ew | 2.74 | 1.8 | 2.34 |
| TIP5P/2018 | 2.78 | 1.86 | 2.4 |
| TIP6P | 2.76 | 1.82 | 2.32 |
| TIP6P-Ew | 2.76 | 1.82 | 2.32 |
| SWM6 | 2.74 | 1.8 | 2.36 |
| TIP7P | 2.78 | 1.86 | 2.42 |

**Table S8** $N_{OH}$ coordination numbers (calculated up to the first minima) at different temperatures.

| model | 254 K | 268 K | 284 K | 295 K | 324 K | 343 K | 366 K |
|---|---|---|---|---|---|---|---|
| SPC | 1.91 | 1.87 | 1.90 | 1.88 | 1.87 | 1.74 | 1.70 |
| TIP3P | 1.89 | 1.85 | 1.87 | 1.82 | 1.79 | 1.75 | 1.70 |
| SPC/E | 1.95 | 1.92 | 1.89 | 1.92 | 1.85 | 1.80 | 1.75 |
| TIP3P-FB | 1.96 | 1.93 | 1.92 | 1.91 | 1.90 | 1.86 | 1.81 |
| SPC/ε | 1.99 | 1.97 | 1.97 | 1.92 | 1.89 | 1.85 | 1.84 |
| OPC3 | 1.95 | 1.95 | 1.93 | 1.90 | 1.84 | 1.85 | 1.82 |
| TIP3P-LJOPT | 1.94 | 1.92 | 1.93 | 1.90 | 1.84 | 1.80 | 1.81 |
| ML-TIP3P | 1.97 | 1.89 | 1.87 | 1.97 | 1.79 | 1.84 | 1.74 |
| TIP3P-ST | 1.98 | 1.97 | 1.93 | 1.92 | 1.91 | 1.85 | 1.79 |
| OPTI-1T | 1.95 | 1.93 | 1.92 | 1.87 | 1.87 | 1.81 | 1.77 |
| OPTI-3T | 1.94 | 1.92 | 1.92 | 1.92 | 1.86 | 1.86 | 1.78 |
| SPC/Fw | 1.96 | 1.94 | 1.92 | 1.89 | 1.87 | 1.79 | 1.80 |
| FBA/ε | 1.94 | 1.95 | 1.94 | 1.91 | 1.86 | 1.91 | 1.82 |
| TIP3P-Buck | 1.92 | 1.93 | 1.88 | 1.86 | 1.82 | 1.83 | 1.77 |
| OPC3-POL | 1.75 | 1.76 | 1.72 | 1.74 | 1.69 | 1.67 | 1.66 |
| SWM3 | 1.71 | 1.71 | 1.71 | 1.70 | 1.65 | 1.63 | 1.59 |
| TIP4P | 1.94 | 1.94 | 1.92 | 1.90 | 1.84 | 1.76 | 1.80 |
| TIP4P-Ew | 1.98 | 1.95 | 1.93 | 1.91 | 1.90 | 1.88 | 1.86 |
| TIP4P/2005 | 1.97 | 1.96 | 1.94 | 1.93 | 1.89 | 1.91 | 1.79 |
| TIP4Q | 1.97 | 1.96 | 1.94 | 1.92 | 1.89 | 1.88 | 1.79 |
| HUANG | 1.97 | 2.00 | 1.93 | 1.91 | 1.88 | 1.85 | 1.75 |
| OPC | 1.97 | 1.96 | 1.94 | 1.91 | 1.89 | 1.86 | 1.83 |
| TIP4P/ε | 1.98 | 1.98 | 1.95 | 1.92 | 1.91 | 1.88 | 1.83 |
| TIP4P-FB | 1.97 | 1.96 | 1.94 | 1.96 | 1.92 | 1.86 | 1.80 |
| TIP4P-D | 1.98 | 1.98 | 1.95 | 1.94 | 1.93 | 1.90 | 1.91 |
| TIP4P-LJOPT | 1.95 | 1.92 | 1.87 | 1.87 | 1.83 | 1.76 | 1.67 |
| TIP4P-ST | 1.98 | 1.96 | 1.96 | 1.93 | 1.90 | 1.85 | 1.83 |
| TIP4P-BG | 1.96 | 1.96 | 1.95 | 1.93 | 1.87 | 1.84 | 1.77 |
| ECCw2024 | 1.98 | 1.96 | 1.93 | 1.93 | 1.89 | 1.91 | 1.81 |
| TIP4P/2005f | 1.97 | 1.97 | 1.97 | 1.94 | 1.91 | 1.87 | 1.83 |
| TIP4P/εFlex | 1.97 | 1.96 | 1.90 | 1.95 | 1.88 | 1.85 | 1.78 |
| TIP4P-Buck | 1.94 | 1.91 | 1.90 | 1.87 | 1.82 | 1.79 | 1.72 |
| SWM4-DP | 1.96 | 1.95 | 1.94 | 1.91 | 1.87 | 1.86 | 1.76 |
| SWM4-NDP | 1.94 | 1.94 | 1.92 | 1.91 | 1.83 | 1.81 | 1.76 |
| SWM4-dip | 1.97 | 1.97 | 1.93 | 1.92 | 1.89 | 1.83 | 1.76 |
| SWM4-HD | 1.99 | 1.98 | 1.97 | 1.95 | 1.90 | 1.87 | 1.88 |
| SWM4-HLJ | 1.95 | 1.98 | 1.91 | 1.94 | 1.92 | 1.86 | 1.86 |
| TIP5P | 1.97 | 1.95 | 1.87 | 1.87 | 1.80 | 1.77 | 1.72 |
| TIP5P-Ew | 1.96 | 1.93 | 1.89 | 1.89 | 1.86 | 1.81 | 1.72 |
| TIP5P/2018 | 1.97 | 1.95 | 1.95 | 1.92 | 1.89 | 1.83 | 1.78 |
| TIP6P | 1.97 | 1.96 | 1.96 | 1.91 | 1.91 | 1.91 | 1.84 |
| TIP6P-Ew | 1.99 | 1.95 | 1.96 | 1.96 | 1.92 | 1.88 | 1.79 |
| SWM6 | 1.99 | 1.95 | 1.95 | 1.94 | 1.85 | 1.78 | 1.79 |
| TIP7P | 1.98 | 1.98 | 1.95 | 1.92 | 1.88 | 1.83 | 1.88 |

**Figures**

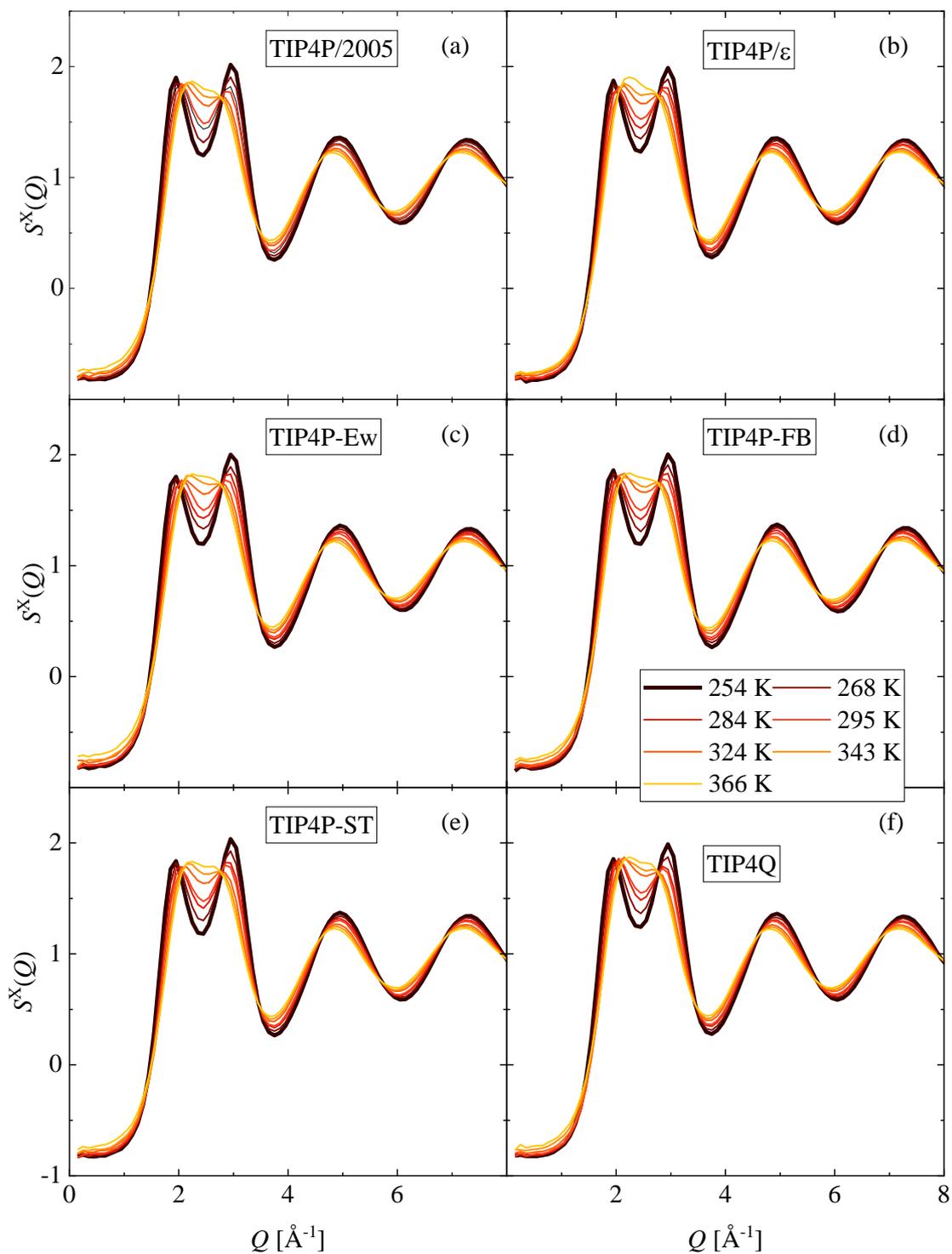

**Figure S1** Temperature dependence of simulated XRD total structure factor for models (a) TIP4P/2005, (b) TIP4P/ε, (c) TIP4P-Ew, (d) TIP4P-FB, (e) TIP4P-ST, and (f) TIP4Q (similar models).

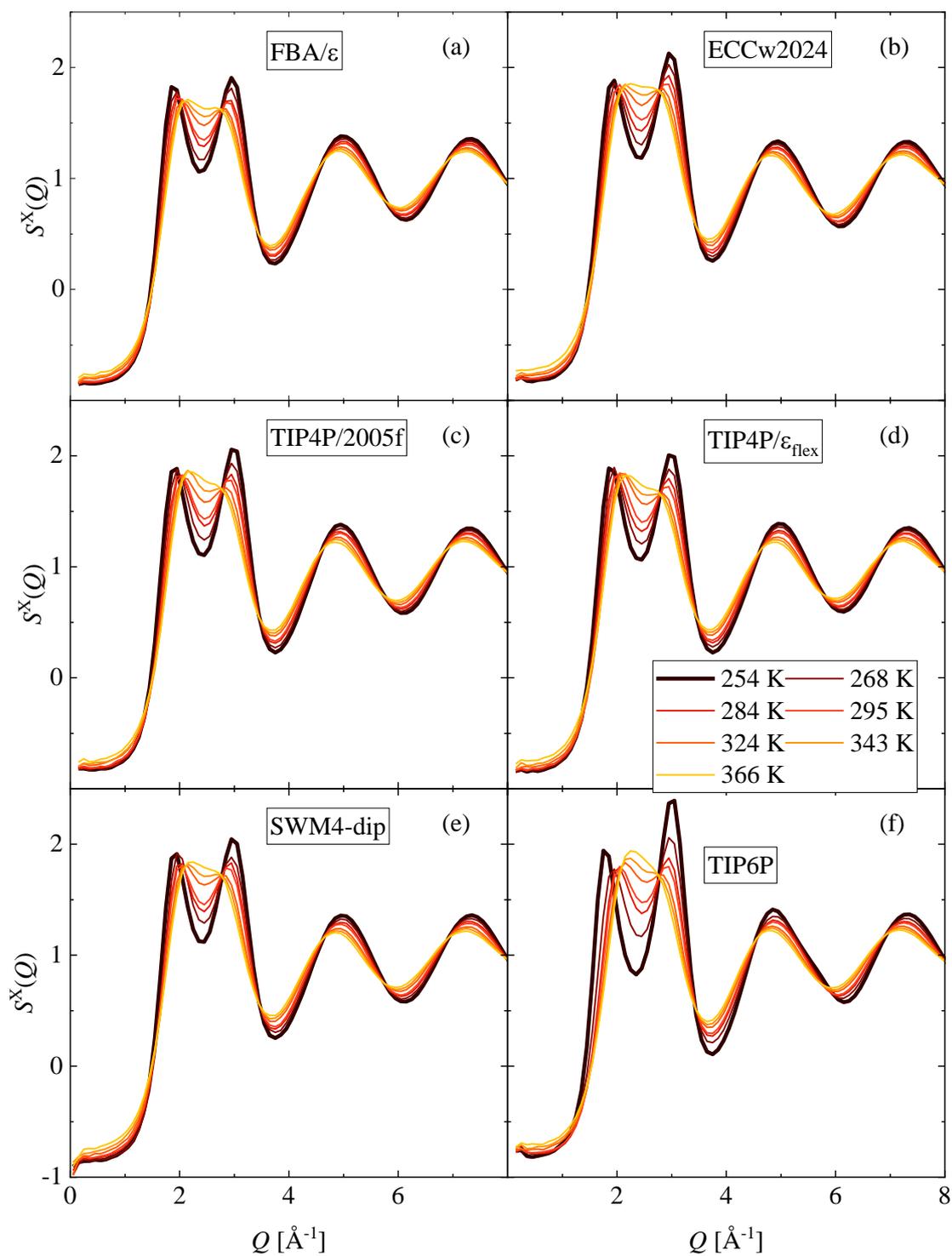

**Figure S2** Temperature dependence of simulated XRD total structure factor for models (a) FBA/ε, (b) ECCw2024, (c) TIP4P/2005f, (d) TIP4P/ε$_{Flex}$, (e) SWM4-dip, and (f) TIP6P models (Group A models).

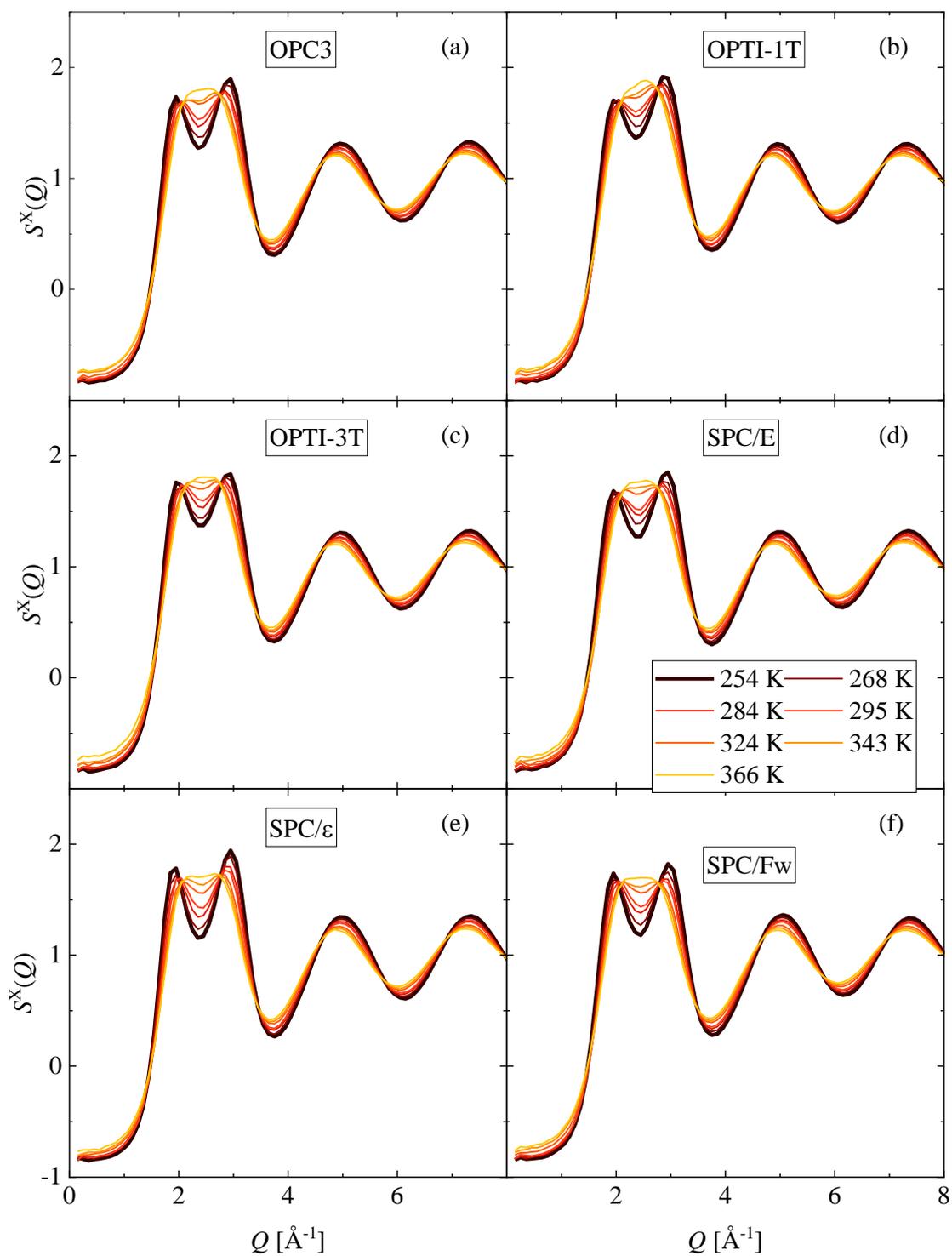

**Figure S3** Temperature dependence of simulated XRD total structure factor for models (a) OPC3, (b) OPTI-1T, (c) OPTI-3T, (d) SPC/E, (e) SPC/ε, and (f) SPC/Fw models (3-site, Group B models).

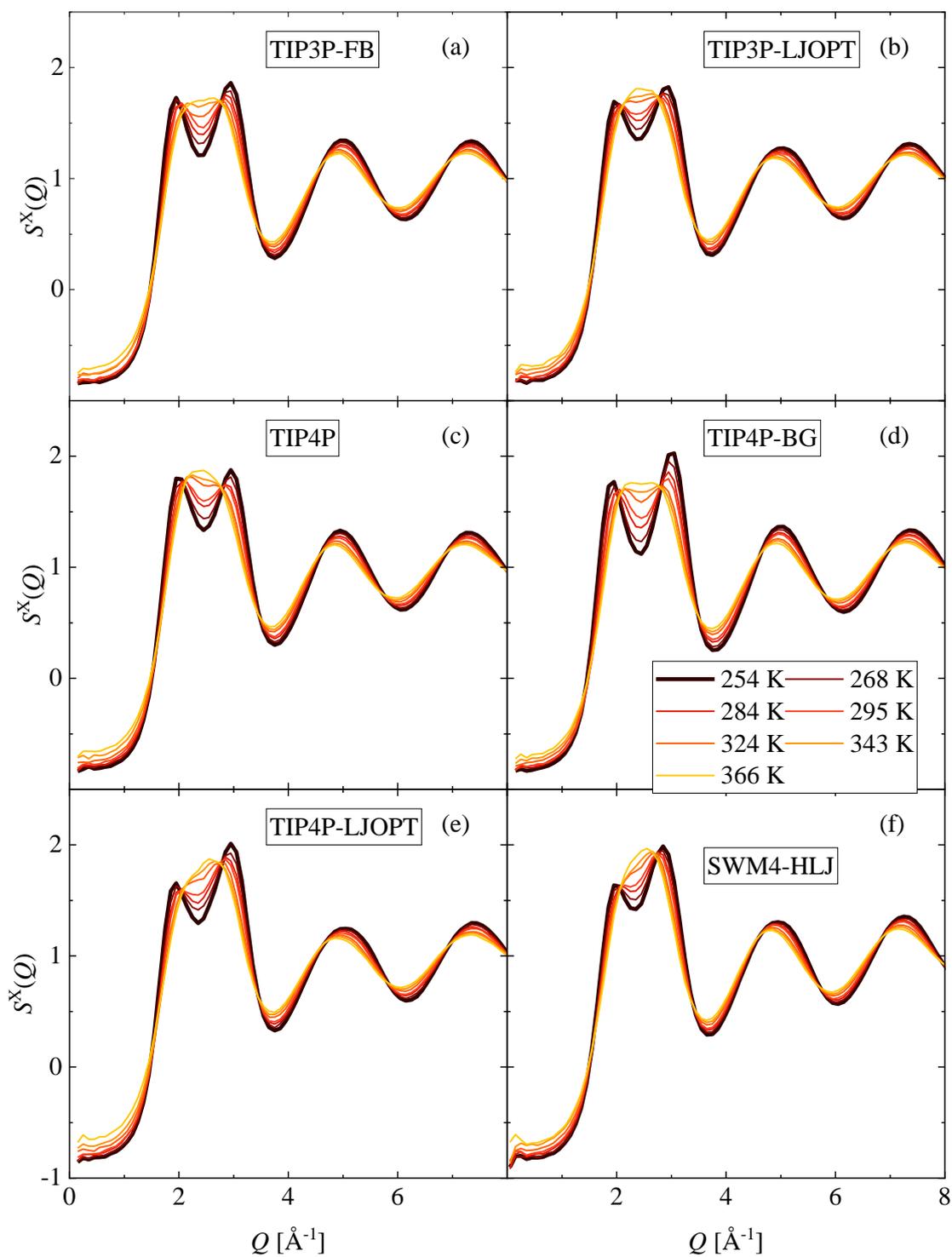

**Figure S4** Temperature dependence of simulated XRD total structure factor for models (a) TIP3P-FB, (b) TIP3P-LJOPT, (c) TIP4P, (d) TIP4P-BG, (e) TIP4P-LJOPT, and (f) SWM-HLJ models (Group B models).

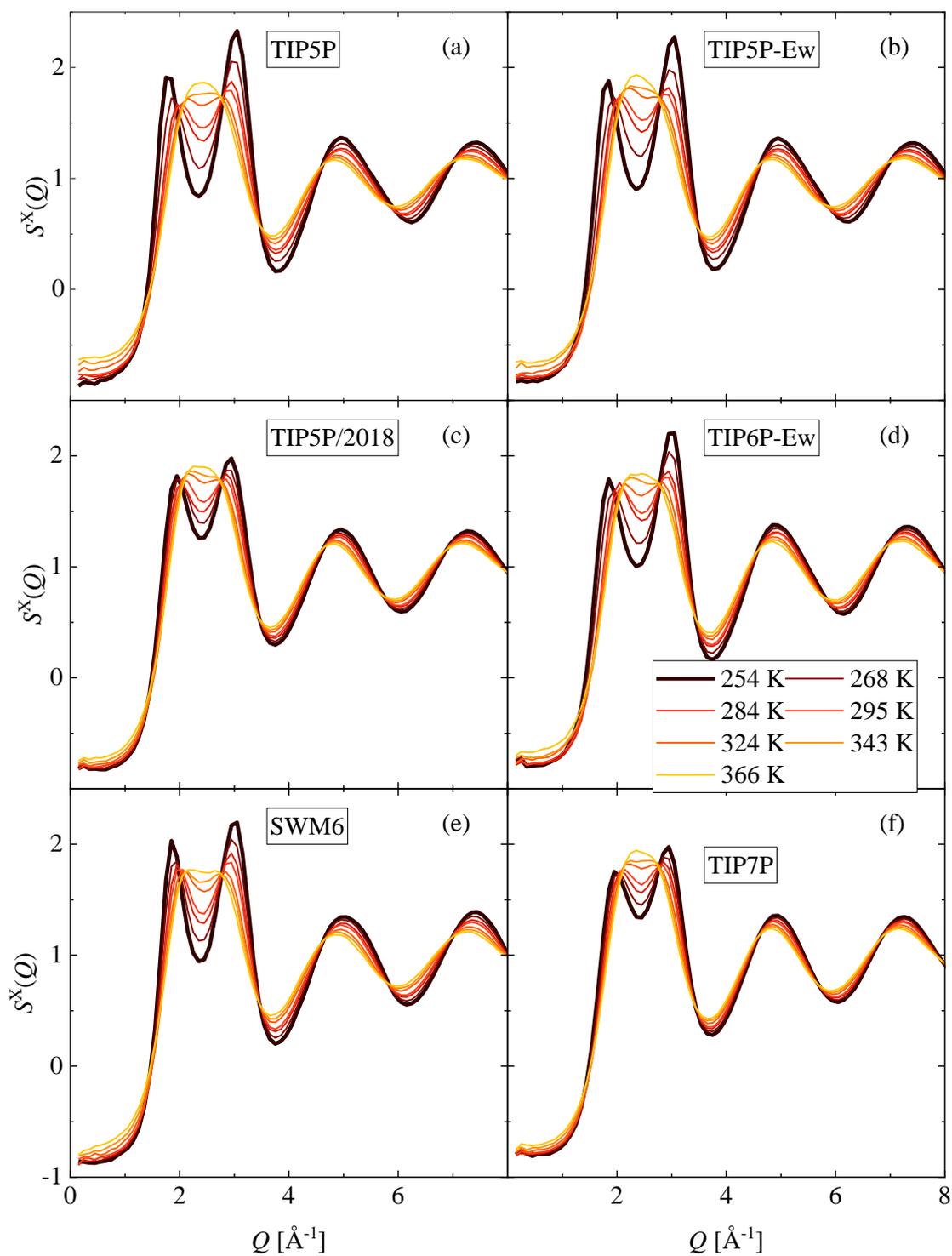

**Figure S5** Temperature dependence of simulated XRD total structure factor for models (a) TIP5P, (b) TIP5P-Ew, (c) TIP5P/2018, (d) TIP6P-Ew, (e) SWM6, and (f) TIP7P models (Group B models).

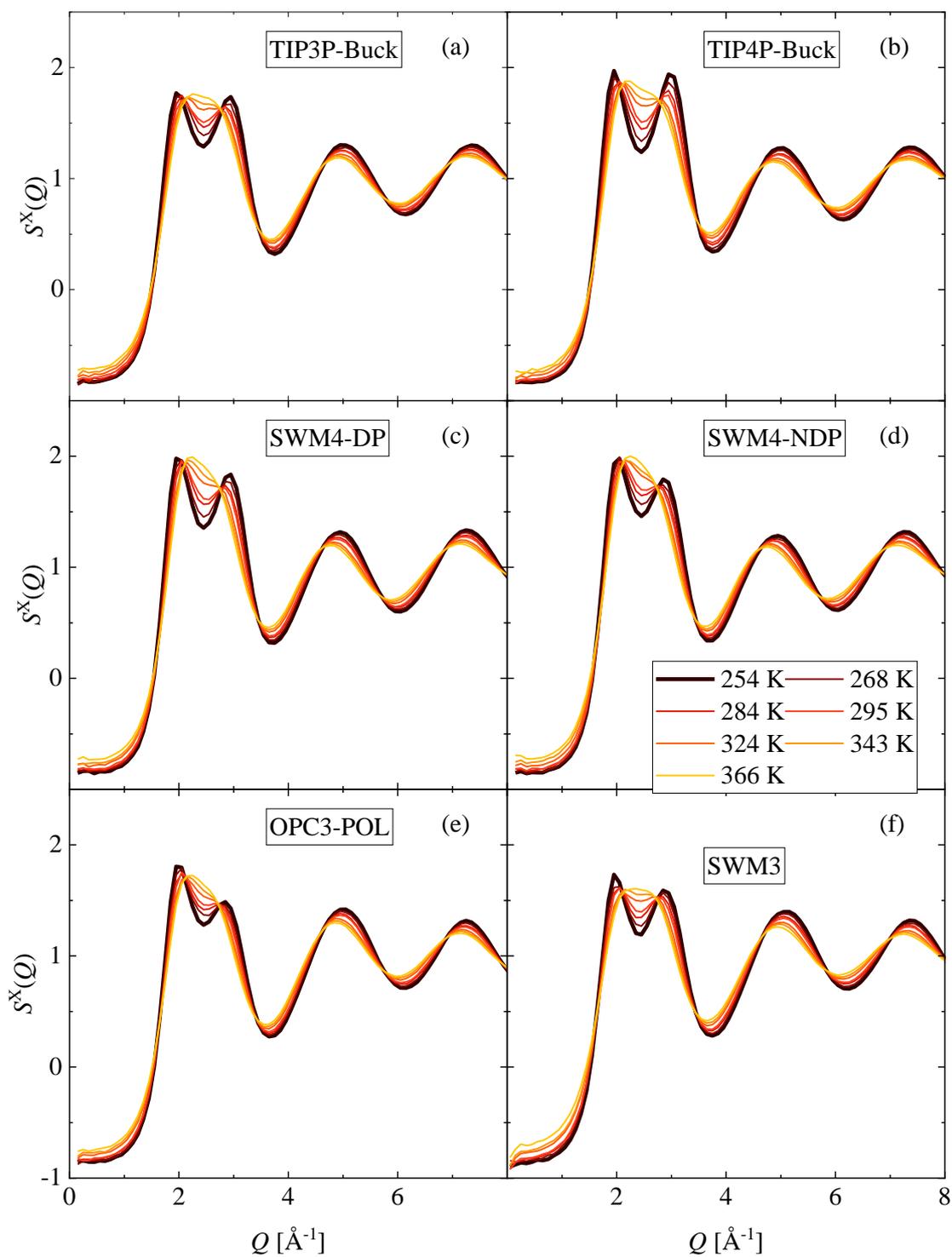

**Figure S6** Temperature dependence of simulated XRD total structure factor for models (a) TIP3P-Buck, (b) TIP4P-Buck, (c) SWM4-DP, (d) SWM4-NDP, (e) OPC3-POL, and (f) SWM3 models (Group C models).

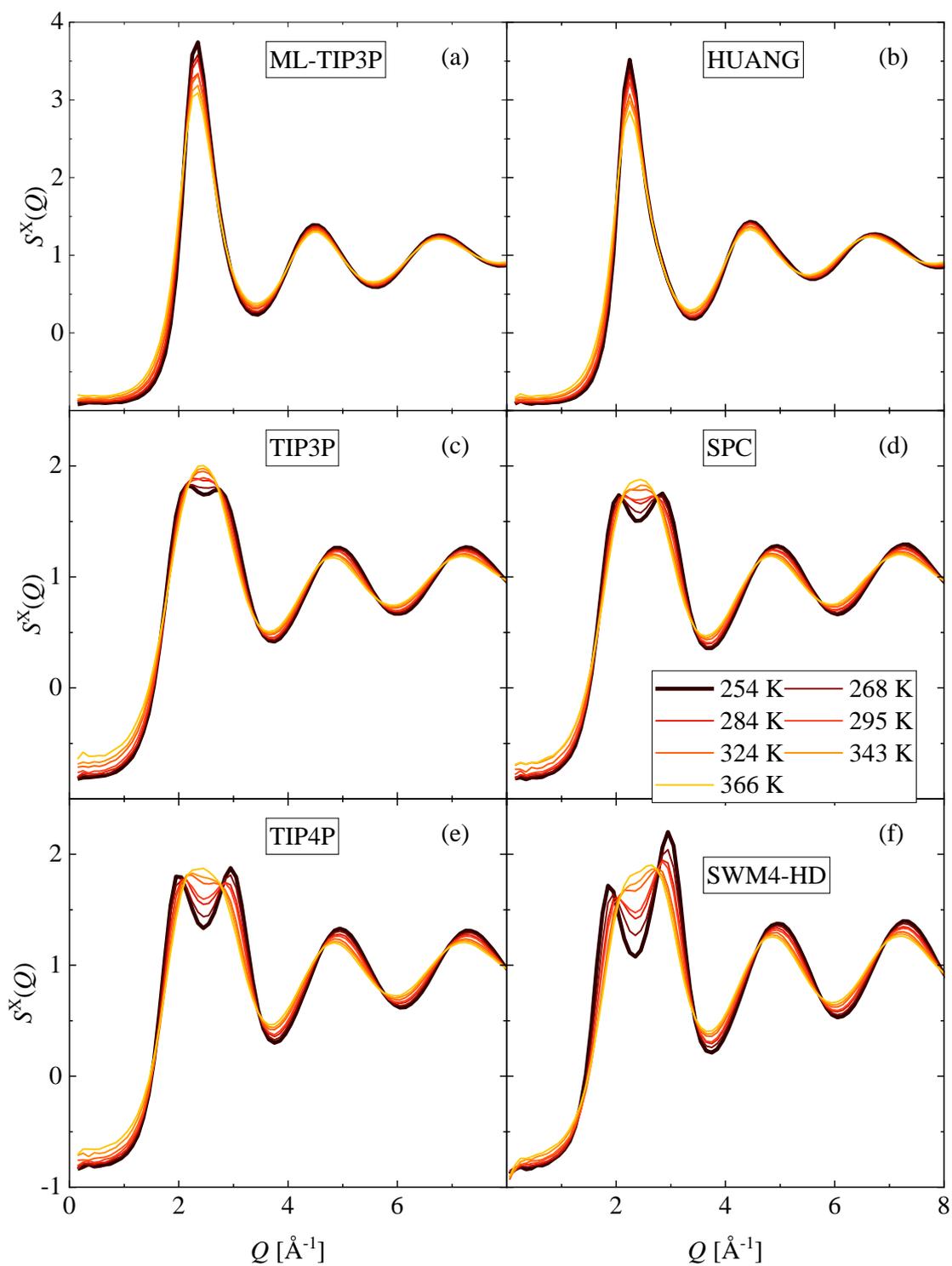

**Figure S7** Temperature dependence of simulated XRD total structure factor for models (a) ML-TIP3P, (b) HUANG, (c) TIP3P, (d) SPC, (e) TIP4P, and (f) SWM4-HD models.

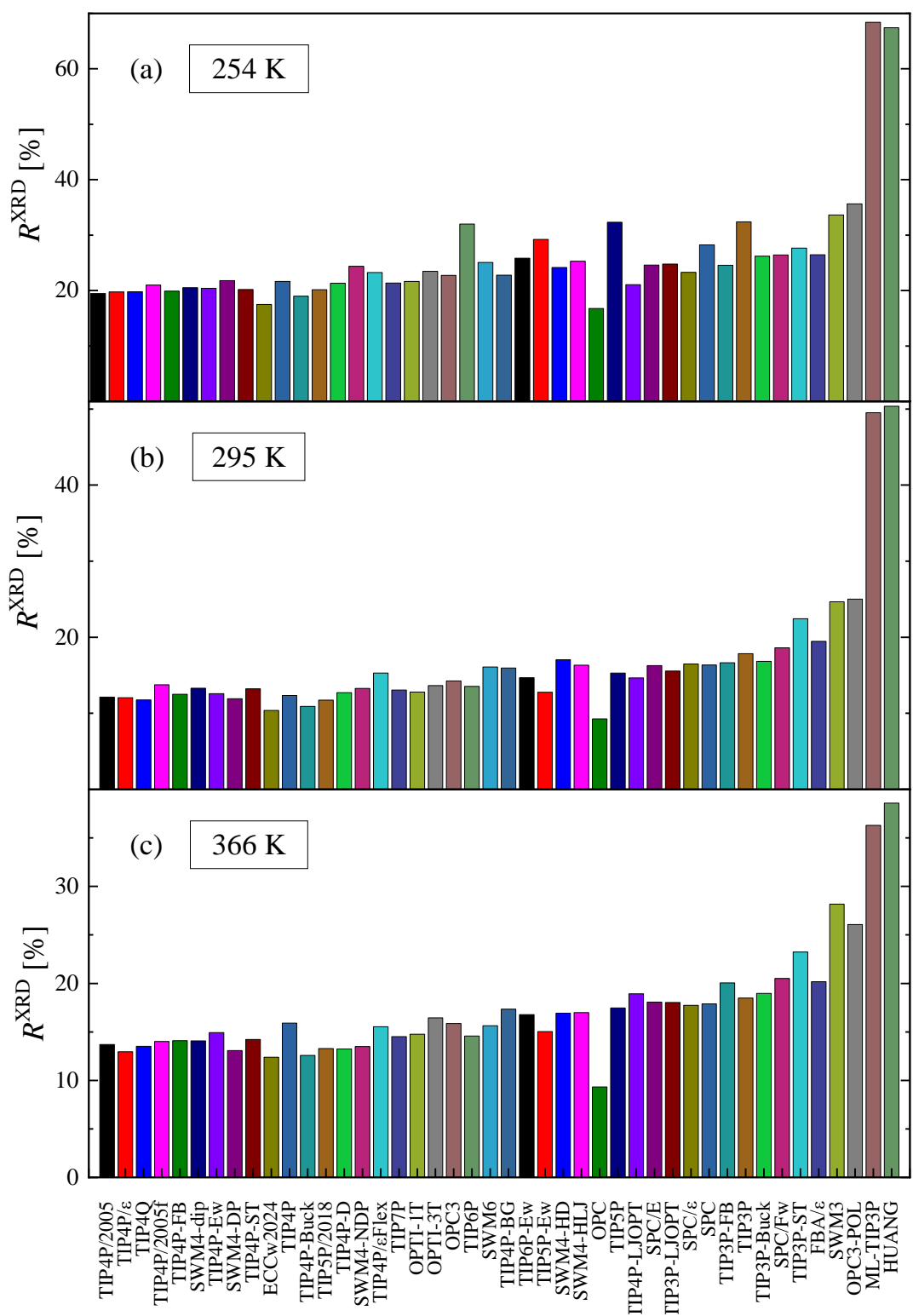

**Figure S8** *R*-factors of the X-ray diffraction total structure factors at (a) 254 K, (b) 295 K, and (c) 366 K, obtained by using different models (For the definition of *R*-factor, see text.)

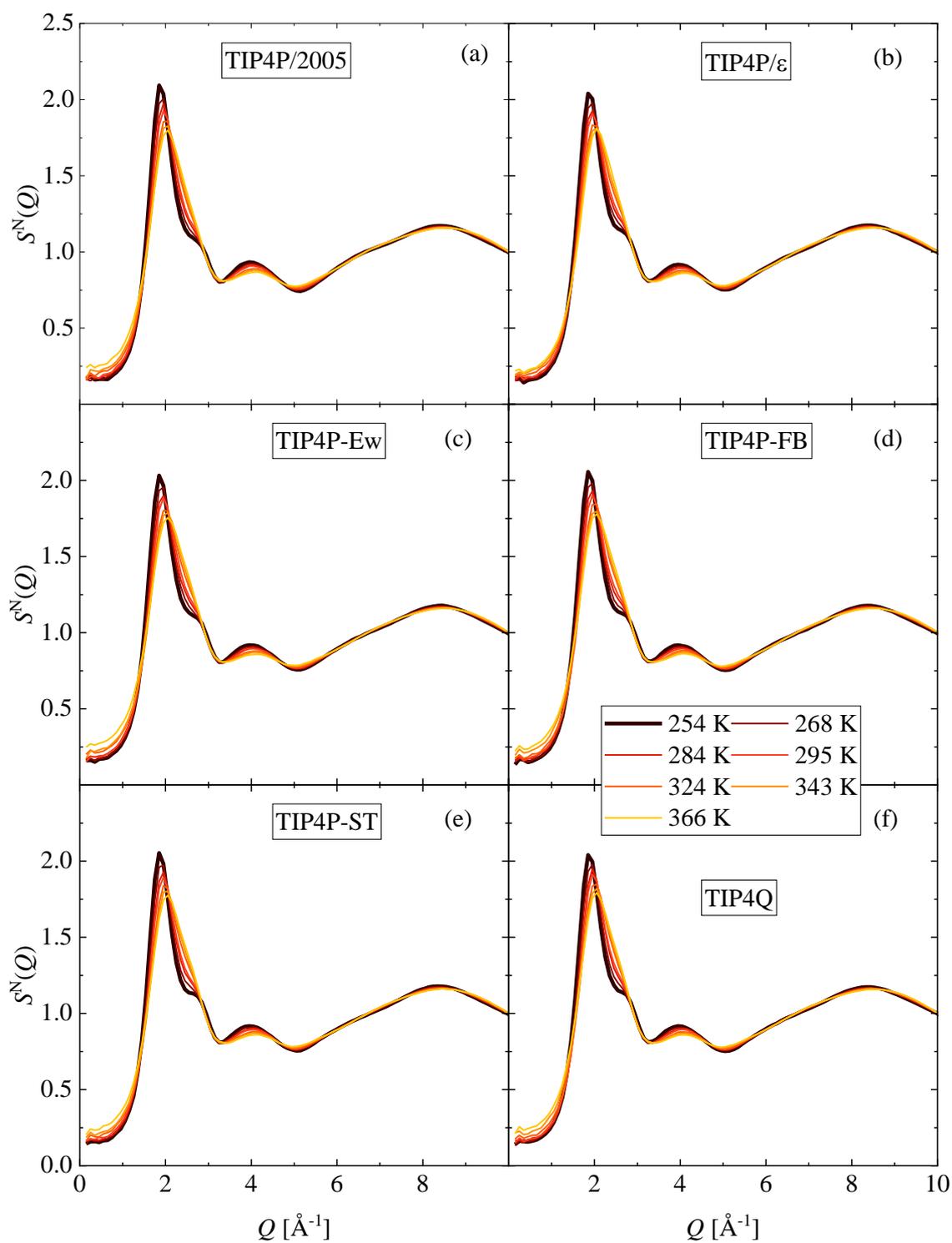

**Figure S9** Temperature dependence of simulated ND total structure factor for models (a) TIP4P/2005, (b) TIP4P/ε, (c) TIP4P-Ew, (d) TIP4P-FB, (e) TIP4P-ST, and (f) TIP4Q (similar models).

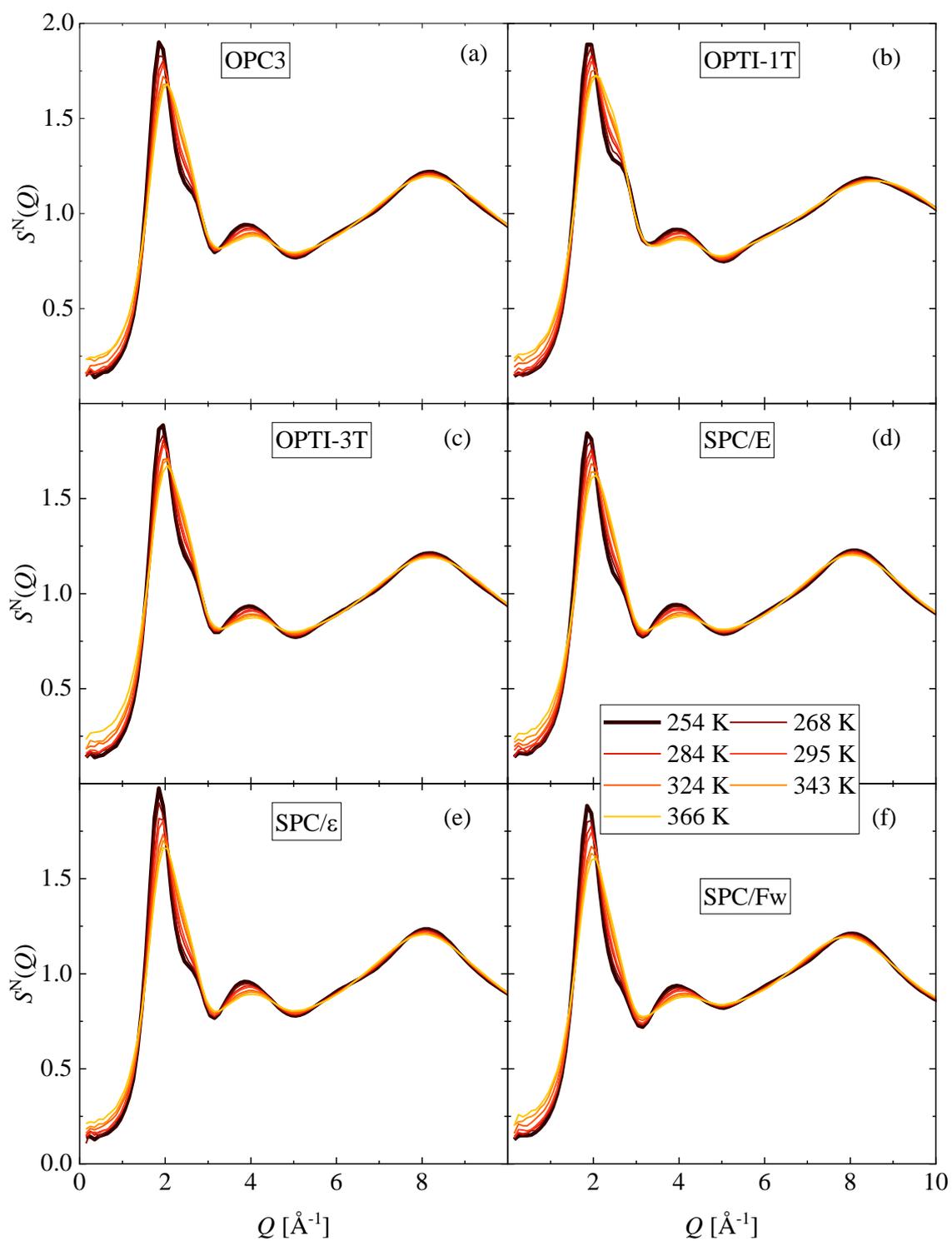

**Figure S10** Temperature dependence of simulated ND total structure factor for models (a) OPC3, (b) OPTI-1T, (c) OPTI-3T, (d) SPC/E, (e) SPC/ε, and (f) SPC/Fw models.

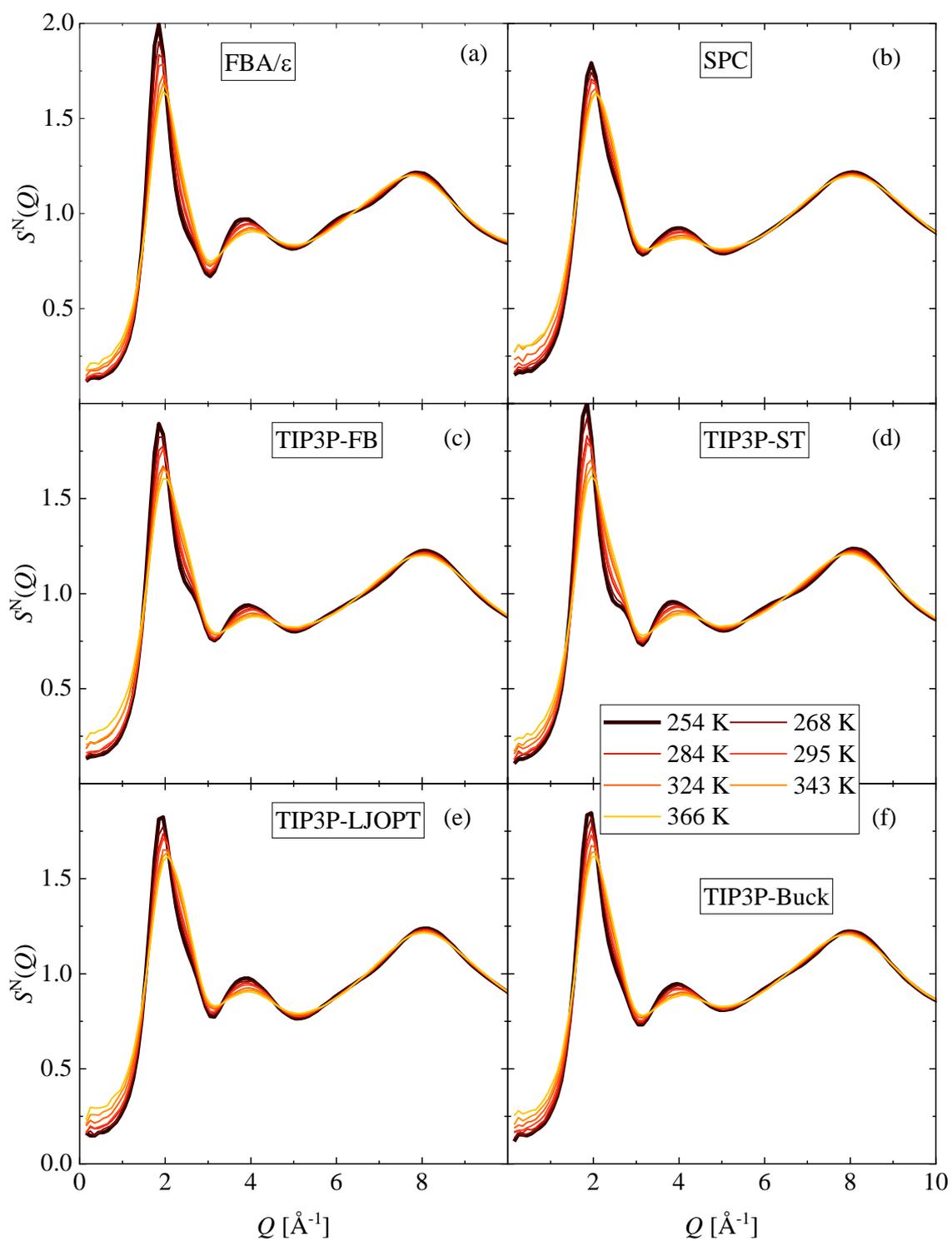

**Figure S11** Temperature dependence of simulated ND total structure factor for models (a) FBA/ε, (b) SPC, (c) TIP3P-FB, (d) TIP3P-ST, (e) TIP3P-LJOPT, and (f) TIP3P-Buck models.

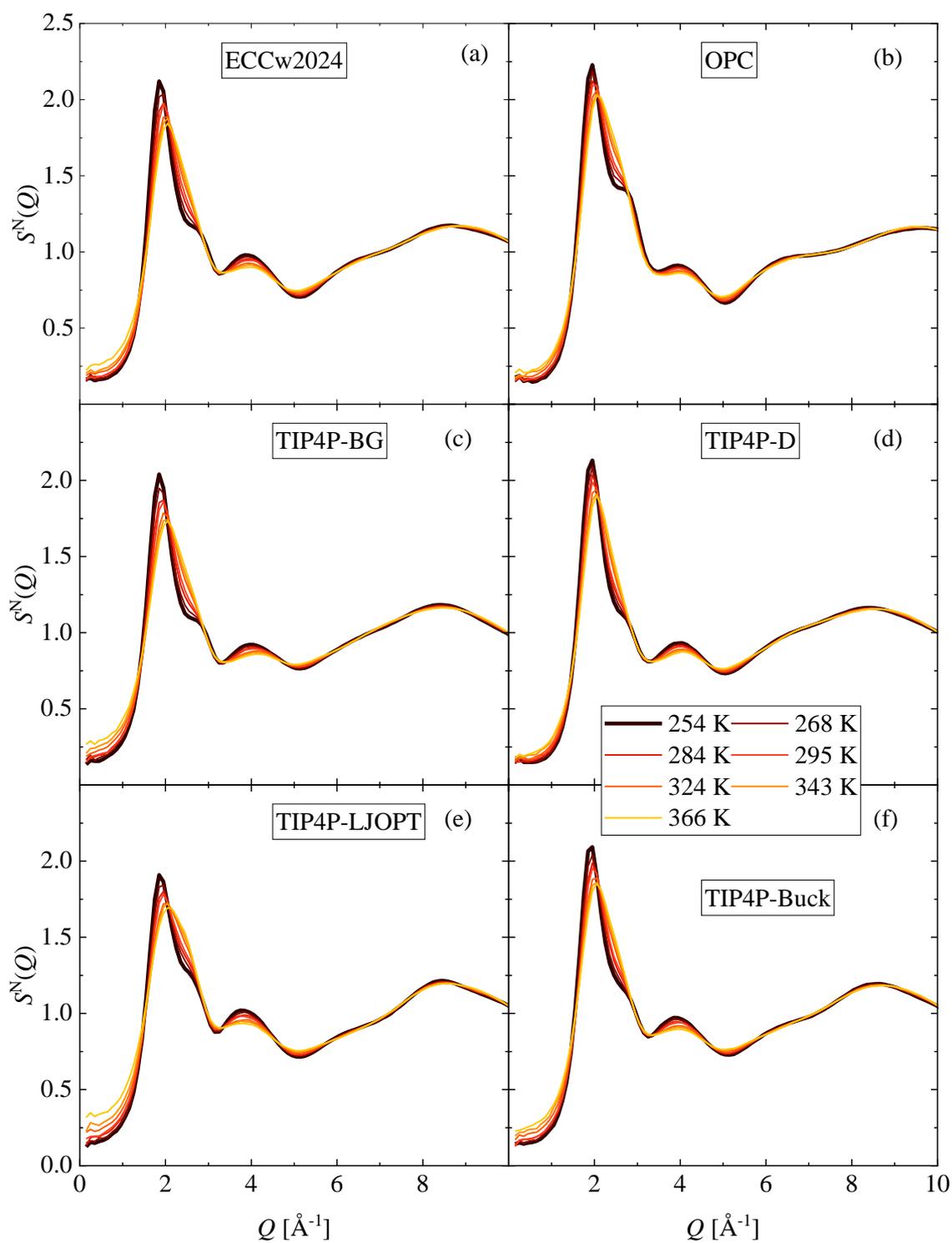

**Figure S12** Temperature dependence of simulated ND total structure factor for models (a) ECCw2024, (b) OPC, (c) TIP4P-BG, (d) TIP4P-D, (e) TIP4P-LJOPT, and (f) TIP4P-Buck models.

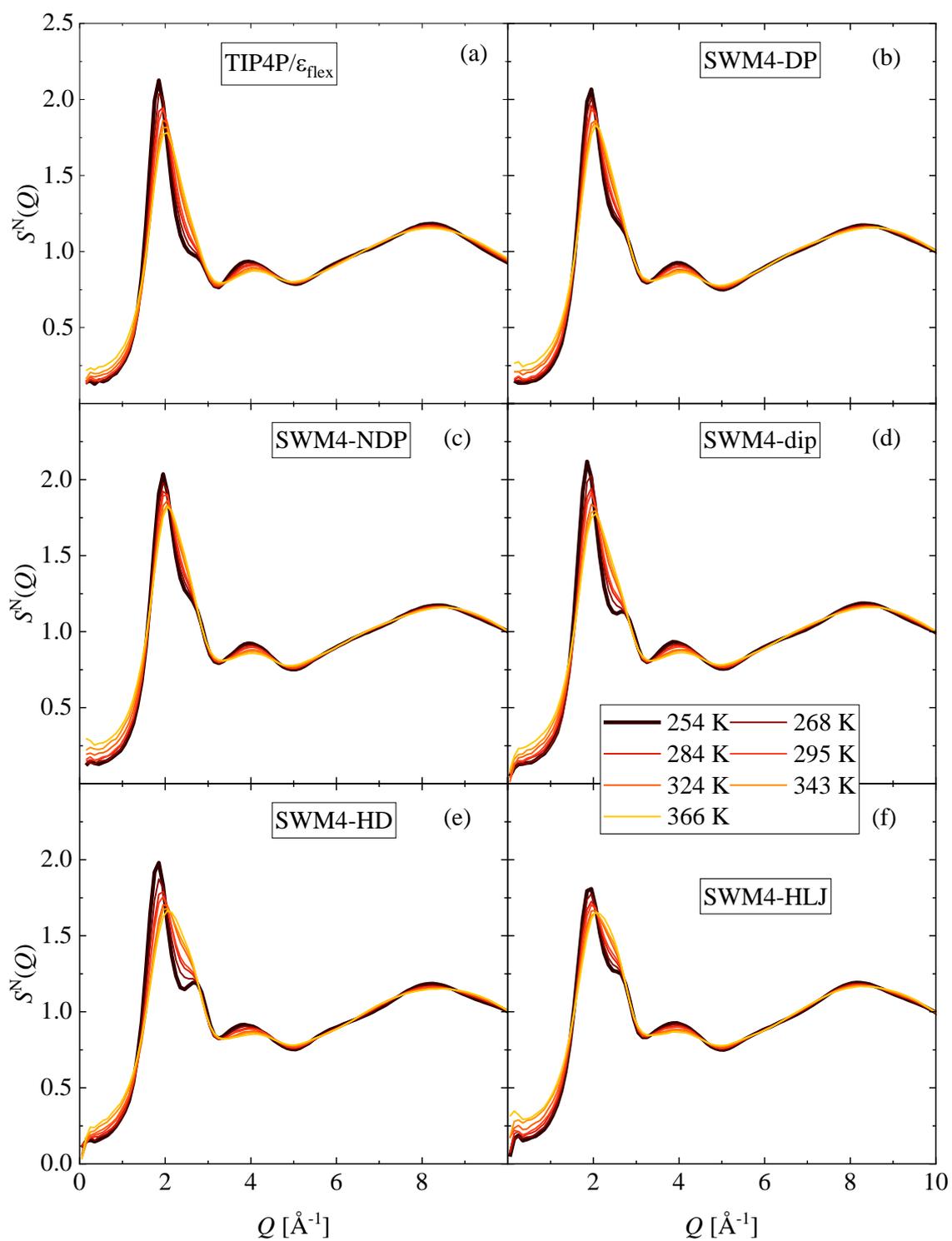

**Figure S13** Temperature dependence of simulated ND total structure factor for models (a) TIP4P/$\varepsilon_{\text{Flex}}$, (b) SWM4-DP, (c) SWM4-NDP, (d) SWM4-dip, (e) SWM4-HD, and (f) SWM4-HLJ models.

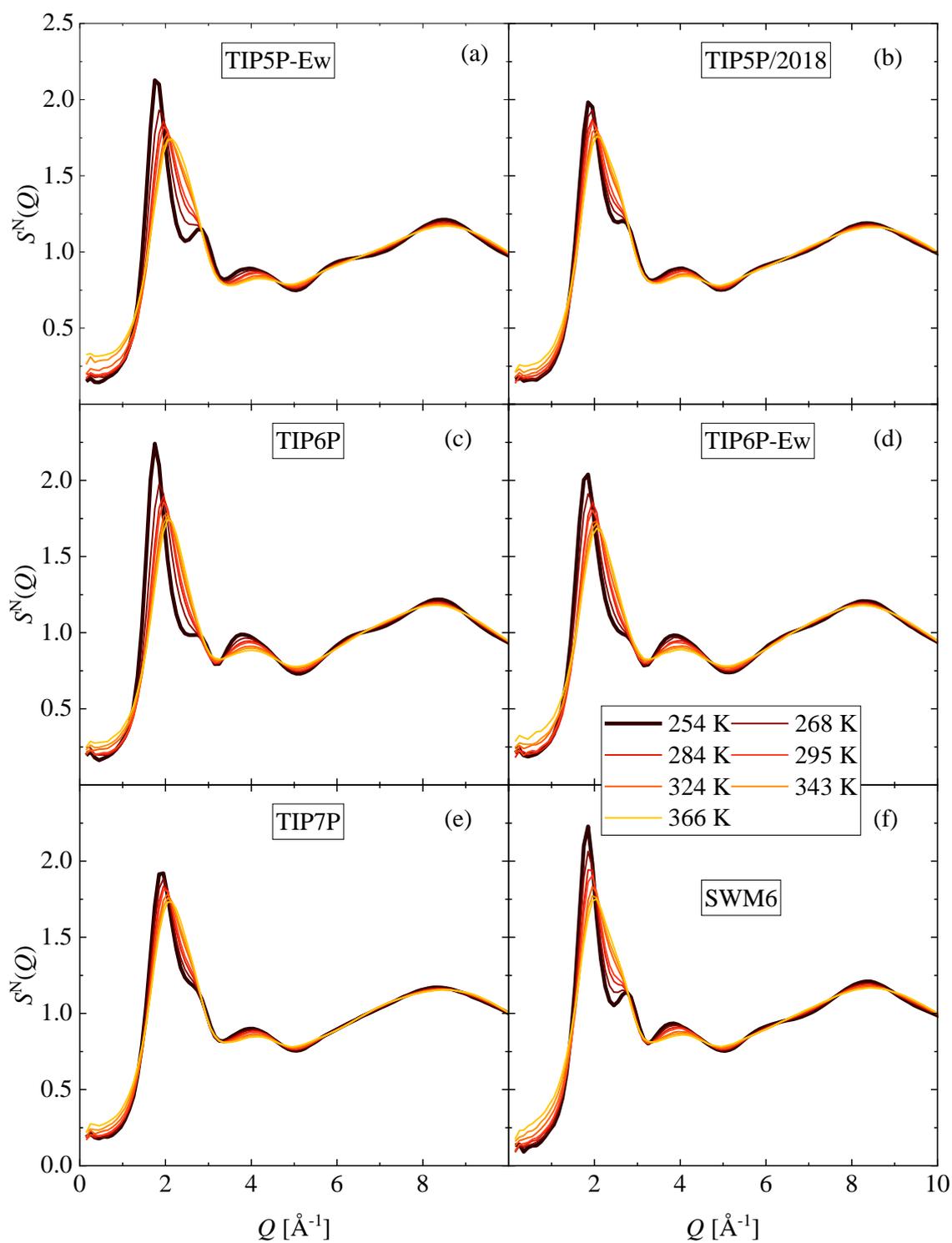

**Figure S14** Temperature dependence of simulated ND total structure factor for models (a) TIP5P-Ew, (b) TIP5P/2018, (c) TIP6P, (d) TIP6P-Ew, (e) TIP7P, and (f) SWM6 models.

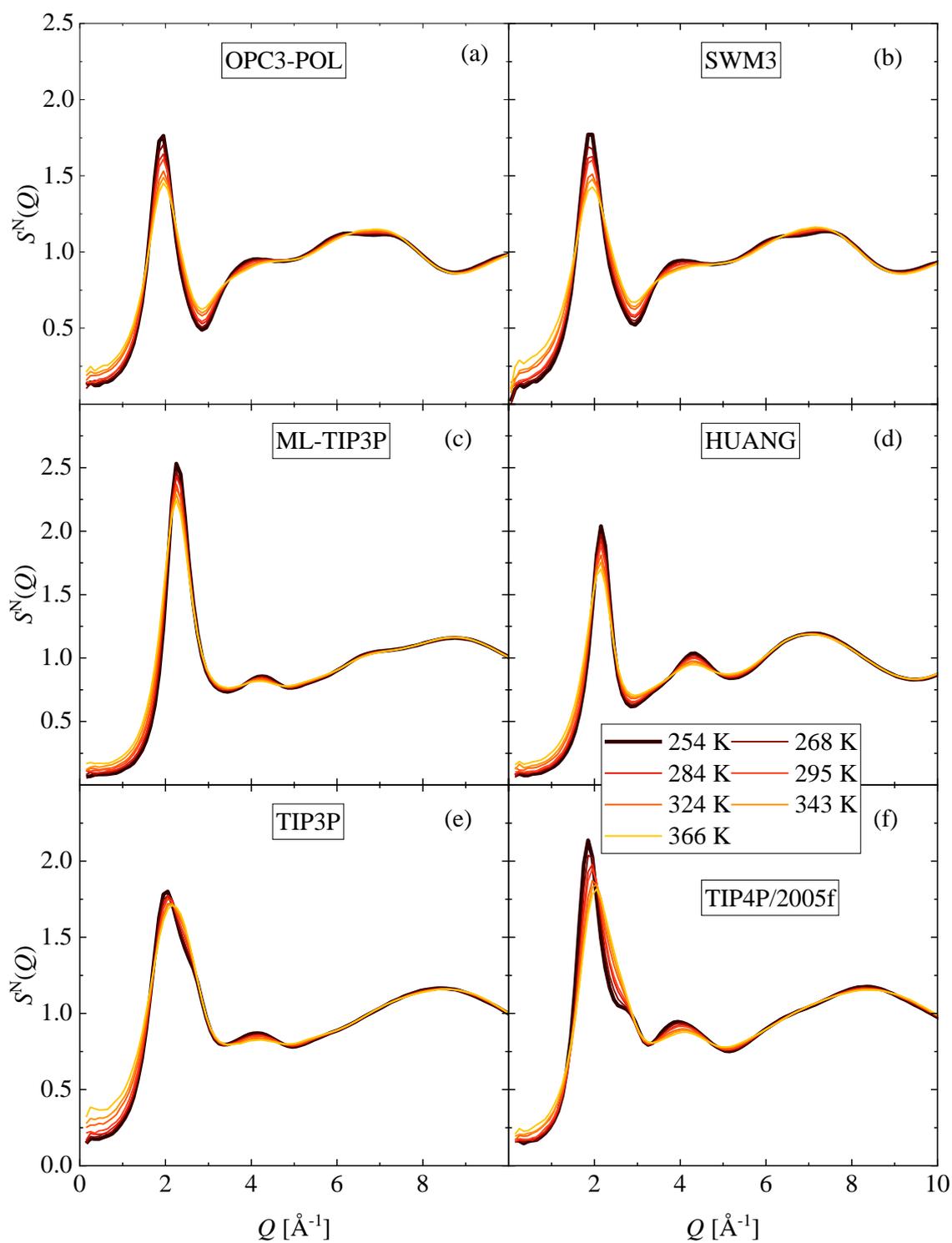

**Figure S15** Temperature dependence of simulated ND total structure factor for models (a) OPC3-POL, (b) SWM3, (c) ML-TIP3P, (d) HUANG, (e) TIP3P, and (f) TIP4P/2005f models.

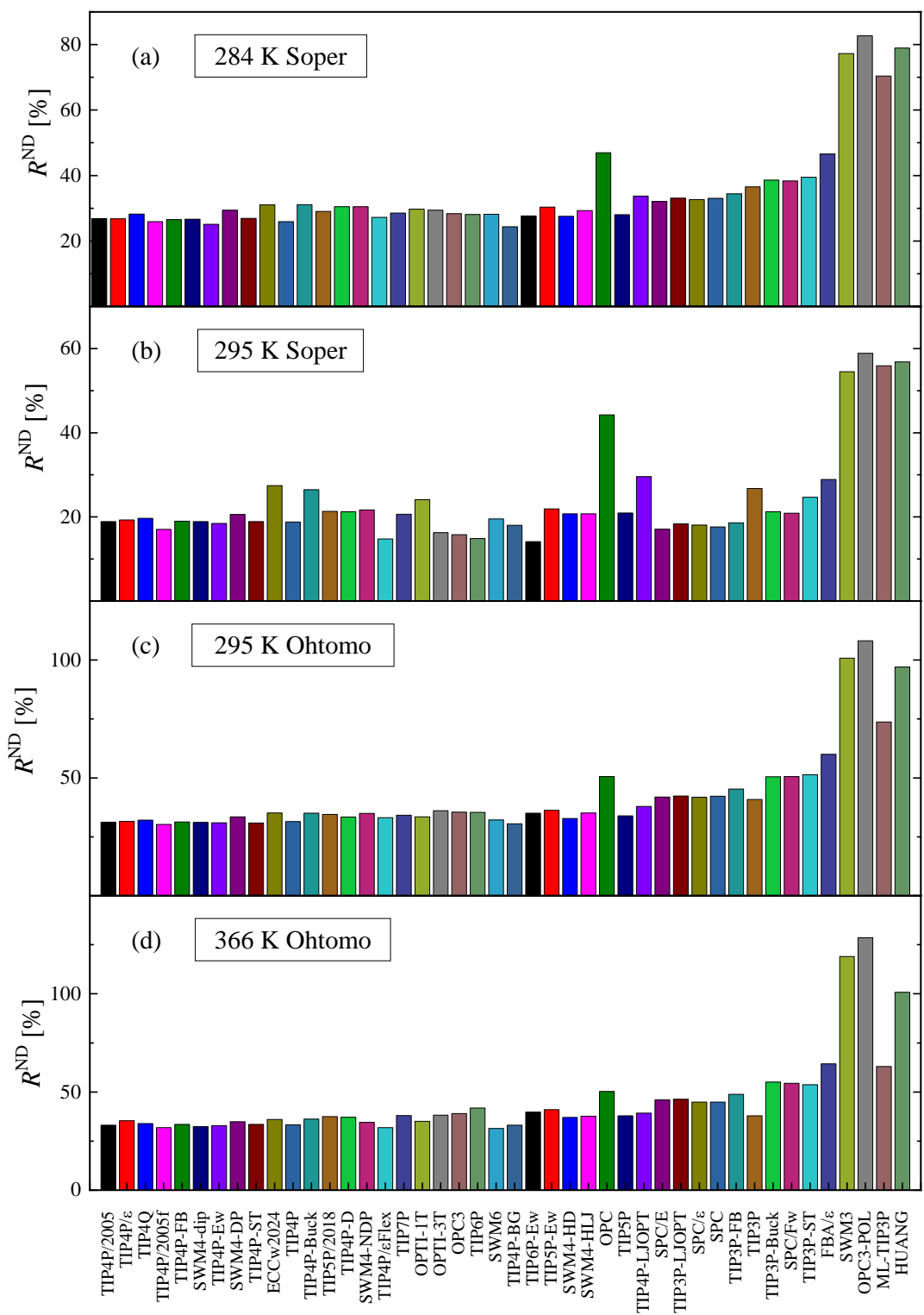

**Figure S16** *R*-factors of the ND diffraction total structure factors at different temperatures, obtained by using different models. (For the definition of *R*-factor, see text.)

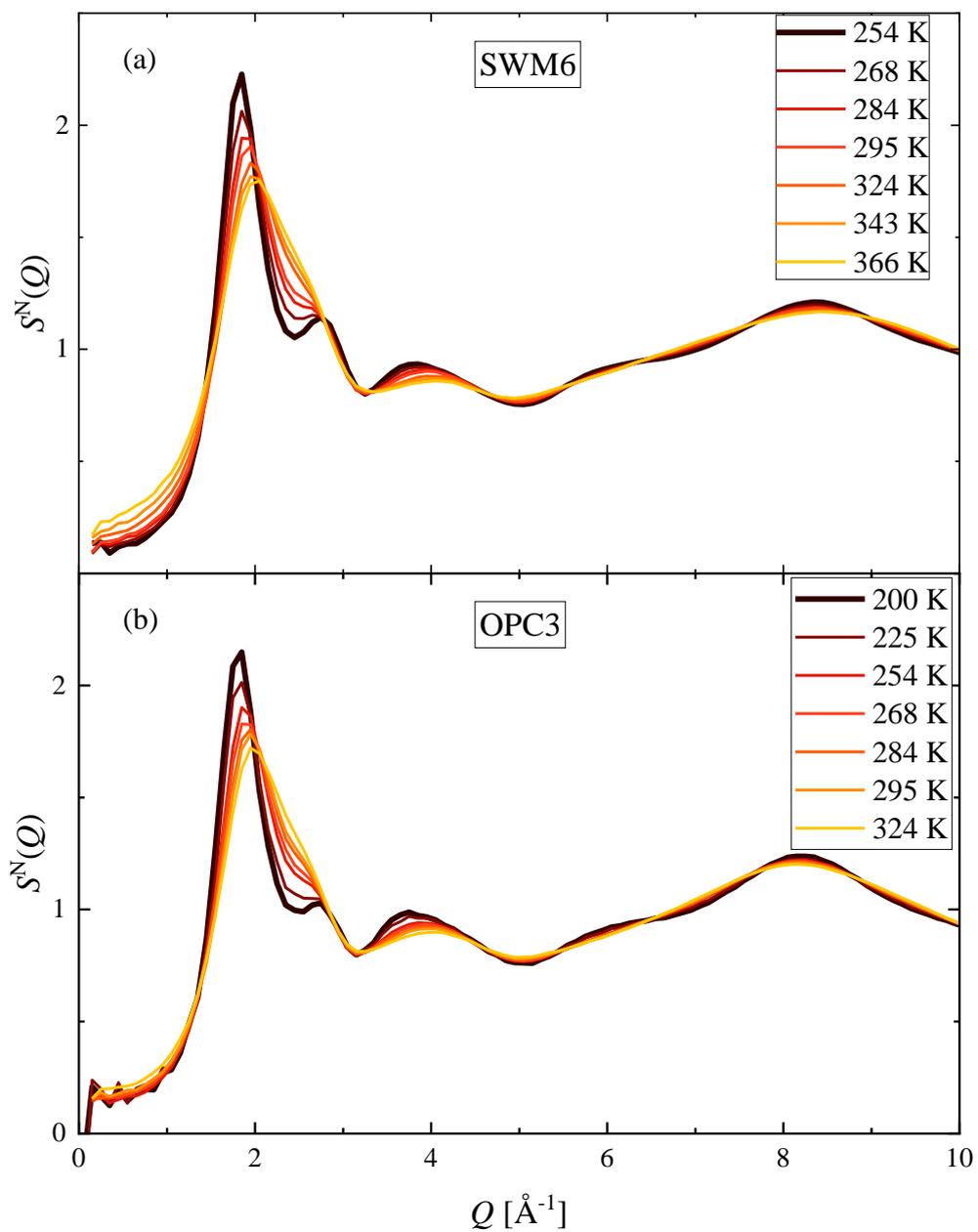

**Figure S17** Comparison of the ND structure factors of (a) SWM6 and (b) OPC3 models. The curves of the OPC3 model are shown at lower temperatures.

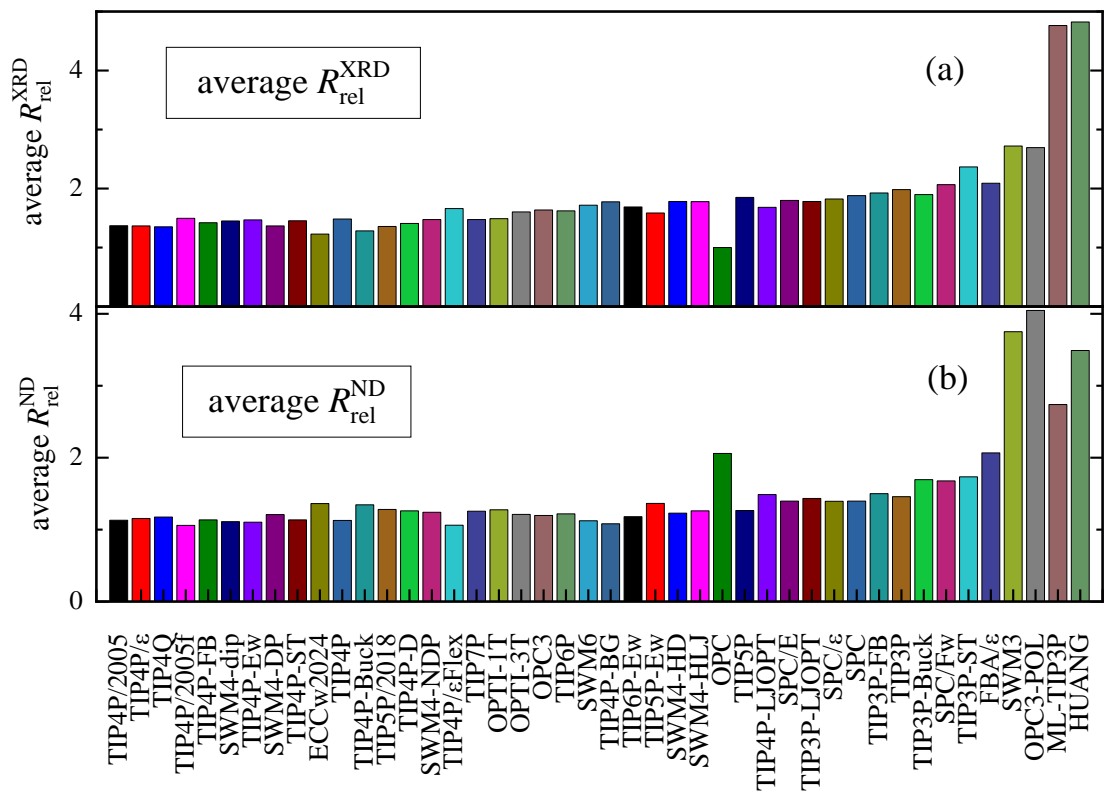

**Figure S18** Average of the relative *R*-factors of the (a) 7 XRD, (b) 5 ND (4 Ohtomo's data and 295 K Soper's data) fits.

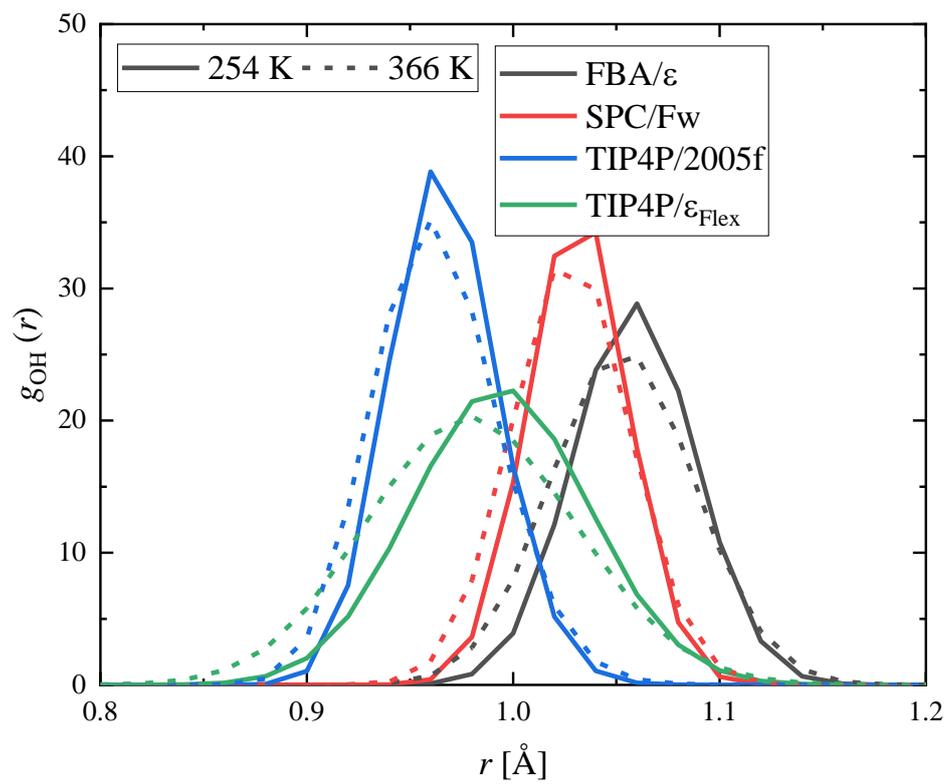

**Figure S19** Intramolecular O-H partial pair distribution functions of the flexible models at the (solid line) lowest and (dashed line) highest investigated temperatures.

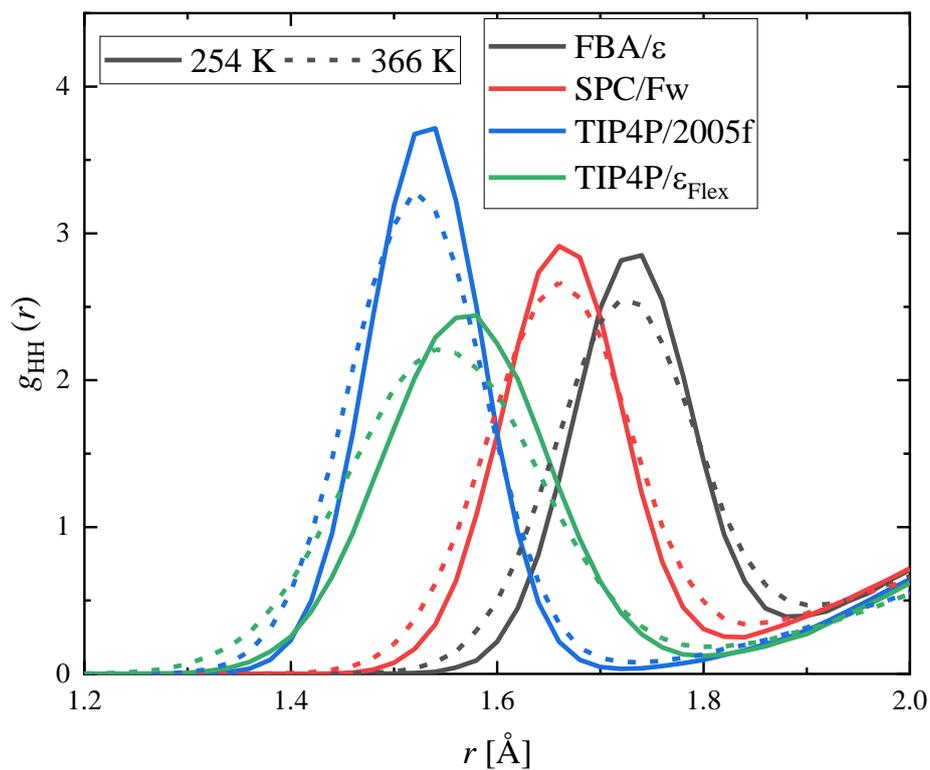

**Figure S20** Intramolecular H-H partial pair distribution functions of the flexible models at the (solid line) lowest and (dashed line) highest investigated temperatures. (The low *r* part of the intermolecular H-H PRDFs are also shown.)

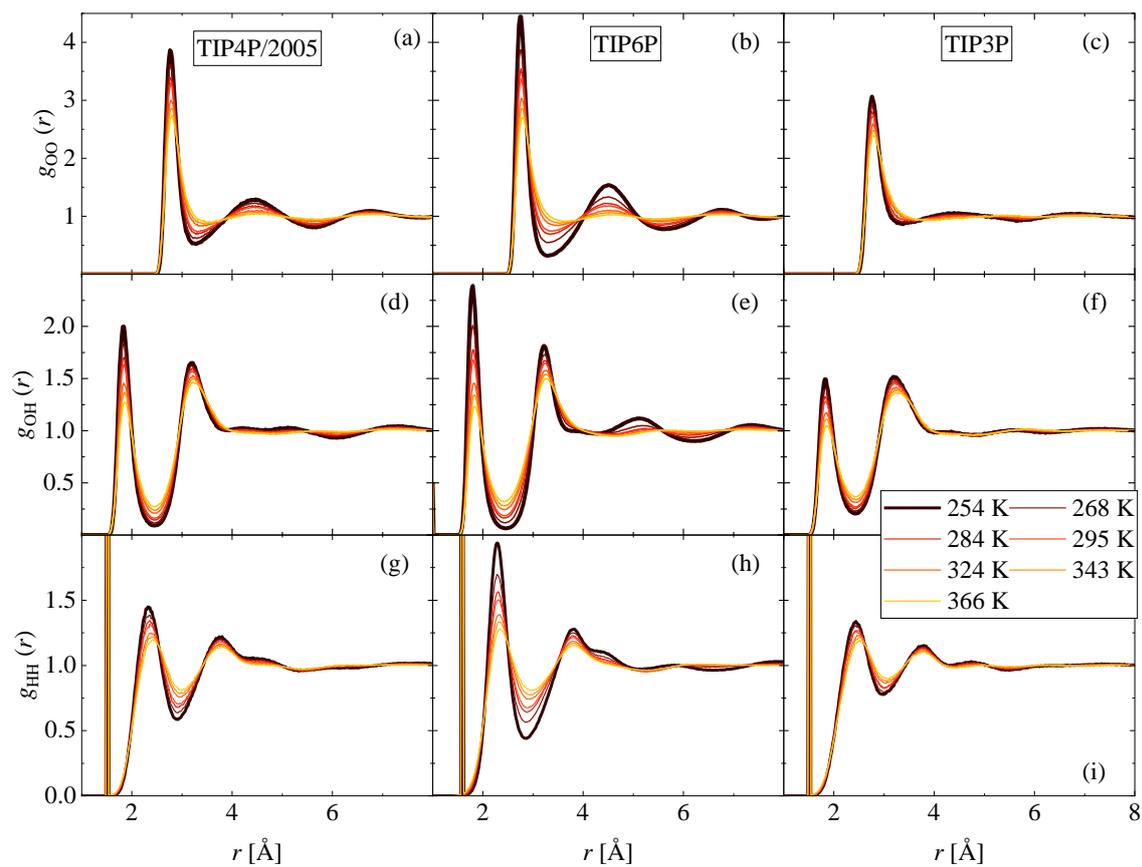

**Figure S21** Partial pair distribution functions at different temperatures obtained by using the (a, d, g) TIP4P/2005, (b, e, h) TIP6P, and (c, f, i) TIP3P models.

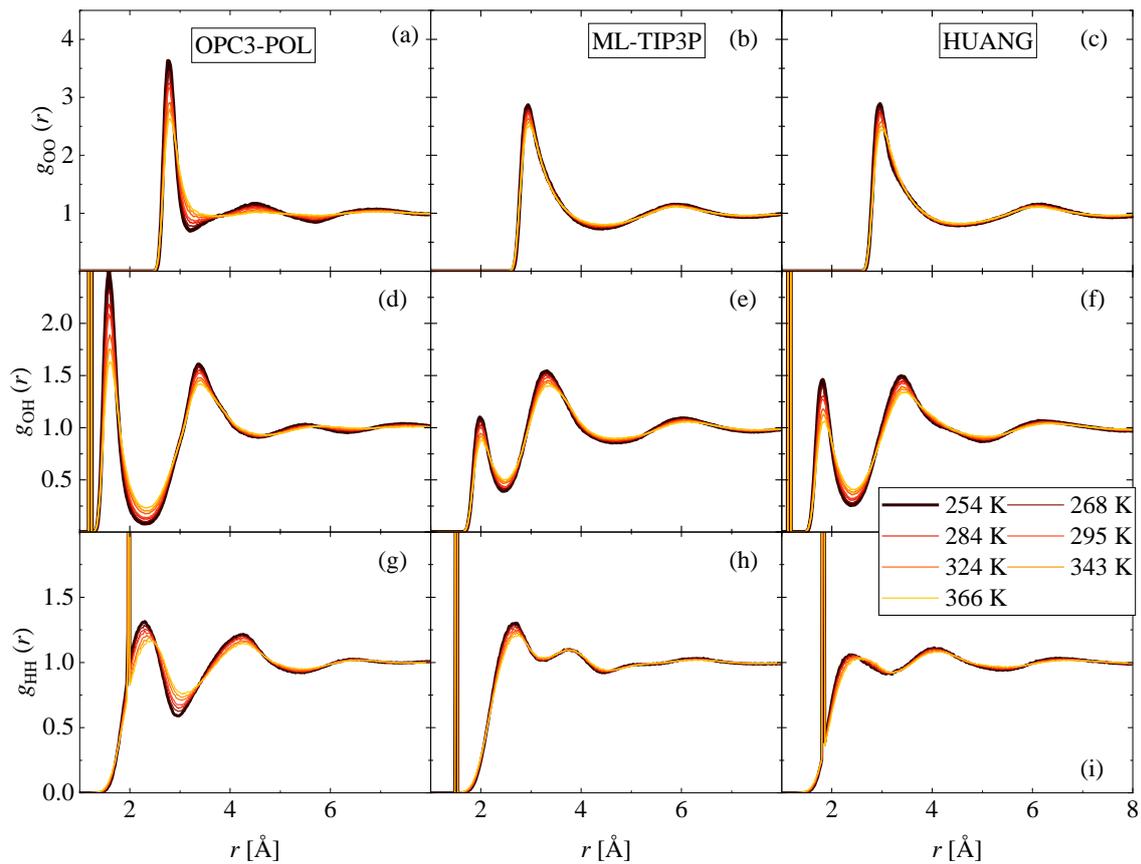

**Figure S22** Partial pair distribution functions at different temperatures obtained by using the (a, d, g) OPC3-POL, (b, e, h) ML-TIP3P, and (c, f, i) HUANG models.

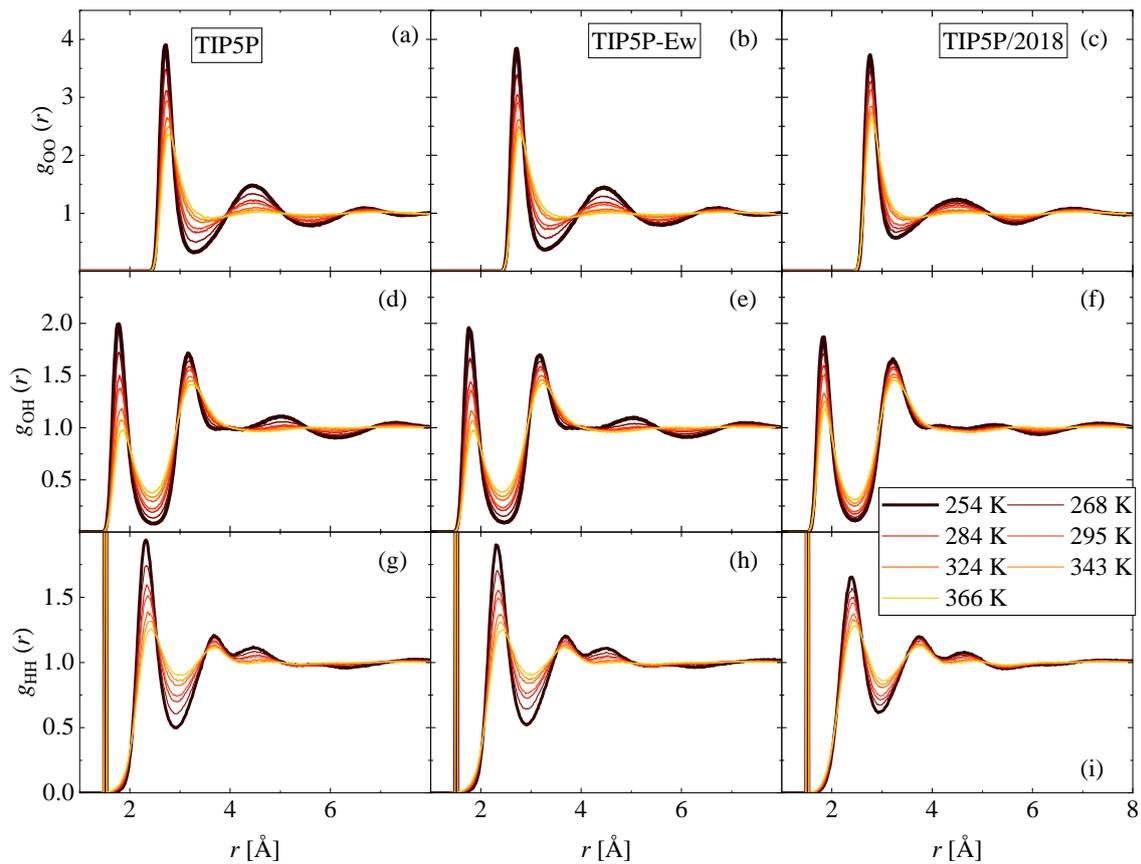

**Figure S23** Partial pair distribution functions at different temperatures obtained by using the (a, d, g) TIP5P, (b, e, h) TIP5P-Ew, and (c, f, i) TIP5P/2018 models.

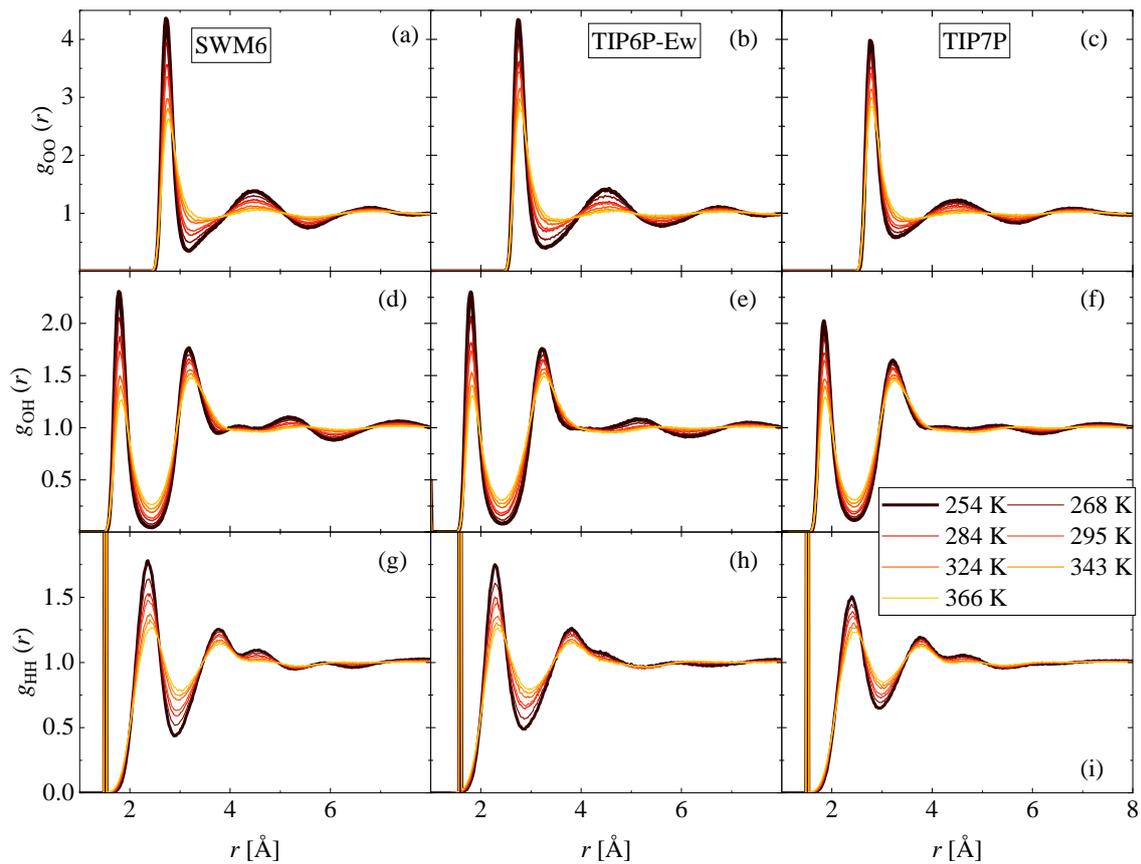

**Figure S24** Partial pair distribution functions at different temperatures obtained by using the (a, d, g) SWM6, (b, e, h) TIP6P-Ew, and (c, f, i) TIP7P models.

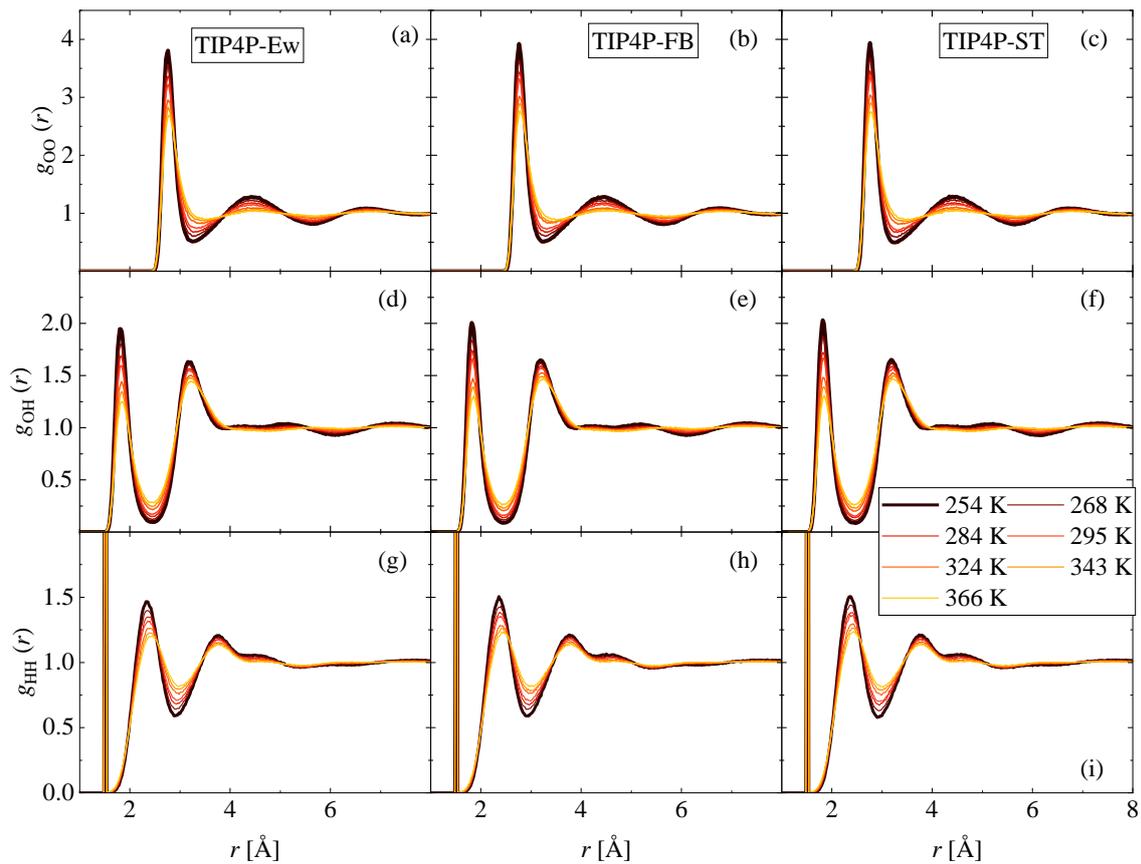

**Figure S25** Partial pair distribution functions at different temperatures obtained by using the (a, d, g) TIP4P-Ew, (b, e, h) TIP4P-FB, and (c, f, i) TIP4P-ST models.

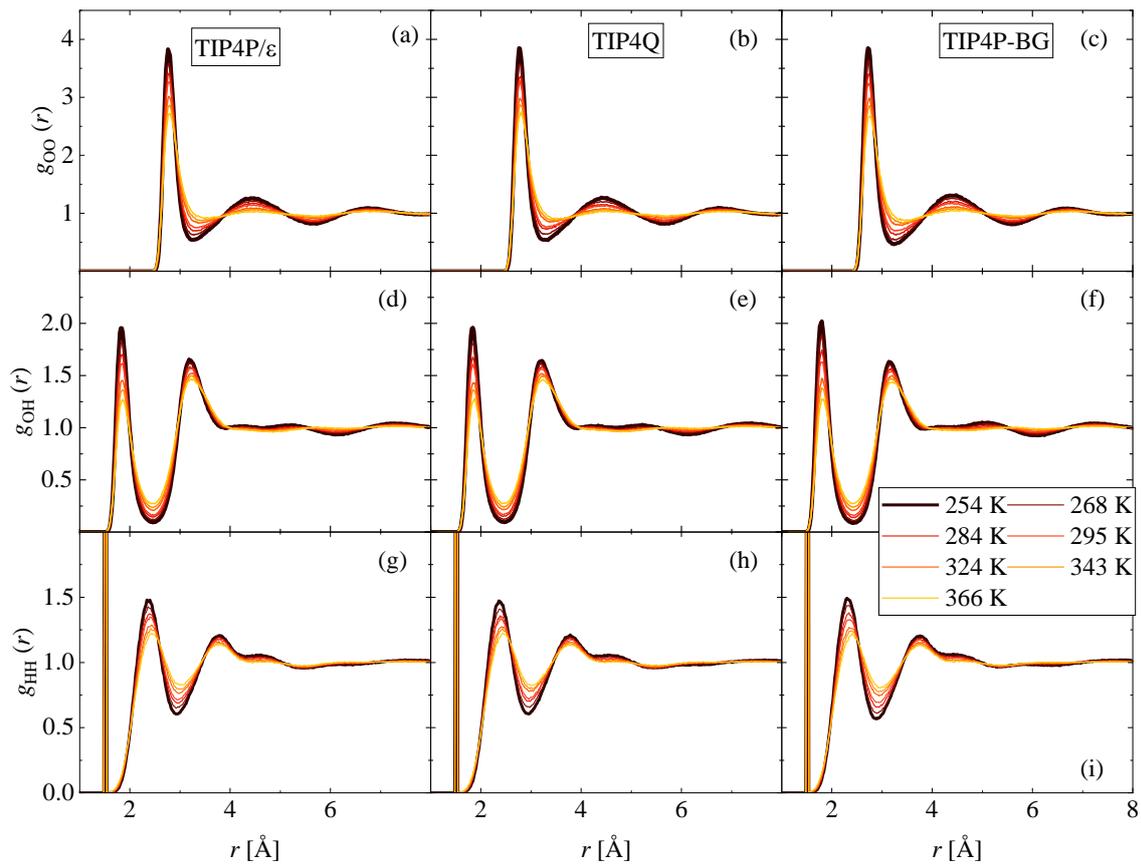

**Figure S26** Partial pair distribution functions at different temperatures obtained by using the (a, d, g) TIP4P/ε, (b, e, h) TIP4Q, and (c, f, i) TIP4P-BG models.

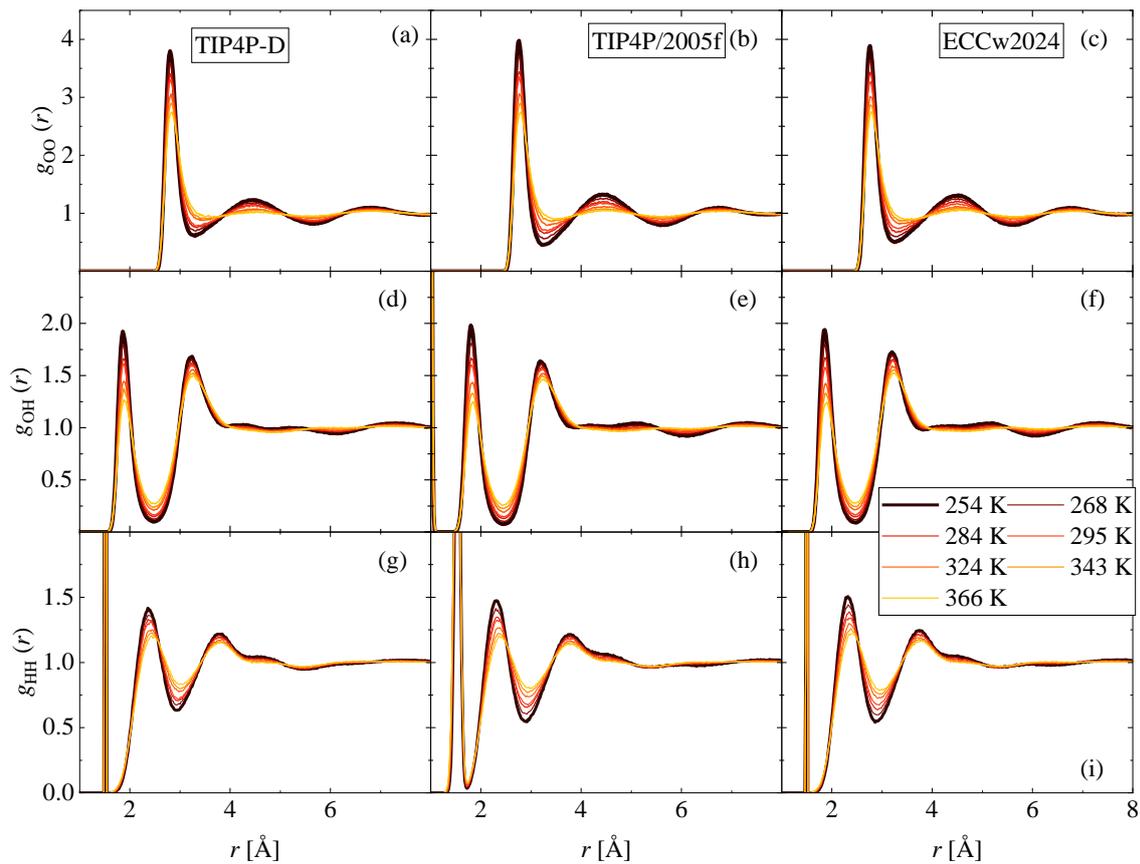

**Figure S27** Partial pair distribution functions at different temperatures obtained by using the (a, d, g) TIP4P-D, (b, e, h) TIP4P/2005f, and (c, f, i) ECCw2024 models.

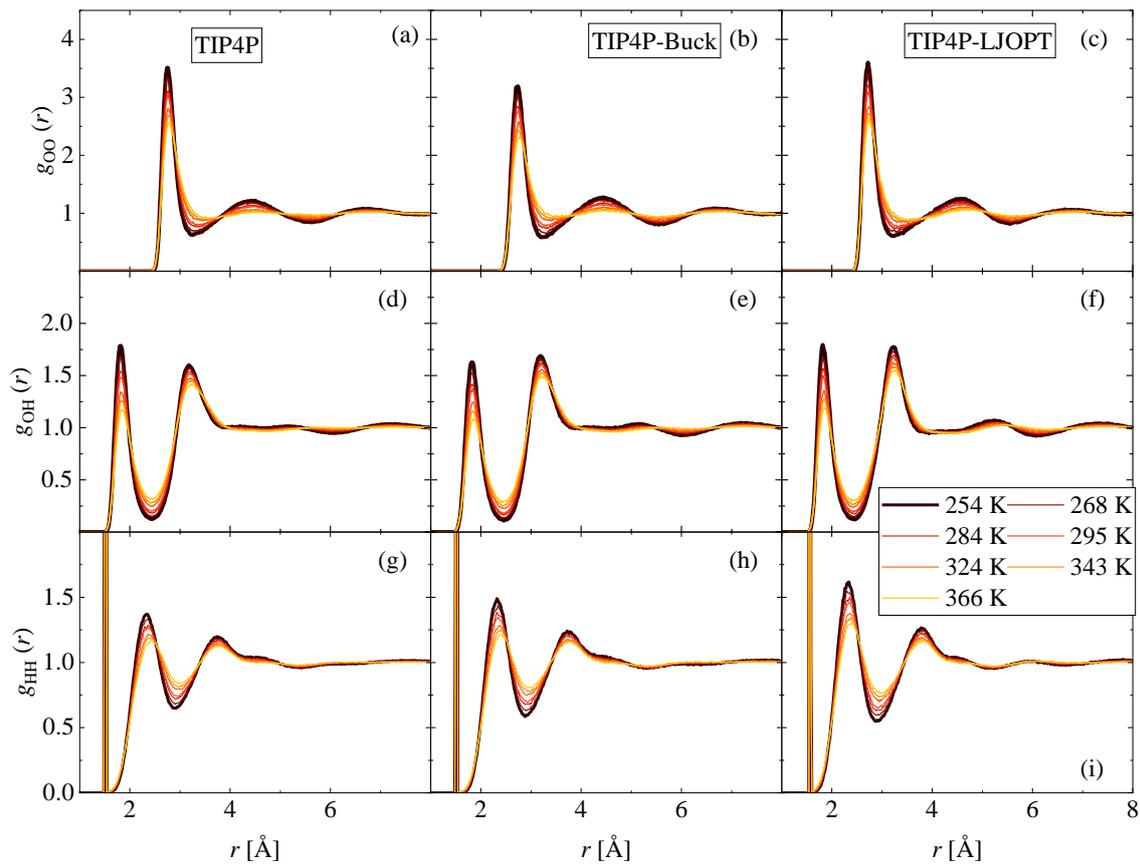

**Figure S28** Partial pair distribution functions at different temperatures obtained by using the (a, d, g) TIP4P, (b, e, h) TIP4P-Buck, and (c, f, i) TIP4P-LJOPT models.

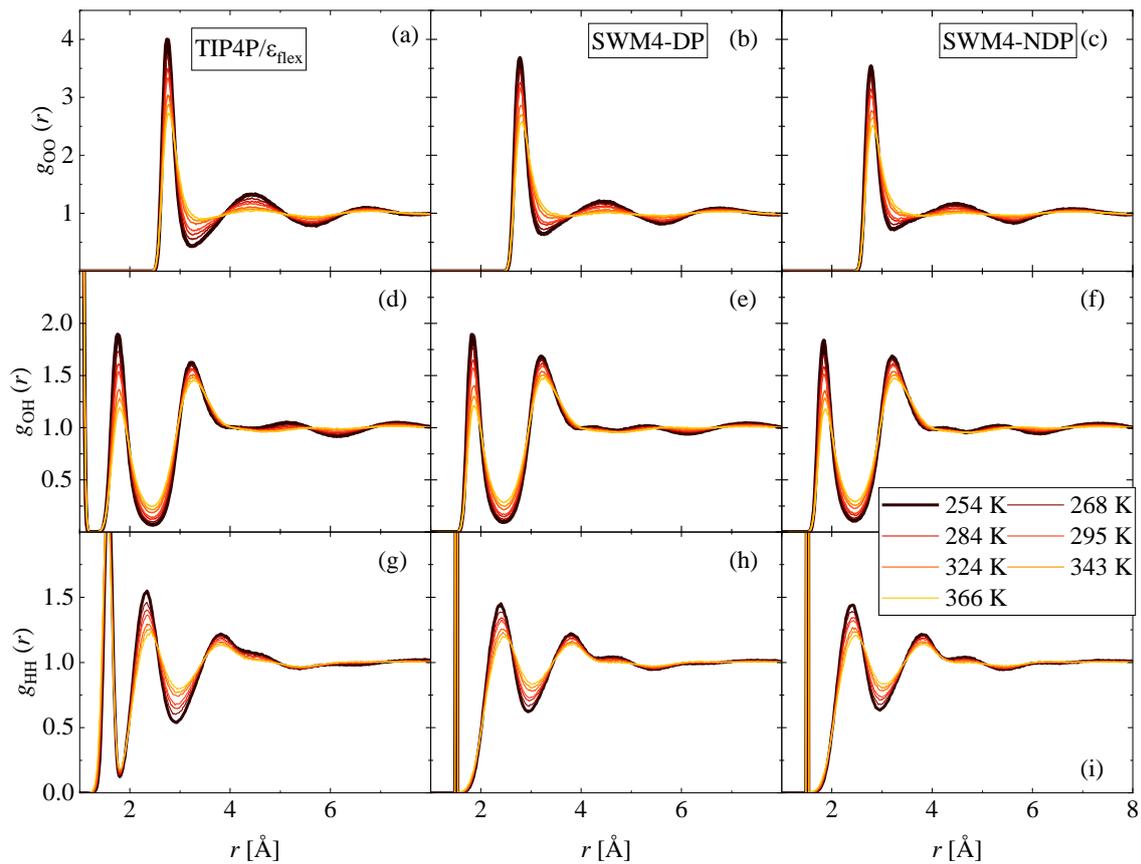

**Figure S29** Partial pair distribution functions at different temperatures obtained by using the (a, d, g) TIP4P/$\varepsilon_{flex}$, (b, e, h) SWM4-DP, and (c, f, i) SWM4-NDP models.

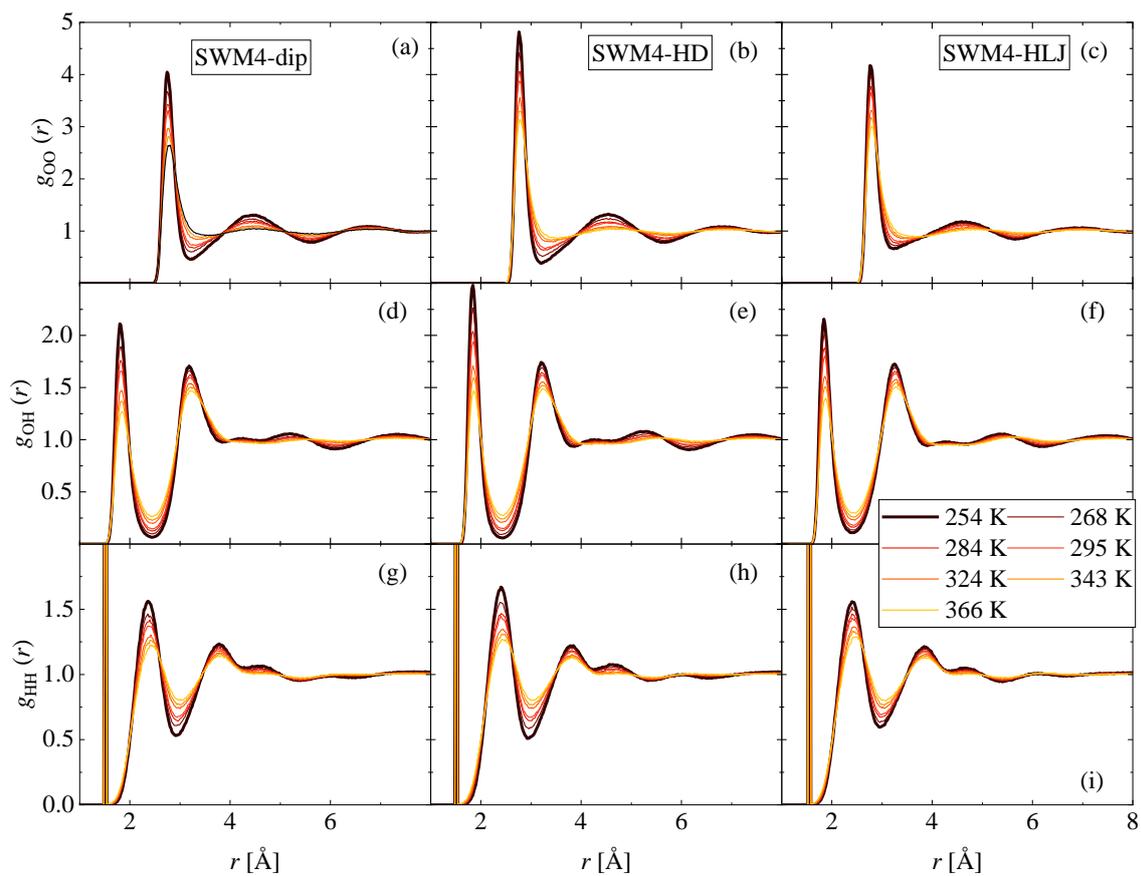

**Figure S30** Partial pair distribution functions at different temperatures obtained by using the (a, d, g) SWM4-dip, (b, e, h) SWM4-HD, and (c, f, i) SWM4-HLJ models.

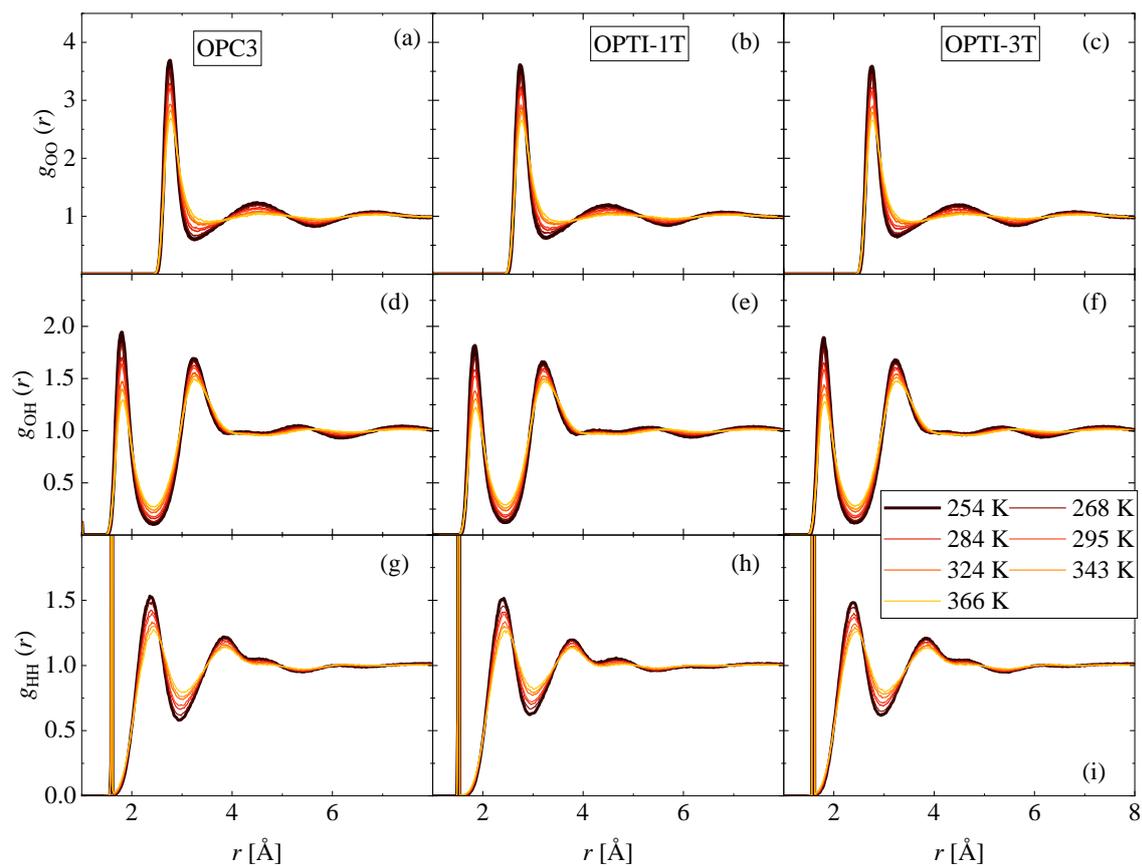

**Figure S31** Partial pair distribution functions at different temperatures obtained by using the (a, d, g) OPC3, (b, e, h) OPTI-1T, and (c, f, i) OPTI-3T models.

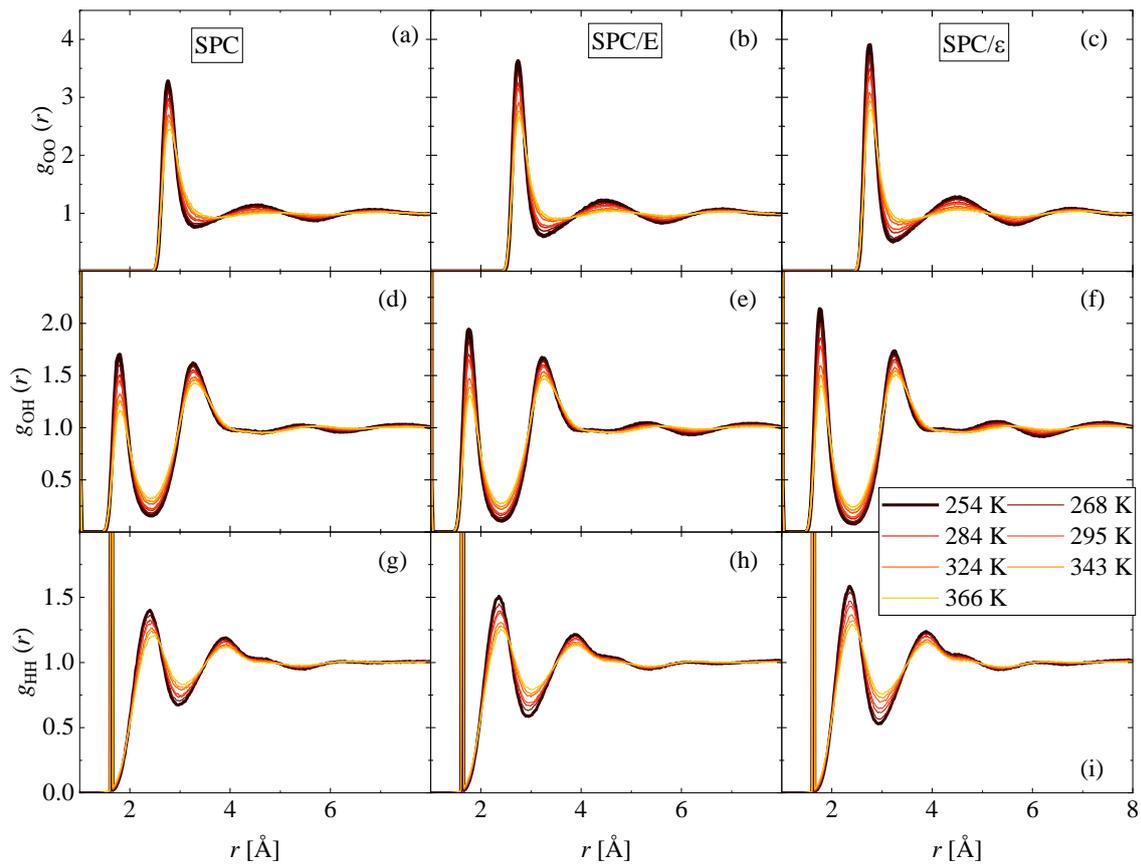

**Figure S32** Partial pair distribution functions at different temperatures obtained by using the (a, d, g) SPC, (b, e, h) SPC/E, and (c, f, i) SPC/ε models.

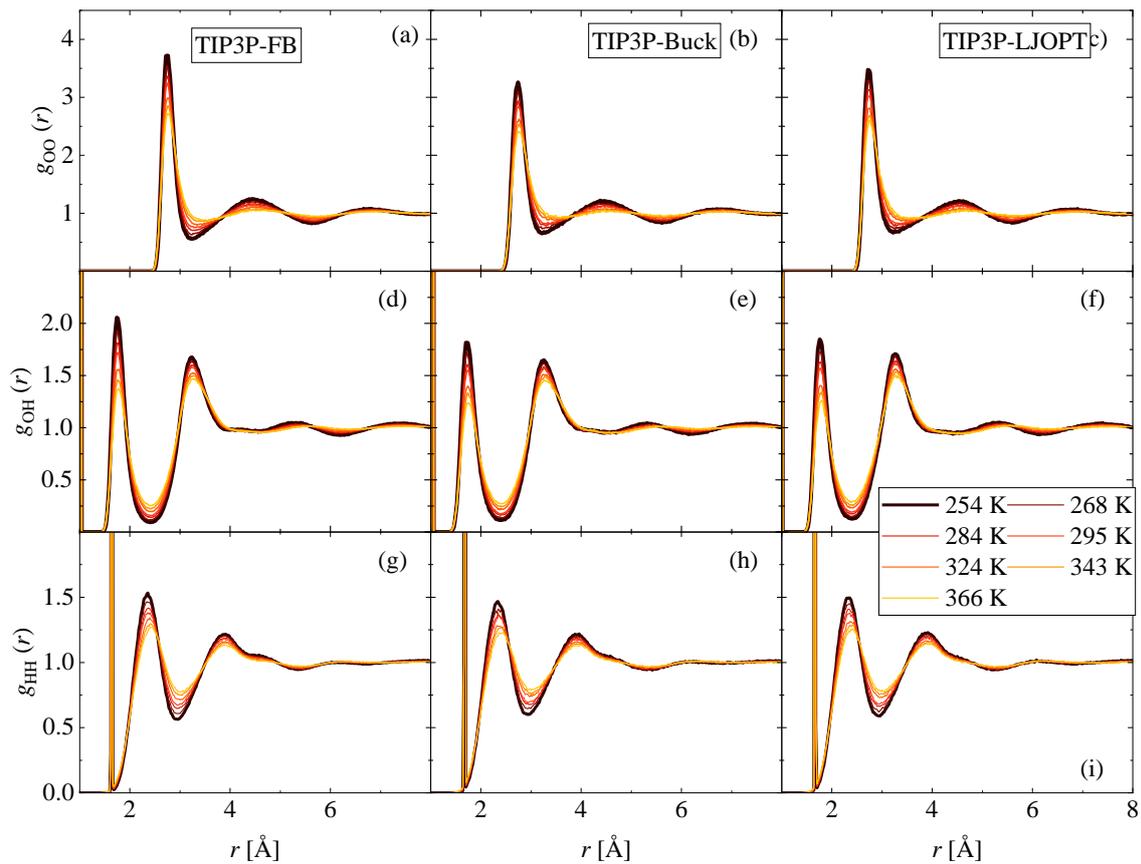

**Figure S33** Partial pair distribution functions at different temperatures obtained by using the (a, d, g) TIP3P-FB, (b, e, h) TIP3P-Buck, and (c, f, i) TIP3P-LJOPT models.

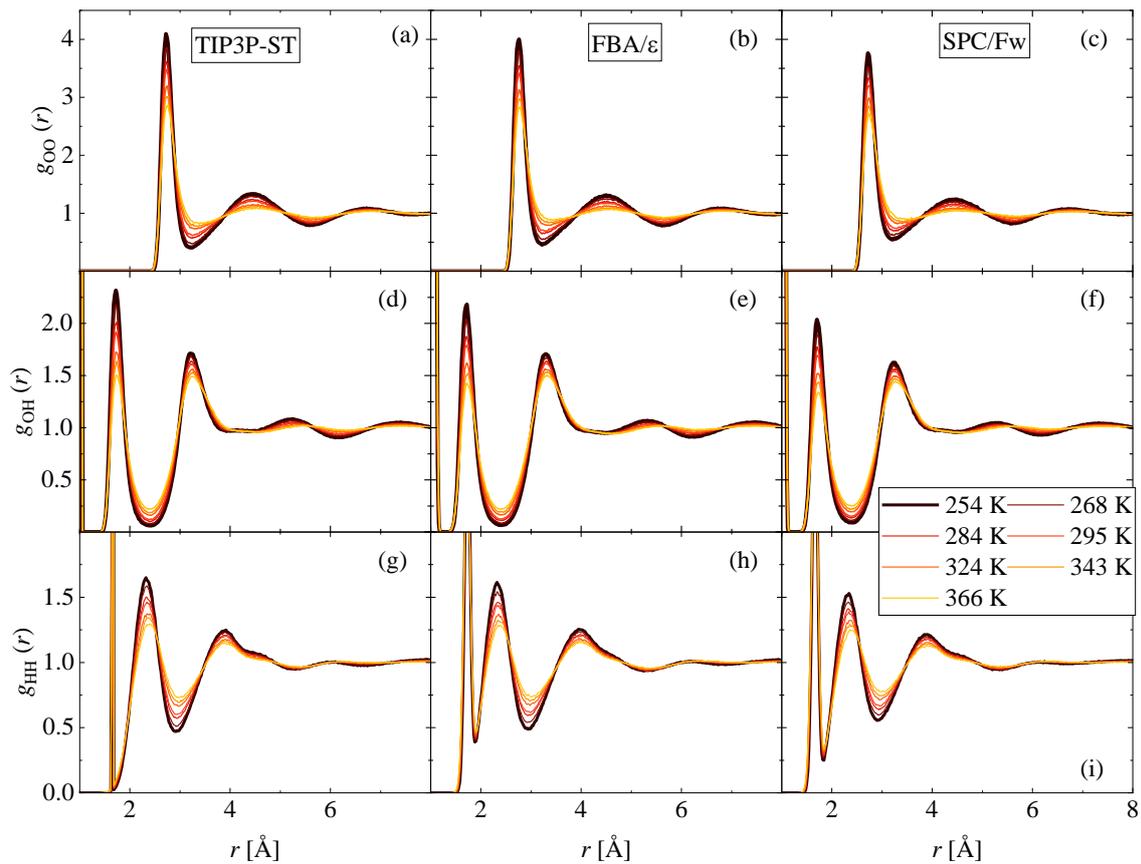

**Figure S34** Partial pair distribution functions at different temperatures obtained by using the (a, d, g) TIP3P-ST, (b, e, h) FBA/ε, and (c, f, i) SPC/Fw models.

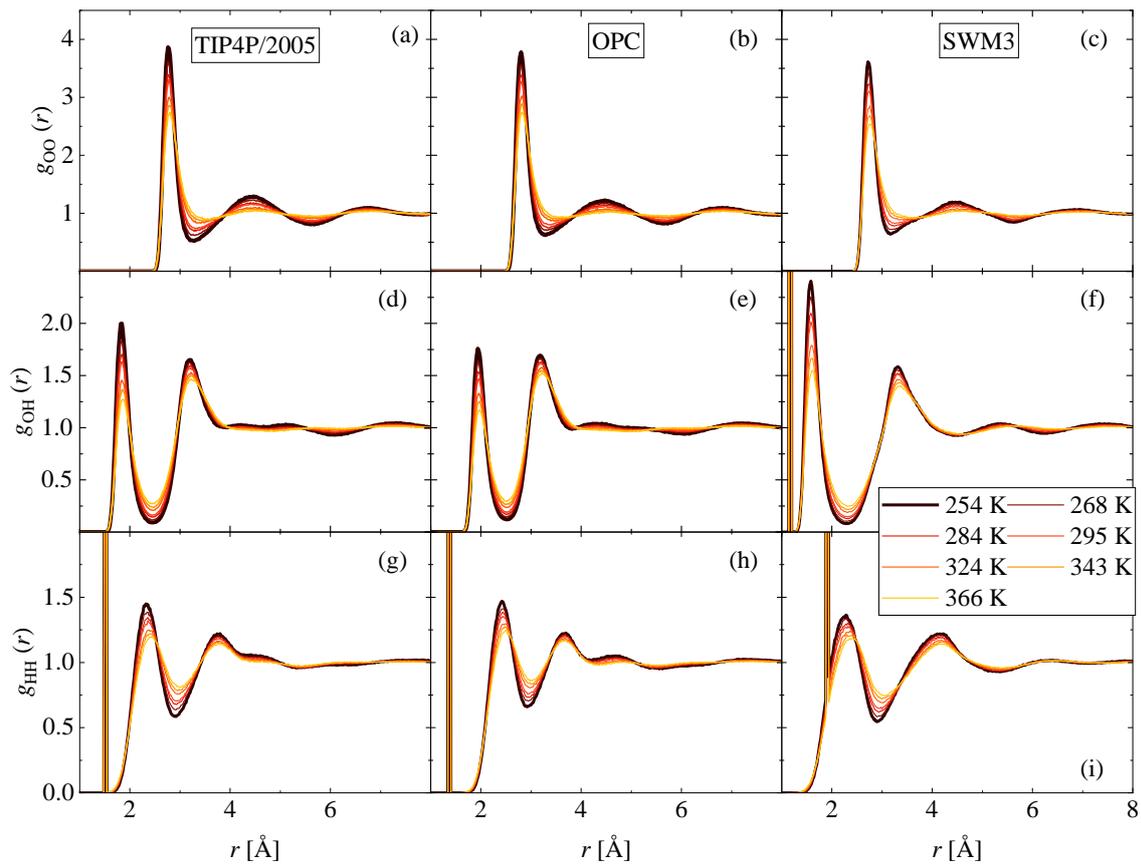

**Figure S35** Partial pair distribution functions at different temperatures obtained by using the (a, d, g) TIP4P/2005, (b, e, h) OPC, and (c, f, i) SWM3 models.

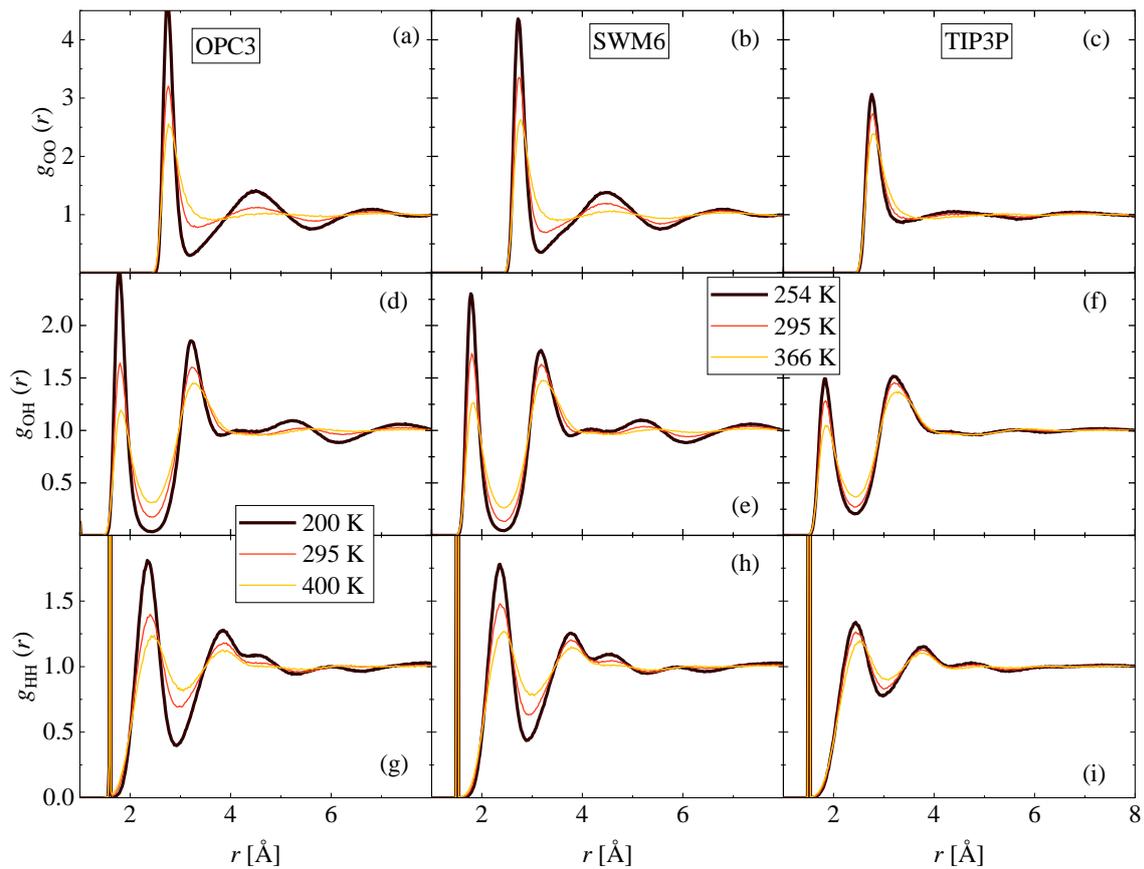

**Figure S36** Partial pair distribution functions at different temperatures obtained by using the (a, d, g) OPC3, (b, e, h) SWM6, and (c, f, i) TIP3P models. The temperature range for OPC3 is broader!

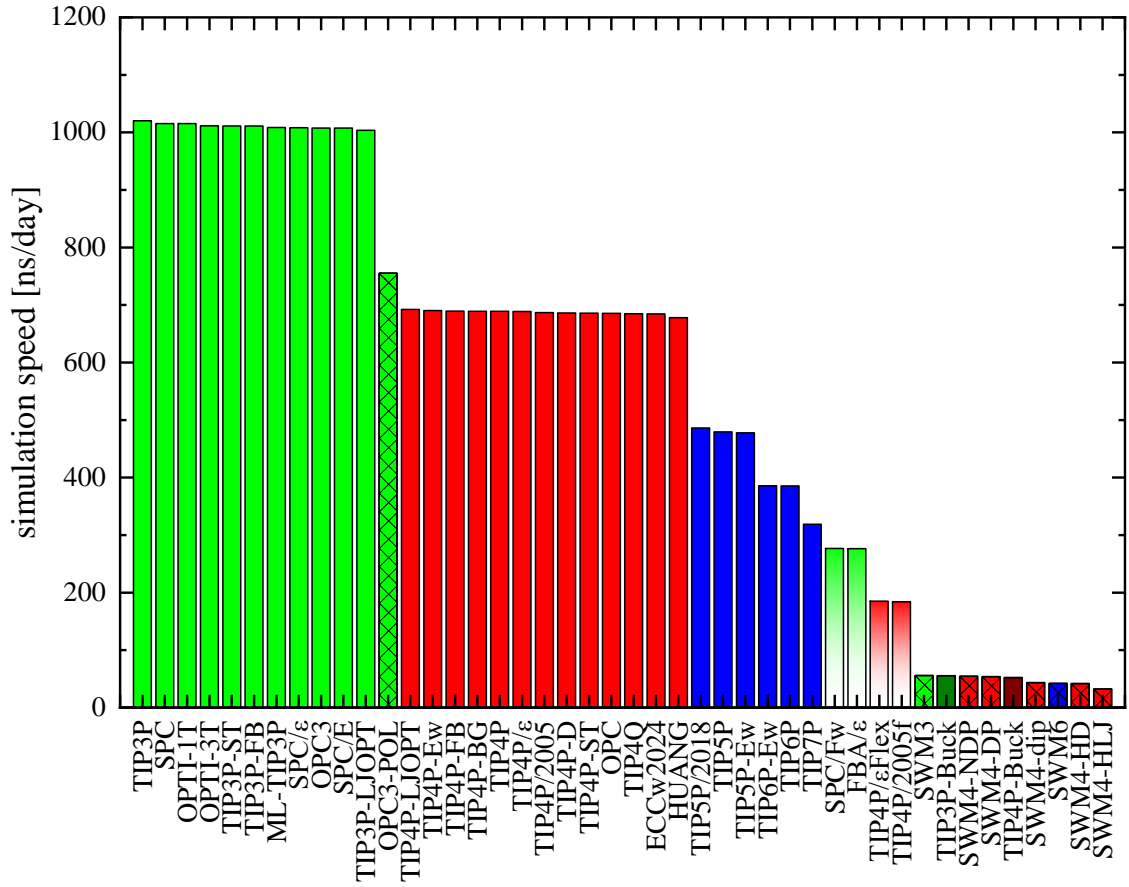

**Figure S37** Simulation speed of the different models.

**References**

[1] G.S. Kell, Density, thermal expansivity, and compressibility of liquid water from 0° to 150°C: Correlations and tables for atmospheric pressure and saturation reviewed and expressed on 1968 temperature scale, J. Chem. Eng. Data 20 (1975) 97. doi:10.1021/je60064a005.

[2] D.E. Hare, C.M. Sorensen, The density of supercooled water. II. Bulk samples cooled to the homogeneous nucleation limit, J. Chem. Phys. 87 (1987) 4840–4845. doi:10.1063/1.453710.

[3] M. Holz, S.R. Heil, A. Sacco, Temperature-dependent self-diffusion coefficients of water and six selected molecular liquids for calibration in accurate 1H NMR PFG measurements, Phys. Chem. Chem. Phys. 2 (2000) 4740–4742. doi:10.1039/b005319h.